\documentclass{aa}  

\usepackage{graphicx}
\usepackage{txfonts}
\usepackage{subcaption} 
\usepackage{siunitx}

\usepackage{lscape}             

\usepackage{placeins}           

\usepackage{bm}
\usepackage{mathrsfs} 
\usepackage{bbold}    

\usepackage{hyperref} 

\usepackage{booktabs}

\usepackage{xcolor}
\usepackage{tikz}
\usetikzlibrary{shapes.geometric,arrows.meta,positioning,fit,backgrounds,calc}
\definecolor{myblue}{RGB}{30,105,220}
\definecolor{procfill}{RGB}{235,244,252}
\definecolor{iofill}{RGB}{255,244,230}
\definecolor{outfill}{RGB}{233,247,235}
\definecolor{wrapfill}{RGB}{247,250,253}
\definecolor{rcolfill}{RGB}{240,247,253}
\definecolor{sideln}{RGB}{90,90,90}

\begin{document}

    \title{Strong Lensing in a non-homogeneous Universe}
   \subtitle{I. Lensing effective distance}
   
   \author{D.~ Bento\inst{1,2}\corrauth{danielbento165@gmail.com} 
   \and I.~Tereno\inst{1,2}
        \and A.~da Silva\inst{1,2}
        \and J.~P. Mimoso\inst{1,2}
        }

   \institute{Departamento de F\'isica, Faculdade de Ci\^encias, Universidade de Lisboa, Edif\'icio C8, Campo Grande, PT1749-016 Lisboa,
Portugal
   \and Instituto de Astrof\'isica e Ci\^encias do Espa\c{c}o, Faculdade de Ci\^encias, Universidade de Lisboa, Campo Grande, PT-1749-016 Lisboa, Portugal}

   \date{Received November 30, 2026}

\abstract{
Strong gravitational lensing is facing a surge of data from space-based high-resolution wide-field observatories, such as {\it Euclid} and {\it Roman}. 
Among a plethora of applications enabled by the current and near-future data is the test of the cosmological principle, one of the fundamental assumptions of the standard cosmological paradigm. We aim to investigate imprints from the non-homogeneous Universe in the Einstein radius strong lensing observable, for the case of a Lema\^{\i}tre--Tolman--Bondi non-homogeneous background with a cosmological constant (a $\Lambda$LTB model).
We develop a strong lensing simulator framework in which the angular diameter distances, including the lens–source distance $D_{\rm LS}$, are derived by integrating the Sachs equations, and apply it for a $\Lambda$LTB model with an observer placed at the centre of the $\Lambda$LTB underdensity.
We find the imprint of the model in the Einstein radius is driven by the lens–source distance $D_{\rm LS}$, which, unlike the distances from the observer, retains a deviation from its homogeneous counterpart, even well-beyond the underdensity radius. We find the optimal source and lens redshifts that maximise the constraining power of the observable and find there is potential to disentangle the models at the precision-level achieved by the space-based data, subject to the introduction of independent constraints on the lens mass.}


   \maketitle

   \nolinenumbers


\section{Introduction}\label{Intro}

Over the years, the standard cosmological paradigm known as the $\Lambda$CDM model, built on the Friedmann--Lema\^{\i}tre--Robertson--Walker (FLRW) solutions of Einstein’s equations under the cosmological principle, has successfully described a variety of cosmological data. It provides a model for the accelerated expansion of the Universe \citep{Riess_1998, Perlmutter_1998}, predicts the observed baryon acoustic oscillation (BAO) scale \citep{Eisenstein_2005}, the primordial abundances of light elements \citep{Cyburt_2016}, and is a remarkably good fit to various cosmic microwave background (CMB) observables \citep{Planck_2020}. Despite all this success, the increased precision of observational data has highlighted statistically significant tensions between results from different cosmological probes when assuming $\Lambda$CDM \citep{Abdalla_2022},  most notably the Hubble tension between the value of the Hubble constant measured using local distance ladder measurements and the one inferred from the CMB \citep{Riess_2016, Riess_2019, Riess_2022}.  Such tensions have motivated the search for alternatives to the standard paradigm \citep[see][for a review]{DiValentino_2021}. Besides the tensions between datasets, the unprecedented amount of high-precision cosmological data, together with 
substantial efforts to understand the associated systematics, made possible with stage-IV dark energy imaging experiments, \textit{Euclid} \citep{Mellier_2025}, \textit{Roman} \citep{rotac} and \textit{Rubin's} LSST \citep{lsst} opens the possibility to explore other aspects of the standard paradigm, including the reassessment of fundamental assumptions about the geometry of spacetime, theory of gravity, and initial conditions, as well as on the nature of dark matter and dark energy. 

Strong gravitational lensing is a valuable tool in astrophysics. The phenomenon arises when a foreground massive object (the lens) creates a differential deflection of the light from a background extended object (the source) producing a distorted image of the source showing arcs or a complete Einstein ring, or when the lens produces multiple images of a point source \citep[see][for an introduction]{Saha_2024}. On galactic scales, strong lensing has been used as a robust method of measuring the total mass and mass profile of galaxies \citep{Kochanek_2006, Treu_2010}, usually applied in combination with dynamical information to break the fundamental mass-sheet degeneracy \citep{Falco_1985}. Since the deflection of light by a foreground mass depends on the angular diameter distances between observer, lens, and source \citep{Bartelmann_2010}, strong lensing also probes the cosmological scenario \citep[see][for a review of recent results and applications]{Treu_2026}. Time delays in multiple images, due to the path length difference and the different gravitational potentials the photons travel through, are used to measure the Hubble constant independently of both the local distance ladder and the early-universe physics \citep{Birrer_2024}. The Einstein radius, being a function of the lens mass and the cosmological distances, is also a cosmological probe. It has been used to constrain cosmological parameters from measurements of one system combined with stellar dynamics data of the lens \citep [e.g.][]{Grillo_2008}, and also combining several lensing systems \citep{Cao_2015}. In order to build a mass-independent estimator, one would ideally need to construct a ratio of Einstein radii from a compound lens system, that is lensing from sources at two or more redshifts lensed by the same lens \citep[as done in][with the first discovered compound system, see also \cite{CollettA_2014}]{Gavazzi_2008}, which nevertheless still suffers from a generalised mass-sheet degeneracy \citep{Schneider_2014}.

In this paper, we investigate the potential of the Einstein radius observable of strong lensing to test the Copernican principle. The Copernican principle \citep{Marra_2011}, the hypothesis that we are not in a special location in the Universe, together with the fact that the Universe appears to be statistically isotropic, leads to the cosmological principle, one of the fundamental assumptions of the standard cosmological paradigm. An extensively used inhomogeneous model \citep[see][for a review of models]{Bolejko_2011} stems from the Lema\^{\i}tre--Tolman--Bondi (LTB) solution \citep{Lemaitre:1933gd, Tolman:1934pn, Bondi_1947}, an exact solution to Einstein’s equations, spherically symmetric and radially inhomogeneous. Combined with a cosmological constant $\Lambda$, it defines the $\Lambda$LTB model, an extension of $\Lambda$CDM with an underdensity embedded in a $\Lambda$CDM background 
\citep{Marra_2010, Redlich_2014}. The underdensity is modelled with two additional parameters, the density contrast at the centre of the inhomogeneity, $\delta_0$, and its size given by the redshift of the inhomogeneity boundary, $z_{\rm B}$ \citep{Valkenburg_2012}.
Incidentally, the scenario of a local underdensity (a void) with a higher expansion rate has also been explored as a way to solve the
Hubble tension. \citet{Mazo_2022} found that a small local void improved the fit to low-redshift Pantheon supernovae when adopting an absolute magnitude consistent with the Planck-inferred $H_0$. Analyses combining a wider range of observations have found limited support for a local void as a solution to the tension \citep{Wu_2017, Kenworthy_2019, Castello_2022,Camarena2022Void}.

Signatures of the $\Lambda$LTB model have been explored with several observables, such as the kinetic Sunyaev--Zel'dovich effect \citep{Bull_2012}, the Compton $y$-distortion of the CMB \citep{Caldwell_2008}, the BAO scales \citep{Zumalac_2012, Camarena_2025}, and galaxy cluster counts \citep{Sakr_2024},
also combined in a joint analysis \citep{Camarena_2021, Camarena_Euclid}. Gravitational lensing studies have focused on the weak regime \citep[see, e.g.,][]{Goto_2011, Fanizza_2015, Adam_2025}, but, to our knowledge, the $\Lambda$LTB universe imprints on strong lensing have not yet been investigated.  Strong lensing systems are rare, requiring an alignment between the observer, lens and source,  which occurs in less than $0.1\%$ of the observed galaxies \citep{Lines_2025}. The high-resolution and wide-field \textit{Euclid} instruments are already revolutionising the field, with the \textit{Euclid} Wide Survey expecting to detect $200\,000$ galaxy-scale strong lenses \citep{Mellier_2025} with a median lens redshift of $z_{\rm L}=0.6$ and the majority of sources in the range $1 < z_{\rm S} < 2$, implying a typical Einstein radius of $\ang{;;0.5}$ \citep{Collett_2015} too small for ground-based resolution. \textit{Euclid} also expects to identify up to 2300 lensed quasars  and lensed supernovae that can be used for time-delay observables. \textit{Euclid}’s first quick data release (Q1) identified around 500 galaxy-scale strong lenses \citep{Euclid_Walmsley_2025} in 0.45\% of \textit{Euclid}’s final surveying area, with the number increasing to a consistent 15,000 \citep{Euclid_Lines_2026} in the first data release (DR1) corresponding to 14\% of the final area, including 84 compound lenses candidates. The identified lenses have a typical Einstein radius of 
$0.85^{+0.44}_{-0.24}\,{\rm arcsec}$.

The Einstein radius depends on the distance ratio $D_{\rm LS} / D_{\rm S}$, so that its practical application is restricted by our ability to model both quantities. In homogeneous models, the lens–source distance $D_{\rm LS}$  is computed combining the distances to the lens and to the source, through the comoving distance. However, in inhomogeneous models that relation does not hold, and the distances must be obtained by propagating the light beam through the optical equations of general relativity \citep{Fleury_2015}. In this work, we derive the system of equations that computes the angular diameter distances in $\Lambda$LTB from the Sachs equations, and implement a strong lensing image simulator that incorporates a distance integrator to compute all relevant distances including $D_{\rm LS}$. With this numerical framework, we search for imprints of the inhomogeneous $\Lambda$LTB  model in strong lensing simulated images. This first paper of a planned series is restricted to scenarios where the observer is at the centre of the inhomogeneity, in which case the observable is the Einstein ring. In this constrained scenario, the goal is not to find a distinctive signature of a breakdown of the Copernican principle, but rather to assess the amplitude of the deviation between the $\Lambda$LTB and $\Lambda$CDM predictions. We look for optimal combinations of lens and source redshifts that maximise this deviation, and find that the two models can be distinguished at the 7$\sigma$ level adopting a 2\% statistical uncertainty in the Einstein radius. From this result, it follows that measurements of galaxy masses from Einstein radii data that assume a $\Lambda$CDM cosmology with the cosmological parameters fixed at standard values, result in a 26\% mass overestimate in the case the Universe is $\Lambda$LTB. This also confirms that in the absence of a distinctive signature, independent knowledge of the mass of the lens is required to construct a decisive homogeneity test using the Einstein radius observable.

The paper proceeds with a description of the $\Lambda$LTB model, and the strong lensing formalism in Sects.~\ref{Framework} and \ref{GL}, respectively. Section~\ref{Formalismo_otico} introduces the problem of distances in non-homogeneous models, deriving the lensing effective distance and comparing it in the two models.  Section~\ref{ch:results} presents the image simulator and applies it to search for optimal lensing configurations, and to find the Einstein rings radii and flux differences. We conclude in Sect.~\ref{ch:conclusions} with a discussion of the results, and future prospects to apply our simulator to the scenario where the observer is no longer at the centre of the underdensity, and to time-delay observables.


\section{The $\Lambda$LTB model}\label{Framework}

We work within general relativity (GR), and the $\Lambda$LTB universe adopted here is an exact solution of Einstein’s equations. This late-time universe is filled with pressureless matter and a cosmological constant, and the observer is located at the centre of the spherically symmetric inhomogeneity. 

Adopting units with $c=1$ throughout, the LTB spacetime metric is given by \citep{Valkenburg_2012}:
\begin{equation}\label{metrica_ltb}
ds^2 = - dt^2 + \frac{R'^2(t,r)}{1-K(r)r^2}dr^2 + R^2(t,r) \left( d\theta^2 + \sin^2\theta\, d\phi^2 \right),
\end{equation}
where a prime denotes the partial derivative with respect to $r$, $K(r)$ is an arbitrary curvature function, and the function $R(t,r)$ generalises the FLRW product $a(t)\,r$ with $a(t)$ being the scale factor. Since both $R$ and $K$ depend on $r$, the LTB model is a generalisation of FLRW, which is recovered in the limit $R(t,r)\to a(t)r$ and $K(r)\to\text{const}$. This radial dependence does not prevent the definition of a cosmic time $t$ common to all comoving observers, but each spherical shell now has its own expansion history.

Consequently, the radial dependence leads us to decompose the scale factor into its transverse and longitudinal components, $a_\perp(t,r)=R/r$ and $a_\parallel(t,r)=R'$, both reducing to $a(t)$ in the FLRW limit. By construction, this decomposition extends to the Hubble expansion rate, which is defined with the respective transverse and longitudinal components:
\begin{equation}\label{H_paralelo}
H_{\perp}(t,r)\equiv \frac{\dot{a}_{\perp}}{a_{\perp}}, \qquad\qquad H_{\parallel}(t,r)\equiv \frac{\dot{a}_{\parallel}}{a_{\parallel}},
\end{equation}
where a dot denotes the partial derivative with respect to $t$. Differentiating $H_\perp$ with respect to $r$, the two expansion components are related by
\begin{equation}\label{H_rel}
H_\parallel=H_\perp+\frac{R}{R'}\frac{\partial H_\perp}{\partial r}.
\end{equation}
This difference between the two rates defines the kinematic shear of the LTB fluid, which is proportional to $H_\parallel - H_\perp$ \citep{Dunsby_2010}.

A light beam propagates along a null geodesic, that is, a curve $x^\mu(\lambda)$ parameterised by an affine parameter $\lambda$, in which the tangent vector $k^\mu\equiv dx^\mu/d\lambda$ satisfies the condition $k^\mu k_\mu=0$. When the geodesic equation is applied to the metric of Eq.~\eqref{metrica_ltb} along radial null geodesics ($d\theta=d\phi=0$), it yields the relations \citep{Camarena_2021}:
\begin{equation}\label{relacao_t_z}
    \frac{dt}{dz} = -\frac{1}{(1+z)H_\parallel},
    \qquad
    \frac{dr}{dz} = \frac{\sqrt{1-K(r)r^2}}{(1+z)a_\parallel H_\parallel}.
\end{equation}
These equations relate $t$, $r$, and $z$ along the observer’s past light cone and will be useful for expressing quantities of the model as functions of redshift.

The $\Lambda$LTB model arises as a generalisation of the LTB model, assuming a cosmological constant $\Lambda$. For this model, $H_\perp$ satisfies a Friedmann-like equation:
\begin{equation}\label{Friedmann_like}
\frac{H_{\perp}^2}{H_{\perp0}^2} = \Omega_{\text{m},0} \left( \frac{a_{\perp0}}{a_{\perp}} \right)^3 + \Omega_{\text{k},0} \left( \frac{a_{\perp0}}{a_{\perp}} \right)^2 + \Omega_{\Lambda,0},
\end{equation}
where $H_{\perp0}\equiv H_\perp(t_0,r)$ and $a_{\perp0}\equiv a_\perp(t_0,r)$, and the $\Omega_{i,0}$ are the density parameters of matter, curvature and the cosmological constant at the present time $t_0$, now expressed as functions of the coordinate $r$ and respecting the normalisation $\Omega_{\text{m},0}(r)+\Omega_{\text{k},0}(r)+\Omega_{\Lambda,0}(r)=1$. Integrating Eq.~\eqref{Friedmann_like} gives, at each shell, the time elapsed since the Big Bang, $t_0-t_{BB}(r)$, where the Big Bang time $t_{BB}(r)$ is another arbitrary function of the model, corresponding to the Big Bang singularity surface at each shell. Since a non-simultaneous Big Bang, $t_{BB}'(r)\neq0$, would introduce inhomogeneity already in the primordial epochs \citep{Camarena_2021}, the condition $t_{BB}(r)=\text{const.}$ is imposed in this work to ensure consistency with the standard inflationary paradigm, so that $t_0-t_{BB}$ is common to all shells.

The $\Lambda$LTB model can thus be parameterised by the arbitrary functions $K(r)$, $m(r)$ and $t_{BB}(r)$, the first two defining the matter and curvature density parameters:
\begin{equation}\label{Omega_m_k}
\Omega_{\text{m},0}(r) = \frac{2m(r)}{H_{\perp0}^2 a_{\perp0}^3 r^3},
\qquad
\Omega_{\text{k},0}(r) = -\frac{K(r)}{H_{\perp0}^2 a_{\perp0}^2},
\end{equation}
where $m(r)\equiv G\,M(r)$, with $M(r)$ the Euclidean mass profile of the inhomogeneity, that is, the mass enclosed within the shell of radius $r$, while that of the cosmological constant follows from the normalisation, $\Omega_{\Lambda,0}(r)=\Lambda/(3H_{\perp0}^2)=1-\Omega_{\text{m},0}(r)-\Omega_{\text{k},0}(r)$.

To model the inhomogeneity, we fix the radial coordinate $r$ such that $m(r) \propto r^3$. This choice fixes a gauge and, together with the condition $t_{BB}(r)=\text{const.}$, makes the curvature profile $K(r)$ the only free function of the model \citep{Sakr_2024}. When considering a spherical inhomogeneity of comoving radius $r_B$, the curvature is parameterised as
\begin{equation}\label{K(r)}
    K(r) = K_B + (K_C-K_B)\,P_3(r/r_B),
\end{equation}
where $K_C$ is the curvature at the centre of the inhomogeneity and $K_B$ is the curvature of the universe outside the inhomogeneity. The radial profile $P_3(x)$, with $x=r/r_B$, ensures a smooth transition to the background and vanishes for $x\ge1$, so that $K(r)=K_B$ outside the inhomogeneity \citep{Camarena_2021}. In order to ensure that the $\Lambda$LTB model recovers the homogeneous background cosmology outside the inhomogeneity, the function $m(r)$ and the constant $K_B$ are fixed by the parameters of the background FLRW model, that is, evaluated at $r\ge r_B$ (hereafter denoted by the superscript ``out''), together with the normalisation $a_\perp(t_0,r_B)=a^{\mathrm{out}}(t_0)=1$. Using the density parameters of Eq.~\eqref{Omega_m_k} at $r=r_B$ and $m\propto r^3$, we obtain:
\begin{equation}\label{m_KB}
m(r) = \frac{\Omega_{\text{m},0}^{\mathrm{out}} \left(H_0^{\mathrm{out}}\right)^2}{2}\,r^3, \qquad \text{and} \qquad K_B = -\Omega_{\text{k},0}^{\mathrm{out}} \left(H_0^{\mathrm{out}}\right)^2 ,
\end{equation}
where $H_0^{\mathrm{out}}$ is the Hubble constant of the background, shared with the reference $\Lambda$CDM model.

Once the mass profile $m(r)$ is fixed, it is possible to define the local mass density associated with each shell,
\begin{equation}\label{rho}
    \rho_m(t,r) = \frac{m'(r)}{4\pi G\, R'(t,r)R^2(t,r)} .
\end{equation}
From the definition in Eq.~\eqref{rho}, the mean density enclosed within the shell of radius $r$ is then defined as
\begin{equation}\label{rho_mean}
\tilde\rho_m(t,r)\equiv\frac{3}{R^3}\int_0^r\rho_m\,R^2R'\,d\bar r=\frac{3\,m(r)}{4\pi G\,R^3}.
\end{equation}
Outside the inhomogeneity, $\rho_m=\tilde\rho_m=\rho_m^{\mathrm{out}}(t)$.

The relation between the mass densities given in Eq.~\eqref{rho} and Eq.~\eqref{rho_mean} and the background density $\rho_m^{\mathrm{out}}$ defines, respectively, the local, $\delta$, and integrated, $\Delta$, mass density contrasts:
\begin{equation}\label{local_density_contrast}
\delta(t,r)\equiv\frac{\rho_m}{\rho_m^{\mathrm{out}}}-1 ,
\qquad\qquad
\Delta(t,r)\equiv\frac{\tilde\rho_m}{\rho_m^{\mathrm{out}}}-1 .
\end{equation}
Similar to the expansion rates in Eq.~\eqref{H_rel}, since $\tilde\rho_m$ is an average of $\rho_m$, the contrasts are related through $\delta=\Delta+R\Delta'/(3R')$, so that the local contrast is greater than the integrated one whenever $\Delta$ grows with $r$. At the centre of the inhomogeneity, $R=0$, so that both contrasts coincide and define the central density contrast, $\delta_0\equiv\delta(t_0,0)=\Delta(t_0,0)$.
Together with the redshift $z_B$ at which the observer’s past light cone reaches $r_B$, obtained from Eq.~\eqref{relacao_t_z}, $\delta_0$ determines the pair $(K_C, r_B)$ (Sect.~\ref{Distance}).

\begin{figure}
    \centering
    \includegraphics[width=1\linewidth]{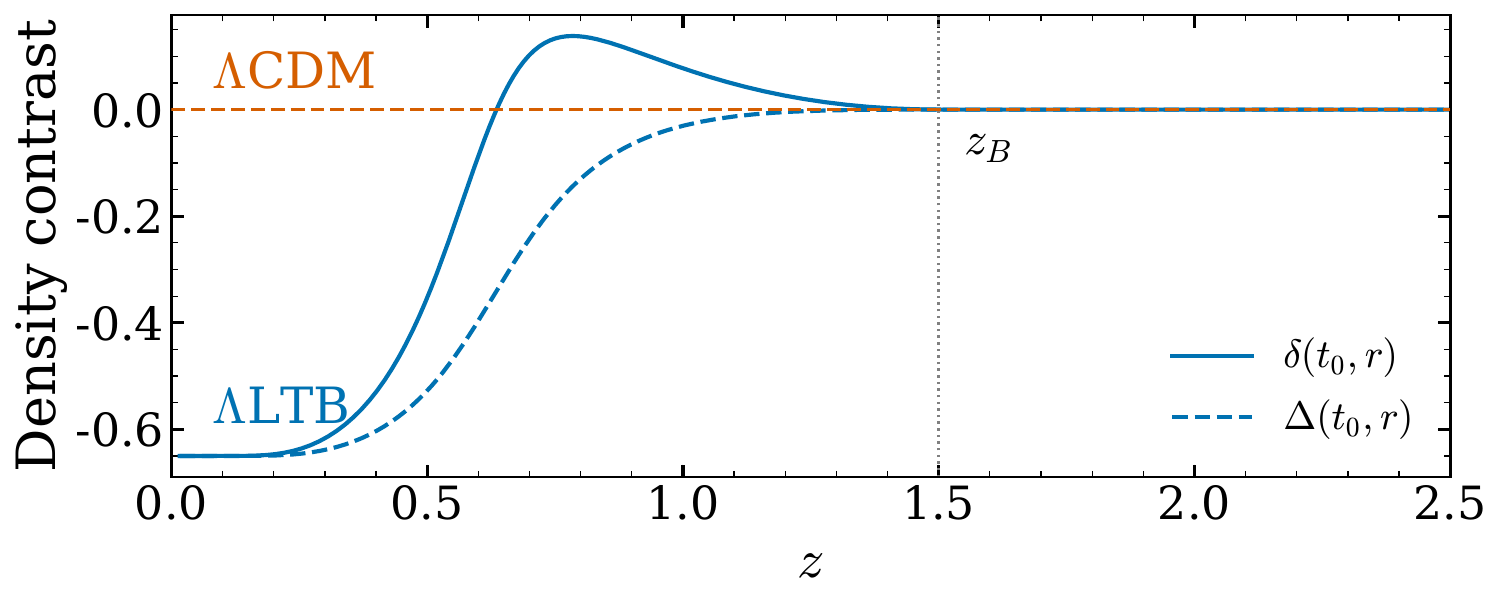}
    \caption{Local, $\delta(t_0,r)$, and integrated, $\Delta(t_0,r)$, density contrasts of the fiducial $\Lambda$LTB model ($\delta_0=-0.65$, $z_B=1.5$) as functions of redshift. The $\Lambda$CDM background corresponds to $\delta=\Delta=0$.}
    \label{fig:density_contrast}
\end{figure}

As illustrated in Fig.~\ref{fig:density_contrast}, both profiles start from $\delta_0$ at the centre and separate as the radius increases. Since this is a compensated model, that is, the curvature profile of Eq.~\eqref{K(r)} ensures that the mass enclosed within the inhomogeneity equals that of the homogeneous background in the same volume, $\Delta$ vanishes at $z_B$, where the LTB and FLRW metrics coincide \citep{Camarena_2022}, remaining zero from there on. For the profile adopted here, $\Delta$ grows monotonically from the centre up to this boundary. In compensation, the matter missing from the central underdense region must accumulate in an outer shell, where the local density exceeds the background one, $\delta>0$, before converging to it at $z_B$.


\section{Gravitational lensing}\label{GL}

The gravitational potential $\Phi$ of the lens acts as a refracting medium for light, working as a first-order perturbation on the background spacetime, which is either the reference Flat-$\Lambda$CDM or the $\Lambda$LTB model of Sect.~\ref{Framework}. In the weak-field limit, $|\Phi|\ll1$, and assuming the Born approximation, which replaces the true photon trajectory with the unperturbed one, the total deflection angle $\hat{\bm\alpha}$ follows from integrating the transverse gradient of $\Phi$ along the line of sight, parameterised by $\ell$ \citep{Bartelmann_2001, LEWIS20061}.

In practice, most gravitational lensing systems have a mass distribution of the lens along the line of sight whose extent is considerably smaller than the distances between the objects that make up the system (observer O, lens L and source S), which justifies the thin-lens approximation. In this approximation, the lens and the source are projected onto planes orthogonal to the line of sight, at angular diameter distances $D_{\rm L}$ and $D_{\rm S}$ from the observer, and the whole deflection occurs in the lens plane \citep{Schneider_1985}. In the coordinates $\bm\xi$ and $\ell$, the deflection angle and the surface mass density of the lens are then
\begin{equation}\label{alpha_xi}
\hat{\bm\alpha}(\bm\xi)=4G\!\int\!\Sigma(\bm\xi')\,\frac{\bm\xi-\bm\xi'}{|\bm\xi-\bm\xi'|^2}\,d^2\xi',\qquad
\Sigma(\bm\xi)=\int\rho(\bm\xi,\ell)\,d\ell .
\end{equation}
The angle $\hat{\bm\alpha}$, however, is not scaled with the real geometry of the system, and to that end it must be related to the distances travelled by the beam, which defines the reduced deflection angle
\begin{equation}\label{alpha_reduzido}
    \bm{\alpha}(\bm{\theta}) = \frac{D_{\rm LS}}{D_{\rm S}}\bm{\hat{\alpha}}(\bm\xi),
\end{equation}
where $D_{\rm LS}$ is the angular diameter distance of the source object as seen from the lens object, while the impact parameter scales with the angular diameter distance to the lens as $\bm{\xi} = D_{\rm L} \bm{\theta}$.

The convergence $\kappa$ is introduced as the ratio between the surface density $\Sigma$ and a critical surface density $\Sigma_{crit}$, such that
\begin{equation}\label{Sigma_crit}
    \kappa(\bm\theta) \equiv \frac{\Sigma(D_{\rm L}\bm\theta)}{\Sigma_{crit}},
    \qquad
    \Sigma_{crit} \equiv \frac{D_{\rm S}}{4\pi G\, D_{\rm L} D_{\rm LS}}.
\end{equation}
This combination of the system distances defines the ``effective lensing distance'' $D_{\rm eff}$,
\begin{equation}\label{Deff}
 D_{\rm eff} \equiv \frac{D_{\rm L} D_{\rm LS}}{D_{\rm S}}.
\end{equation}

In the absence of the lens, the position $\bm{\eta}$ of the source on the source plane would be seen by the observer under an angle $\bm\beta = \bm\eta/D_{\rm S}$, the true angle of the source, which the lens equation relates to the apparent angle of the deflected source, $\bm\theta$:
\begin{equation}\label{Eq.Lente}
    \bm{\beta} = \bm{\theta} - \bm{\alpha}.
\end{equation}

To study the deformation of the images caused by the gravitational lensing effect, one uses the amplification matrix $\bm{A}$, which arises when expanding the lens equation, Eq.~\eqref{Eq.Lente}, in a first-order Taylor series around an image position, $\delta\bm\beta=\bm A\,\delta\bm\theta$, so that $\bm{A}=\partial\bm\beta/\partial\bm\theta$ \citep{Meneghetti_2022}.
This formulation based on the thin-lens approximation allows us to define the effective lens potential, $\psi(\bm{\theta})$, obtained by projecting the three-dimensional gravitational potential, $\Phi$, along the line of sight, to which the reduced deflection angle is related through its gradient:
\begin{equation}\label{pot_eff}
    \psi(\bm{\theta}) = \frac{D_{\rm LS}}{D_{\rm L} D_{\rm S}}\, 2 \int \Phi(D_{\rm L}\bm{\theta}, \ell) \, d\ell,
    \qquad
    \bm\alpha = \nabla_\theta\psi,
\end{equation}
so that, applying Eq.~\eqref{Eq.Lente}, we obtain the decomposition into convergence, shear and rotation:
\begin{equation}\label{amplificacao}
A_{ij} = \frac{\partial \beta_i}{\partial \theta_j} = \delta_{ij} - \frac{\partial \alpha_i}{\partial \theta_j} =
\begin{pmatrix}
    1 - \kappa - \gamma_1 & - \gamma_2+\omega \\
    - \gamma_2-\omega & 1 - \kappa + \gamma_1
\end{pmatrix},
\end{equation}
with the convergence $\kappa=\frac{1}{2}(\psi_{,11}+\psi_{,22})$ being the laplacian of the potential consistently with its definition as a mass density in Eq.~\ref{Sigma_crit}, the shear $\gamma=(\gamma_1,\gamma_2)$ given by $\gamma_1=\frac{1}{2}(\psi_{,11}-\psi_{,22})$ and $\gamma_2=\psi_{,12}$, where the comma denotes the derivative with respect to $\bm\theta$, and $\omega=0$, since $\bm{A}$ is built from the Hessian of $\psi$, which is symmetric.

Since gravitational lensing conserves surface brightness \citep{Bartelmann_2001}, the surface brightness of an image at a point $\bm{\theta}$ is equal to that of the source at the corresponding point, $I(\bm{\theta})=I^{(s)}[\bm{\beta}(\bm{\theta})]$, where $I^{(s)}$ is the surface brightness distribution in the source plane. The observed flux therefore changes only through the apparent area of the image, so that the magnification, the ratio between the areas of the image and of the source, is
\begin{equation}\label{magnificacao}
    \mu = \frac{1}{\det\bm{A}} = \frac{1}{(1-\kappa)^2 -|\gamma|^2}.
\end{equation}

\subsection{Strong lensing}\label{Strong_Lensing}

The thin-lens approximation is used in both the weak and the strong lensing regimes, which differ in the magnitude of convergence and shear. In strong lensing \citep{Kochanek_2006}, when there is a sufficiently close alignment between the observer, the lens and the source, the lens equation provides several solutions, resulting in multiple images of the same source or in strongly distorted images, such as arcs and rings. In this regime, the theoretical magnification diverges at the points where the Jacobian of the transformation vanishes, that is, $\det\bm A(\bm{\theta}) = 0$. The set of these points in the lens plane forms closed and smooth curves, named critical curves. The mapping of these curves onto the source plane, through the lens equation, defines the so-called caustic curves, which need not be smooth and may present cusps. 

The critical curves and caustics determine, respectively, the regions of high magnification and the image multiplicity. The number of observed images only changes when the source crosses a caustic, and always by $\pm 2$, corresponding to the creation or annihilation of a pair of images with opposite parities ($\mu > 0$ and $\mu < 0$) near the critical curve. As a consequence, the odd-number theorem, or Burke's theorem, follows \citep{Burke_1981}; since a source outside the caustics presents a single image, a non-singular lens always produces an odd number of images \citep{McKenzie_1985}, although the odd image is strongly demagnified and too faint to be detected in most systems.

Having summarised the general formalism, we now address the particular case of a lens with circular symmetry. In this case, we can move the origin of the coordinates to the centre of symmetry and reduce the deflection, in general a two-component vector, to a one-dimensional problem, in which the deflection angle points towards the centre of symmetry, with modulus $\hat{\alpha}(\xi)=4GM(\xi)/\xi$, where $M(\xi)$ is the mass enclosed within the radius $\xi$ \citep{narayan_1997}. Combining the scaling relation of Eq.~\eqref{alpha_reduzido} with this deflection angle and substituting into the lens equation, Eq.~\eqref{Eq.Lente}, the lens equation for an arbitrary circularly symmetric mass profile becomes
\begin{equation}\label{lente_circ}
    \beta(\theta) = \theta - \frac{D_{\rm LS}}{D_{\rm L} D_{\rm S}} \frac{4GM(\theta)}{\theta}.
\end{equation}
Owing to the perfect rotational symmetry of this system, a source exactly aligned with the optical axis ($\beta=0$) 
is mapped onto the lens plane, also called the image plane, as a continuous ring, the Einstein ring. Imposing this alignment condition on the previous equation, we isolate the angular radius, obtaining the Einstein radius, $\theta_{\rm E}$:
\begin{equation}\label{theta_E}
    \theta_{\rm E} = \left[4GM(\theta_{\rm E})\frac{D_{\rm LS}}{D_{\rm L} D_{\rm S}}\right]^{\frac{1}{2}},
\end{equation}
which can be expressed on a physical scale using the geometric relation $\xi_E = D_{\rm L}\theta_{\rm E}$, as $\xi_E=[4GM(\xi_E)\,D_{\rm eff}]^{1/2}$, written in terms of the effective lensing distance, $D_{\rm eff}$ defined in Eq.~\eqref{Deff}.

Circular symmetry makes the matrix $\bm{A}$ of Eq.~\eqref{amplificacao} diagonal in the basis formed by the tangential and radial directions, with eigenvalues $\lambda_t=1-\alpha(\theta)/\theta=1-\bar\kappa$ and $\lambda_r=1-d\alpha/d\theta=1-2\kappa+\bar\kappa$, where $\bar\kappa$ is the mean convergence inside $\theta$. The condition $\lambda_t=0$ is equivalent to $\alpha(\theta)=\theta$, that is, to $\bar\kappa(\theta)=1$, and is precisely the one that defines the Einstein radius in Eq.~\eqref{theta_E}. Thus, the Einstein ring coincides with the tangential critical curve, a circle of radius $\theta_{\rm E}$. Mapping it onto the source plane with Eq.~\eqref{lente_circ}, we see that the corresponding caustic degenerates into a single point at $\beta=0$, which explains the formation of a continuous ring. The radial critical curve, $\lambda_r=0$, is a circle of radius $\theta_{\rm r}<\theta_{\rm E}$, mapped onto a circular radial caustic, and for profiles with a core it only exists if the central convergence satisfies $\kappa_0>1$.

\subsection{Lens and source profiles}\label{perfis_lente_fonte}

In this work, although other profiles were tested, the ``softened (non-singular) isothermal sphere'' lens profile, hereafter referred to as NIS, and the S\'ersic brightness profile for the source were chosen for a case study. Since the NIS is a softening of the ``singular isothermal sphere'' (SIS), it is convenient to define it: for a constant one-dimensional velocity dispersion $\sigma_v$, the SIS has an effective potential $\psi(\theta)=4\pi\sigma_v^2\,(D_{\rm LS}/D_{\rm S})\,\theta$, which, through Eq.~\eqref{pot_eff}, subsequently enables us to define its Einstein radius
\begin{equation}\label{theta_E_SIS}
    \theta_{\rm E}=4\pi\sigma_v^2\frac{D_{\rm LS}}{D_{\rm S}}.
\end{equation}
This profile is, however, singular at the centre, where the density diverges. To circumvent this singularity, the NIS introduces a core of radius $r_{\rm c}$, inserted directly into the effective lens potential \citep[Table~2]{narayan_1997}, such that
\begin{equation}\label{psi_NIS}
    \psi(\theta)=\theta_E\sqrt{\theta^2+r_{\rm c}^2},
\end{equation}
where $\theta_{\rm E}$ is the Einstein radius of the SIS with the same $\sigma_v$, Eq.~\eqref{theta_E_SIS}, and the SIS is recovered for $r_{\rm c}\to0$. Through Eq.~\eqref{pot_eff} the tangential critical curve $\theta_{\rm t}=\sqrt{\theta_{\rm E}^2-r_{\rm c}^2}$ is obtained, which we denote by $\theta_{\rm crit}$. The resulting ring is therefore slightly smaller than $\theta_{\rm E}$. Physically, this profile represents the total mass distribution (the sum of stellar mass and dark matter) of massive elliptical galaxies. These have a density profile that is on average isothermal in the region where the ring forms \citep{Koopmans_2009}, which is why the model is suited for studying lens galaxies.

For the source, the surface brightness is described by the S\'ersic profile \citep{Sersic_1963, Graham_2005},
\begin{equation}\label{sersic}
    I^{(s)}(R)=I_e\exp\left\{-b_n\left[\left(\frac{R}{R_\mathrm{eff}}\right)^{1/n}-1\right]\right\},
\end{equation}
where $R=|\bm\beta-\bm\beta_{\rm s}|$ is the distance to the centre of the source $\bm\beta_{\rm s}$, $R_\mathrm{eff}$ is the effective radius containing half the light, $I_{\rm e}$ is the brightness at that radius, $n$ is the S\'ersic index controlling the concentration of the profile, and $b_n$ is a constant fixed by the definition of $R_\mathrm{eff}$. For indices $n>1$ the brightness profile is concentrated in a narrow central peak, which would make the image peak dependent on the grid resolution. To circumvent this effect, we soften the central profile by including a core, such that $R \to \sqrt{R^2+r^2_{\rm core}}$.

Merely by varying the index $n$, this profile is capable of describing the majority of the surface brightness profiles of galaxies, from galactic discs ($n=1$) up to elliptical galaxies ($n=4$), the latter being adopted in this work.


\section{Angular diameter distances in $\Lambda$LTB}\label{Formalismo_otico}

\subsection{The angular diameter distance and the lens--source problem}\label{Distance}

As shown in Sect.~\ref{GL}, the background spacetime enters the thin-lens description only through the angular diameter distances $D_{\rm L}$, $D_{\rm S}$ and $D_{\rm LS}$, combined in the effective lensing distance $D_{\rm eff}$ (Eq.~\eqref{Deff}). For models that aim to relax the cosmological principle, light rays propagate along geodesics defined in non-homogeneous and/or anisotropic spacetimes, where the cross-sectional area of the light beam undergoes gravitational focusing and geometric distortions different from those in FLRW. This translates into a distance--redshift relation different from the one predicted by Friedmann models. 

 The distances depend on the background dynamics of the model, which in $\Lambda$LTB we compute using the version of \texttt{vd2020}\footnote{Available at: \url{https://github.com/valkenburg/vd2020}} incorporated into \texttt{monteLLTB}\footnote{Available at: \url{https://github.com/davidcato/monteLLTB}} by \citet{Camarena_2022}, which builds on the Fortran implementation developed by \citet{Valkenburg_2012}. The independent input parameters relevant to the background evolution are $h$, $\omega_{\text{b}}$, $\omega_{\text{cdm}}$, and $\Omega_{\text{k}}\equiv\Omega_{\text{k},0}^{\mathrm{out}}$, with the radiation density neglected and $\Omega_{\text{m}}\equiv\Omega_{\text{m},0}^{\mathrm{out}}$, and $\Omega_{\Lambda}$ as derived quantities, while the inhomogeneity is defined by the pair $(\delta_0, z_B)$. The fiducial values used in this work are listed in Table~\ref{tab:fiducial_parameters}. From these, the code fixes $m(r)$ and $K_B$ through Eq.~\eqref{m_KB} and computes the pair $(K_C, r_B)$ by requiring the central shell to have the same age as the background and the past light cone of the central observer, Eq.~\eqref{relacao_t_z}, to reach $r=r_B$ exactly at $z=z_B$.
\begin{table}
\centering
\caption{Input parameters and derived quantities used in the \texttt{vd2020} calculations.}
\begin{tabular}{lcc}
\toprule
Parameter & Symbol & Value \\
\midrule
Reduced Hubble parameter & $h$ & 0.67 \\
Physical baryon density & $\omega_{\text{b}}$ & 0.02225 \\
Physical cold dark matter density & $\omega_{\text{cdm}}$ & 0.121398 \\
Matter density parameter & $\Omega_{\text{m}}$ & 0.3200 \\
Curvature density parameter & $\Omega_{\text{k}}$ & 0 \\
Dark energy density parameter & $\Omega_{\Lambda}$ & 0.6800 \\
\midrule
Central density contrast & $\delta_0$ & $-0.65$ \\
Inhomogeneity boundary redshift & $z_B$ & 1.5 \\
\bottomrule
\end{tabular}
\label{tab:fiducial_parameters}
 \tablefoot{The upper block corresponds to the parameters of the asymptotic FLRW background, common to the $\Lambda$CDM and $\Lambda$LTB models, while the lower block contains the parameters that characterise the LTB inhomogeneity.}
\end{table}
As output, \texttt{vd2020} provides the background and past-light-cone quantities that characterise the dynamics of the $\Lambda$LTB model, namely the radial and transverse scale factors and Hubble expansion rates, the radial coordinate $r$ and the cosmic time $t$ as functions of $z$, and the angular diameter and luminosity distances. However, \texttt{vd2020} only computes angular diameter distances from the central observer, $D_{\rm A}(0,z)$, which strongly limits its usage in the study of gravitational lensing systems, since these require the lens--source angular diameter distance, $D_{\rm LS}\equiv D_{\rm A}(z_{\rm L},z_{\rm S})$.

In general, an angular diameter distance is not additive, that is, $D_{\rm LS}\neq D_{\rm S}-D_{\rm L}$ \citep{Rasanan_2015}. However, in a model that allows the definition of a comoving distance, as is the case of the Flat-$\Lambda$CDM model, comoving distances along the same radial line are additive. There, the condition of homogeneity and isotropy results in homogeneous hypersurfaces of simultaneity, on which the radial distance between the observer and a source at fixed $r$ is simply $a(t)\,r$, so that integrating along the past light cone defines the comoving distance,
\begin{equation}\label{D_C}
    D_{\rm C}(z)\equiv a(t_0)\,r=\int_0^z\frac{dz'}{H(z')},
\end{equation}
with $H(z)=H_0\left[\Omega_{\rm m}(1+z)^3+\Omega_\Lambda\right]^{1/2}$. This factorisation between temporal and radial quantities is what underlies the understanding of this distance as ``comoving'', and gives $D_{\rm A}=D_{\rm C}/(1+z)$ and $D_{\rm LS}=[D_{\rm C}(z_{\rm S})-D_{\rm C}(z_{\rm L})]/(1+z_{\rm S})$, that is,
\begin{equation}\label{Approx_D_Ls}
    D_{\rm LS} = D_{\rm S} - \frac{1 + z_{\rm L}}{1 + z_{\rm S}}D_{\rm L}.
\end{equation}
In $\Lambda$LTB, in contrast, the scale factors $a_\perp=R/r$ and $a_\parallel=R'$ depend on $r$, which prevents $R(t,r)$ from being factorised in the form $a(t)f(r)$. This behaviour is due to the kinematic shear of the dust, proportional to $H_\parallel-H_\perp\neq0$ (Sect.~\ref{Framework}), which prevents a common rescaling of the spatial slices, so that no comoving distance can be defined and Eq.~\eqref{Approx_D_Ls} fails\footnote{For a lens at $z_{\rm L}=0.3$, Eq.~\eqref{Approx_D_Ls} deviates from the numerical integration of Sect.~\ref{DA_integracao} by up to $2.3\%$ ($\sim31$~Mpc) in $\Lambda$LTB, against $1.4\times10^{-4}$~Mpc in Flat-$\Lambda$CDM (Appendix~\ref{app:DLS_approx}).}. Therefore, $D_{\rm LS}$ needs to be computed explicitly.

\subsection{Sachs equations for the angular diameter distance}\label{Sachs_formalism}

Computing $D_{\rm LS}$ in $\Lambda$LTB requires tracking the evolution of a light beam through the fully covariant optical equations of general relativity. By merely describing the trajectory of a single light ray, a null geodesic alone is unable to convey how a given spacetime deforms an image. To this end, we consider a set of light rays connecting each point of the source to the observer, which defines a light beam that, when the rays are close together, can be treated as a geometric object in its own right \citep{Fleury2015thesis}. Its shape is described by the transverse separation $\zeta^A$ between two neighbouring geodesics, measured on the screen plane of an observer with four-velocity $u^\mu$, spanned by an orthonormal basis $(s_1^\mu,s_2^\mu)$, known as the Sachs basis, which is orthogonal to both $k^\mu$ and $u^\mu$ and is parallel-transported along the beam \citep{Seitz_1994,Fleury2015thesis}. Projecting the geodesic deviation equation, also known as the Jacobi equation, onto this basis yields $\ddot\zeta^A=\mathcal{R}^A_B\,\zeta^B$, where $\mathcal{R}^A_B$ is the curvature projected onto the screen space and a dot denotes the derivative with respect to the affine parameter $\lambda$, chosen such that $\lambda=0$ at the observer and $k_\mu u^\mu=1$ \citep[\S2.1.1]{Fleury2015thesis}. Its linearity in $\zeta^A$, together with the initial condition $\zeta^A(0)=0$, allows the solutions to be written as $\zeta^A(\lambda)=\mathcal{D}^A_B(\lambda)\,\dot\zeta^B(0)$, where $\mathcal{D}^A_B$ is the Jacobi matrix and $\dot\zeta^B(0)$ corresponds to the angular separation observed between the rays.

By definition, the Jacobi matrix $\mathcal{D}^A_B(\lambda)$ satisfies the Jacobi equation in matrix form,
\begin{equation}\label{eq_jacobi_matricial}
    \ddot{\mathcal{D}} = \mathcal{R}\,\mathcal{D}, \qquad \dot{\mathcal{D}}(0)=\mathbb{1},\quad \mathcal{D}(0)=0.
\end{equation}
Thus, it is the matrix $\mathcal{D}$ and not $\zeta^A$ that carries the information about the beam's morphology along the optical path, and the amplification matrix of Sect.~\ref{GL} follows from the ratio between the real Jacobi matrix, $\mathcal{D}_{\text{real}}$, perturbed by the presence of a lens mass, and the background Jacobi matrix, $\mathcal{D}_{\text{bg}}$, both evaluated at the source, $\bm{A} \equiv \mathcal{D}_{\text{bg}}^{-1}(\lambda_{\rm S})\,\mathcal{D}_{\text{real}}(\lambda_{\rm S})$ \citep[\S2.2.2]{Fleury2015thesis}. Defining the deformation rate matrix $S\equiv\dot{\mathcal{D}}\,\mathcal{D}^{-1}$, Eq.~\eqref{eq_jacobi_matricial} yields a matrix Riccati-type equation, $\dot S+S^2=\mathcal{R}$, where $\mathcal{R}$ can be decomposed into its trace and symmetric traceless parts: the isotropic part $\mathscr{R}$, arising from the Ricci tensor $R_{\mu\nu}$, and the anisotropic part $\mathscr{W}=\mathscr{W}_1+i\mathscr{W}_2$, arising from the Weyl tensor $C_{\mu\nu\alpha\beta}$,
\begin{equation}\label{escalares_RW}
\mathscr{R} \equiv -\frac{1}{2} R_{\mu\nu} k^\mu k^\nu, \qquad \mathscr{W} \equiv -\frac{1}{2} C_{\mu\nu\alpha\beta} (s^\mu_1 - i s^\mu_2) k^\nu k^\alpha (s^\beta_1 - i s^\beta_2).
\end{equation}

On the other hand, $S$ can be decomposed into its trace, $\Theta\equiv\frac{1}{2}\text{tr}(S)$, the optical expansion scalar, its symmetric traceless part, the optical shear $\sigma\equiv\sigma_1+i\sigma_2$, and its antisymmetric part, the vorticity $\omega$. Since, by construction, the null geodesic congruences converge to a single point at the observer, the light beam is always orthogonal to a family of null hypersurfaces, which by Frobenius's theorem guarantees $\omega=0$ \citep{R_s_nen_2010}. Regrouping each component of the Riccati equation, we obtain the Sachs equations,
\begin{equation}\label{sachs}
    \dot{\Theta} + \Theta^2 + |\sigma|^2 = \mathscr{R}, \qquad \dot{\sigma} + 2\Theta\sigma = \mathscr{W}.
\end{equation}
Although $\Theta$ and $\sigma$ fully describe the beam, the quantity entering the lensing configuration is the angular diameter distance itself, and the optical expansion diverges at the observer, $\Theta\sim1/\lambda$ \citep[\S2.3.3]{Fleury2015thesis}, which makes it convenient to rewrite the Sachs equations in terms of $D_{\rm A}$. Assuming an orthonormal Sachs basis ($s_A^\mu s_{B\mu}=\delta_{AB}$), the cross-sectional area of the beam per unit solid angle is related to the angular diameter distance by $A=D_{\rm A}^2$, which, through the definition of the optical expansion, leads to $\Theta=\dot A/(2A)=\dot D_{\rm A}/D_{\rm A}$ \citep[\S2.3.2]{Fleury2015thesis}, and thus $\dot\Theta=\ddot D_{\rm A}/D_{\rm A}-\Theta^2$. Inserting this result in the first of Eqs.~\eqref{sachs} and multiplying the second by $D_{\rm A}^2$, we obtain the system
\begin{equation}\label{Sachs_DA}
\left\{
\begin{aligned}
    &\frac{d^2D_{\rm A}}{d\lambda^2} = \left(\mathscr{R} - |\sigma|^2 \right)D_{\rm A} , \\
    &\frac{d(D_{\rm A}^2\sigma)}{d\lambda} = D_{\rm A}^2\mathscr{W},
\end{aligned}
\right.
\end{equation}
which describes the evolution of the angular diameter distance along the beam, determined by the Ricci focusing and, through the shear, by the Weyl curvature. To apply Eq.~\eqref{Sachs_DA} to $\Lambda$LTB, the scalars of Eq.~\eqref{escalares_RW} must be written explicitly for the LTB metric of Eq.~\eqref{metrica_ltb}. Since $k^\mu$ is null, Einstein's equations give $R_{\mu\nu}k^\mu k^\nu=8\pi G\,T_{\mu\nu}k^\mu k^\nu$, so that the cosmological constant does not contribute and, for the pressureless matter of the model, $T_{\mu\nu}=\rho_m u_\mu u_\nu$,
\begin{equation}\label{R_3}
    \mathscr{R}=-4\pi G\,\rho_m(t,r)\,(k^0)^2,
\end{equation}
where $\rho_m$ is the local density of Eq.~\eqref{rho} and $k^0=dt/d\lambda$. Unlike in Flat-$\Lambda$CDM, where it vanishes by isotropy, the Weyl contribution is, in general, non-zero in $\Lambda$LTB. Projecting the Weyl tensor onto the Sachs basis yields \citep[Eq.~2.24]{Sarma_2026}
\begin{equation}\label{Weyl_scalar_LTB_rho}
    \mathscr{W}=4\pi G\left(\frac{J}{R}\right)^2\left(\rho_m-\tilde\rho_m\right),
\end{equation}
where $\tilde\rho_m$ is the mean density of Eq.~\eqref{rho_mean} and $J=R(t_0,r_0)\sin\psi$ is the angular momentum of the photon per unit momentum, conserved along the ray, with $\psi$ the angle between the line of sight of an observer at $r_0$ and the direction of the centre of the inhomogeneity. It follows that the Weyl contribution vanishes for radial rays, $J=0$, or when $\rho_m=\tilde\rho_m$, as in an FLRW spacetime, the condition $J=0$ being of particular interest for an observer at the centre of the inhomogeneity \citep{Ghassemi_2009}. Since $D_{\rm A}=0$ at the vertex, the second of Eqs.~\eqref{Sachs_DA} then gives $\sigma\equiv0$, reducing the system to its Ricci part. To apply the same reduced equation to the lens--source segment, the source, the lens and the observer are taken as perfectly aligned along a radial line, with a circular lens profile. The deflection then occurs at the Einstein radius, at arcsecond scales from the radial direction, and the axial symmetry of the system ensures that a single $D_{\rm LS}$ describes the whole ring, the closest configuration to the radial case. 

The reduced equation for $D_{\rm A}$ is defined in terms of the affine parameter $\lambda$, which is not a quantity provided by the \texttt{vd2020} code. 
Since redshift increases monotonically along the radial null geodesics considered here, it can be used as the independent variable in the numerical integration.
With the normalisation $k_\mu u^\mu=1$ at the observer's position, and knowing that $u^\mu=(1,0,0,0)$ and $g_{00}=-1$ in the LTB metric, the definition of redshift can be rewritten as $k^0=-(1+z)$. Applying the geodesic equation to the temporal component and using the null condition of the radial geodesic, we define the operator $d/d\lambda=(dz/d\lambda)\,d/dz=H_\parallel(1+z)^2\,d/dz$, which, applied to the reduced Eq.~\eqref{Sachs_DA}, gives an ordinary differential equation (ODE) for the angular diameter distance as a function of redshift:
\begin{equation}\label{eq_DA}
    D_{\rm A}''(z) + \left(\frac{H_\parallel'}{H_\parallel}+\frac{2}{1+z}\right)D_{\rm A}'(z)+\frac{4\pi G\rho_{m}(z)}{H_\parallel^2(1+z)^2}D_{\rm A}(z)=0.
\end{equation}
Henceforth, a prime applied to $D_{\rm A}$, $H_\parallel$, or any other quantity written as a function of redshift denotes the derivative with respect to $z$, distinguishing it from the previous use of the prime for $\partial/\partial r$. In Flat-$\Lambda$CDM, where $H_\parallel=H(z)$ and $\rho_m=\rho_m^{\rm out}$, Eq.~\eqref{eq_DA} is a pure focusing equation, where the cosmological constant does not locally focus light, since $g_{\mu\nu}k^\mu k^\nu=0$, affecting $D_{\rm A}$ only through the expansion history, and whose solution relates directly to the comoving distance, $D_{\rm A}=D_{\rm C}/(1+z)$. Finally, $D_{\rm LS}$ is obtained by integrating the same equation from the lens, which becomes the new vertex of the beam.

\subsection{Numerical integration and results}\label{DA_integracao}

To solve Eq.~\eqref{eq_DA}, we developed a dedicated code, 
described in Appendix~\ref{app:impl}.
It takes as input the redshifts of the lens and source objects and the background quantities returned by \texttt{vd2020} for the parameters of Table~\ref{tab:fiducial_parameters}, namely $H_\parallel$, $R$, $R'$, the present-day profiles $a_{\perp0}(r)$ and $H_{\perp0}(r)$, and the relation between $r$ and $z$ along the past light cone.

\begin{figure*}
    \centering    \includegraphics[width=\textwidth]{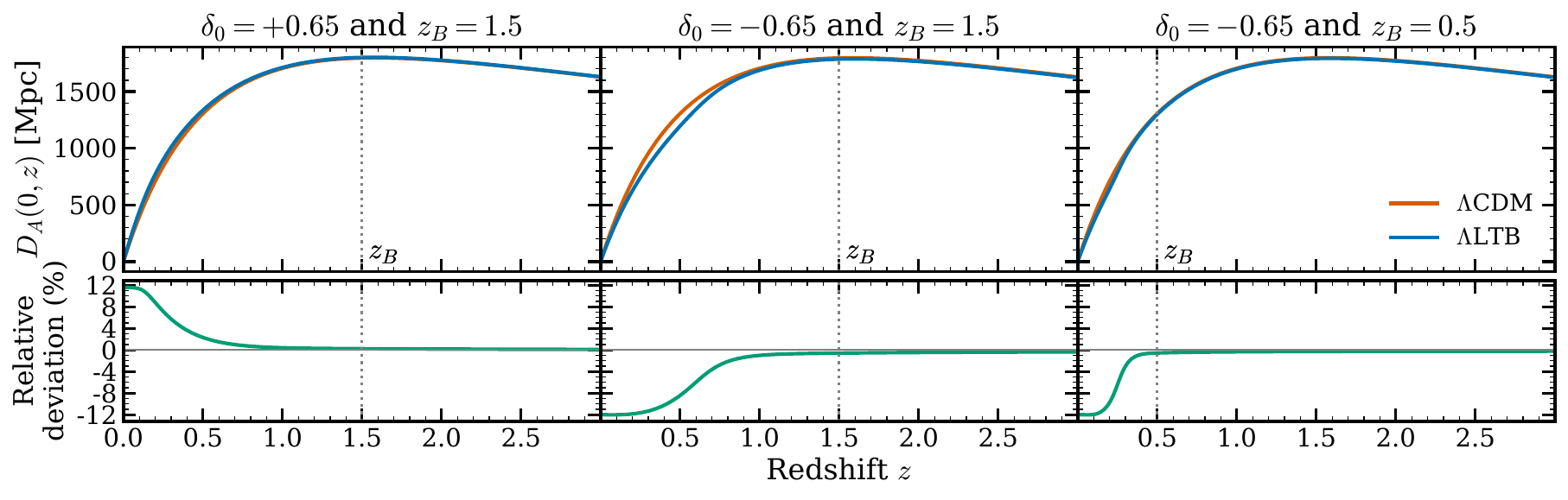}
    \caption{Angular diameter distance measured from the observer, $D_{\rm A}(0,z)$, for three inhomogeneities: a large overdensity ($\delta_0=0.65$, $z_B=1.5$; {\em left panel}), the adopted large underdensity ($\delta_0=-0.65$, $z_B=1.5$; {\em central panel}), and a smaller underdensity ($\delta_0=-0.65$, $z_B=0.5$; {\em right panel}). {\em Top panels}: Distances in Mpc for the $\Lambda$CDM (orange) and $\Lambda$LTB (blue) models. {\em Bottom panels}: Percentage relative deviation of the $\Lambda$LTB model with respect to the $\Lambda$CDM reference. The dotted vertical line marks the boundary of the inhomogeneity in each case.}
    \label{fig:DA_os}
\end{figure*}

In Fig.~\ref{fig:DA_os} we present the resulting angular diameter distance measured from the observer, $D_{\rm A}(0,z)$, comparing the values from $\Lambda$CDM with those from $\Lambda$LTB for three inhomogeneities, the central panel corresponding to the adopted one. We note that a change in the sign of the density contrast corresponds to a change in the sign of the deviation. Indeed, in an overdensity $D_{\rm A}$ is larger than in the homogeneous case, whereas in an underdensity it is smaller \citep[confirming the findings of][]{Tomita_2000}. We also note that near the observer the relative deviation depends only on $\delta_0$, with both underdense models reaching a $-11.9\%$ difference despite their $z_B$ values differing by a factor of $3$. This is a result of, in the limit $z\to0$, the distance being fixed by the central expansion rate, which is determined by $\delta_0$ through the condition $t_{BB}(r)=\text{const.}$ (Sect.~\ref{Framework}). In contrast, it is $z_B$ that determines the deviation further from the observer, with the maximum difference between the models increasing from $\sim69$~Mpc for $z_B=0.5$ to $\sim118$~Mpc for $z_B=1.5$.
For the adopted inhomogeneity, the relative deviation reaches a maximum of $-9.9\%$ at $z\approx0.42$, where the absolute difference is $\sim118$~Mpc. This difference decreases to $-0.57\%$ ($\sim10$~Mpc) as the function approaches the boundary $z_B$, and continues to decrease beyond it, reaching $-0.15\%$ at $z\sim20$. The difference does not  converge to zero because any radial geodesic that reaches the central observer must cross the underdensity. Thus, sources located beyond the boundary present a very similar although not identical angular diameter distance in both models, suggesting that measurable differences must result primarily from the distances to the lens and from the lens--source path.
\begin{figure}
    \centering
    \includegraphics[width=\linewidth]{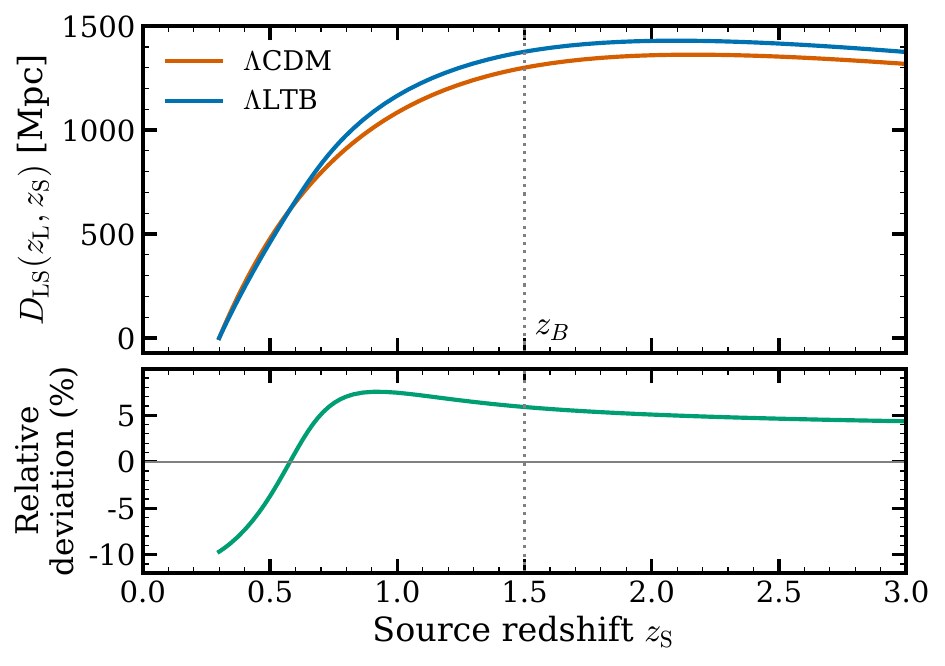}
    \caption{Lens--source angular diameter distances $D_{\rm LS}(z_{\rm L},z_{\rm S})$ for a fixed lens at $z_{\rm L}=0.3$ as a function of the source redshift $z_{\rm S}$. {\em Top panel}: Absolute distance in Mpc for the $\Lambda$CDM (orange) and $\Lambda$LTB (blue) models. {\em Bottom panel}: Percentage relative deviation of the $\Lambda$LTB model with respect to the $\Lambda$CDM reference. The dotted vertical line marks the boundary of the spherical underdensity at $z_B=1.5$.}
    \label{fig:D_LS}
\end{figure}

Figure~\ref{fig:D_LS} presents the comparison of the distances $D_{\rm LS}$ between the two models. In contrast to what is observed with $D_{\rm A}(0,z_{\rm S})$ in Fig.~\ref{fig:DA_os} (central panel), the $\Lambda$LTB model is above $\Lambda$CDM from $z_{\rm S}\approx0.58$ onward, while for sources closer to the lens the ordering is reversed. This sign change comes from the fact that near the lens the distance $D_{\rm LS}$ is determined mostly by the local longitudinal expansion rate $H_\parallel$ of Eq.~\eqref{H_paralelo}, which inside the underdensity exceeds the expansion rate of the homogeneous universe by about $10\%$ at $z_{\rm L}$. As the source approaches the overdense shell that compensates the underdensity (Fig.~\ref{fig:density_contrast}), the steeper decrease of the transverse expansion rate with $r$ makes $H_\parallel$, by Eq.~\eqref{H_rel}, smaller than $H_\perp$ and, from $z\approx0.45$ onwards, also smaller than the $\Lambda$CDM rate, with a minimum of about $81\%$ of it at $z\approx0.67$, producing a greater contribution to $D_{\rm LS}$ for higher redshifts. Since by construction, beyond the boundary $z_B$, the two models share the same background, with $H_\parallel=H_\perp=H$, no further difference in the distances between them is generated, leaving the excess already accumulated inside the underdensity, whose relative weight decreases slowly as the distances grow, without vanishing. The absolute difference reaches a maximum of $\sim82$~Mpc at $z_{\rm S}\approx1.09$ and decreases slowly to $\sim57$~Mpc at $z_{\rm S}=3$. Similar to what occurred for $D_{\rm A}(0,z)$, the peak of the absolute difference does not coincide with the peak of the relative deviation, which reaches its maximum of $7.5\%$ at $z_{\rm S}\approx0.91$ and decreases to $5.9\%$ at the boundary $z_B$, beyond which it keeps falling, though ever more slowly, down to $4.4\%$ at $z_{\rm S}=3$. This slow decay evidences that, in contrast to what is observed for $D_{\rm L}$ or $D_{\rm S}$ (Fig.~\ref{fig:DA_os}), the implications of the inhomogeneous geometry remain significant in the lens--source distance even for sources well outside the underdensity, and therefore in the ratio $D_{\rm LS}/D_{\rm S}$ on which the lensing configuration depends.

The presence of the underdensity thus introduces deviations of significant magnitude in the distances that define the gravitational lens system, with different regimes according to the separation between the system objects, as illustrated by the sign change of $D_{\rm LS}$, which can influence the value of the effective lensing distance (Eq.~\eqref{Deff}). It is important to note, however, that these relative deviations, of up to 
about $7.5\%$ for $D_{\rm LS}$, remain within the relative uncertainty currently achieved in observational measurements of angular diameter distances 
\citep{Wong_2019}. This fact reinforces the assertion that $D_{\rm A}$ and $D_{\rm LS}$ as observables are not, on their own, sufficient for the distinction we seek between the $\Lambda$LTB and $\Lambda$CDM models, which justifies our choice to resort to the strong lensing phenomenon as a cosmological discrimination strategy.

\section{Imprints of $\Lambda$LTB on Strong Lensing}\label{ch:results}

\subsection{Strong lensing image simulator}\label{Lens_Ger}

To generate the images of the strong lensing phenomenon, we developed a light-mapping code, the strong lensing image simulator, 
described in  
Appendix~\ref{app:implimg}. 
The code receives three groups of inputs: the angular diameter distances of the gravitational lens system, calculated by the integrator of Sect.~\ref{DA_integracao}, through which the cosmology of the $\Lambda$LTB and $\Lambda$CDM models enters indirectly; the mass distribution profile of the lens, from which the effective lens potential of Eq.~\eqref{pot_eff} is extracted in angular coordinates; and the surface brightness profile of the source. 

The lens is modelled by the NIS profile of Eq.~\eqref{psi_NIS}, with $\sigma_v=250$~km\,s$^{-1}$ and $r_c=0.05''$, and the source by the S\'ersic profile of Eq.~\eqref{sersic}, with $R_{\rm eff}=0.8''$, $n=4$, and $r_{\rm core}=0.05''$. These values identify both the source and the lens as massive elliptical galaxies, and the lensing system considered in this work is thus a perfectly aligned ($\bm\beta_s=0$) galaxy--galaxy lens system. 

At the end of the process, the output of the simulator consists of the images of the lens and source planes, the respective critical and caustic curves, the $\kappa$ and $\gamma$ maps, and, if applied, the image convolved with a Gaussian point spread function (PSF). These images are idealised without statistical noise or instrumental effects, allowing the isolation of the purely cosmological geometry effects on the strong lensing phenomenon.

 The code was validated against the analytical and semi-analytical solutions for this circular lens and source (Appendix~\ref{app:implimg}), reproducing the reference solutions with high numerical precision, with all deviations below $10^{-3}$ and without relevant dependence on discretisation.
 
\subsection{Effective lensing distance and selection of optimal configurations}\label{Deff_res}

In order to explore the imprint of the non-homogeneous model on lensing through the angular diameter distances discussed in Sect.~\ref{DA_integracao}, we consider the effective lensing distance defined in Eq.~\eqref{Deff}. This quantity has a direct physical interpretation. On the one hand, by the definition of the critical surface density, Eq.~\eqref{Sigma_crit}, the larger $D_{\rm eff}$ is, the smaller the surface density required for the lens to produce multiple images. On the other hand, for a spherically symmetric lens, the physical Einstein radius in the lens plane, $\xi_E$ (Sect.~\ref{Strong_Lensing}), increases with $D_{\rm eff}$. In both cases, this distance can be interpreted as an indicator of the efficiency of the gravitational lensing effect.

\begin{figure}
    \centering
    \includegraphics[width=\linewidth]{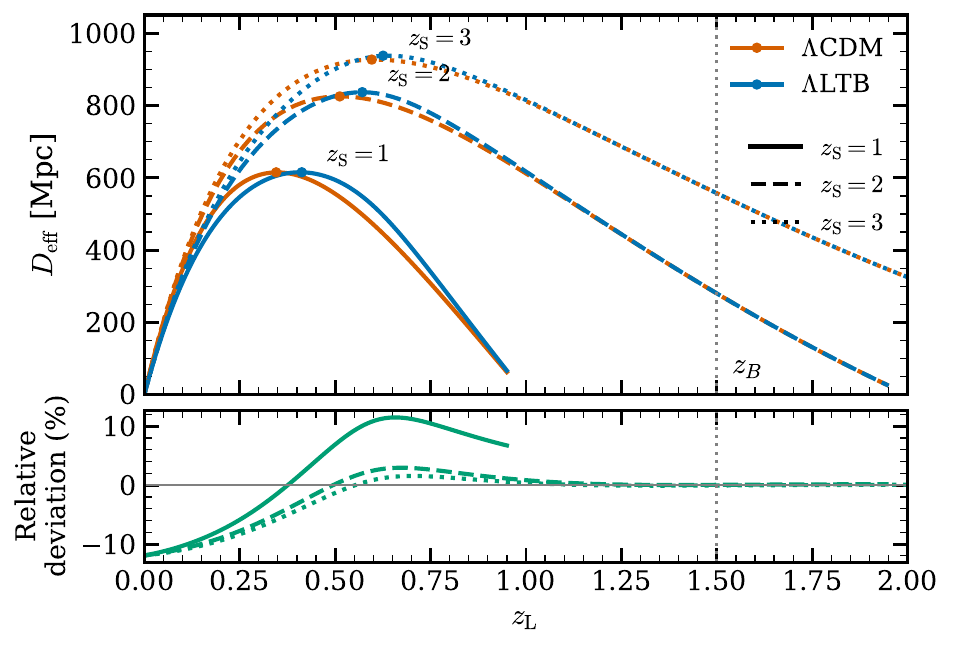}
    \caption{Effective lensing distance $D_{\rm eff}=D_{\rm L}D_{\rm LS}/D_{\rm S}$, defined in Eq.~\eqref{Deff}, as a function of lens redshift $z_{\rm L}$, for three illustrative sources at $z_{\rm S}=1$ (solid line), $2$ (dashed), and $3$ (dotted). {\em Top panel:} $D_{\rm eff}$ in Mpc for the $\Lambda$CDM (orange) and $\Lambda$LTB (blue) models, with the maximum of each curve marked by a dot. {\em Bottom panel}: Percentage relative deviation of the $\Lambda$LTB model with respect to the $\Lambda$CDM reference. The dotted vertical line marks the boundary of the underdensity at $z_B=1.5$.}
    \label{fig:Deff_zs}
\end{figure}

Figure~\ref{fig:Deff_zs} shows $D_{\rm eff}$ as a function of $z_{\rm L}$ for three illustrative sources, at $z_{\rm S}=1$, $2$, and $3$. For a fixed source at a given $z_{\rm S}$, $D_{\rm eff}$ vanishes at both ends of the optical path, that is, for $D_{\rm L}\to0$ and $D_{\rm LS}\to0$, with a peak between these two extreme lens positions\footnote{Note that this maximum value relates to the physical scale of the ring, since the angular Einstein radius $\theta_{\rm E}=\xi_E/D_{\rm L}$ decreases monotonically with $z_{\rm L}$.}. In all cases, the maximum for the $\Lambda$LTB model occurs at a higher lens redshift than the $\Lambda$CDM maximum, with an offset that decreases as the source moves away from the underdensity. The largest differences between the two models occur close to the observer, about $-12\%$, reproducing the deviation of $D_{\rm A}(0,z)$ (Fig.~\ref{fig:DA_os}, central panel), since $D_{\rm eff}\to D_{\rm L}$ when $z_{\rm L}\to0$, while away from the observer the deviation changes sign and peaks at $z_{\rm L}\approx0.7$.

The behaviour of $D_{\rm eff}$ indicates that there will be a combination of lens and source redshifts that maximises the difference between the lensing signal of the two models without compromising the efficiency of the lens. To find this optimal combination, a sweep of the $(z_{\rm S},z_{\rm L})$ parameter space was performed, with the angular diameter distances calculated on a grid with $z_{\rm L}\in[0.05,\,2]$ (100 values) and $z_{\rm S}\in[0.3,\,10]$ (200 values). This parameter space was restricted by the condition $z_{\rm L}<z_{\rm S}-0.05$, because the calculation of $D_{\rm LS}$ requires integrating Eq.~\eqref{eq_DA} between the lens and the source, and for excessively short redshift intervals the integrator produces insufficient points for the interpolation of the results. The adopted margin of $0.05$ guarantees this integration in all the curves of Fig.~\ref{fig:Deff_zs}, and the excluded strip has no relevance to the search\footnote{It was verified that reducing the margin to zero on the grid does not alter any of the obtained candidates.}, since $D_{\rm eff}\to0$ when $D_{\rm LS}\to0$. Each $D_{\rm eff}(z_{\rm L})$ curve of the type shown in Fig.~\ref{fig:Deff_zs} corresponds to a column of the grid, and the set of peaks of all columns defines an efficiency ridge, $z_{\rm L}^{\max}(z_{\rm S})$, across the $(z_{\rm S},z_{\rm L})$ plane. The $\Lambda$LTB ridge is located at higher $z_{\rm L}$ than the $\Lambda$CDM one, as infered from the maxima in Fig.~\ref{fig:Deff_zs}, with both ridges converging at high $z_{\rm S}$. It should be noted that $D_{\rm eff}$ only depends on the geometry of the lens system, meaning no specific profile is assumed for the lens throughout this entire process.

The searching process combines two metrics. The first is the relative ``efficiency of the lens'', which we define as the normalised $D_{\rm eff}$, either by the global maximum of the entire grid, defining the global efficiency parameter $E$, or by the local maxima per column, that is, for each source $z_{\rm S}$, defining the positional efficiency $\epsilon$, which equals 1 on the ridge and measures how well placed the lens is in the system. Both are calculated for $\Lambda$LTB, since it is the test model whose signature is to be compared with that of the reference $\Lambda$CDM. The second metric is the difference between the cosmological models, since nothing guarantees that the most efficient configuration is the most appropriate one for distinguishing between models. This is measured by the difference $\Delta D_{\rm eff}\equiv D_{\rm eff}^{\Lambda\mathrm{LTB}}-D_{\rm eff}^{\Lambda\mathrm{CDM}}$, whose sign indicates in which of the models the lens is more efficient, so that the search was conducted separately for both the $\Delta D_{\rm eff}>0$ and $\Delta D_{\rm eff}<0$ regimes.

We adopted five methods to select an optimal configuration. The methods are different combinations of the two metrics. They are defined in Appendix~\ref{app:selecao}, where the selection process is also detailed.
The five methods resulted in 16 different criteria, each one providing its top candidate. Running the image simulator, they were ranked according to the difference in critical radius between the $\Lambda$LTB and $\Lambda$CDM predictions (Appendix \ref{app:selecao}, Table~\ref{tab:decisao}). The overall best candidate, with $\Delta\theta_{\rm crit}=+0.123''$, is a configuration with the source at $z_{\rm S}=1.031$ and the lens at $z_{\rm L}=0.542$. We note that the six top candidates differ by less than one pixel. At this redshift, the lens is close to the maximum efficiency position for that source redshift, with $\epsilon=0.938$, being slightly shifted towards the source side. Both the lens and the source are inside the underdensity. These values are within the photometric redshift distribution of lenses identified in the recent \textit{Euclid} Data Release 1 (DR1), which ranges from $0.13$ to $1.98$ with a median of $0.68$ \citep{Euclid_Nersesian_2026}.

\subsection{Strong lensing simulation of the optimal configuration}\label{SL_map}

\begin{figure}
    \centering
    \includegraphics[width=\linewidth]{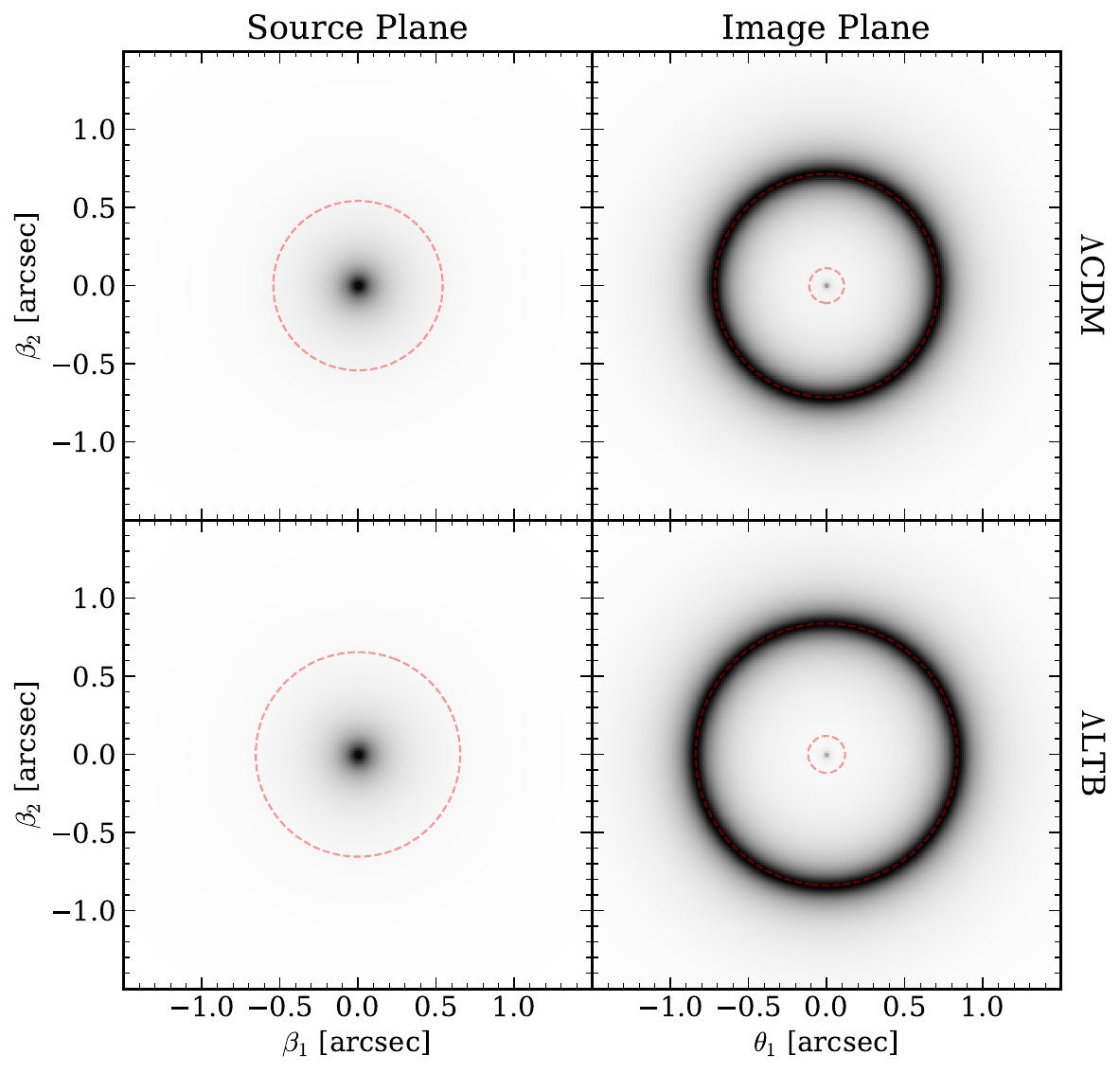}
    \caption{Gravitational lens mapping simulation for the selected configuration ($z_{\rm S}=1.031$, $z_{\rm L}=0.542$) for the $\Lambda$CDM ({\em top}) and $\Lambda$LTB ({\em bottom}) models. {\em Left panels}: Source plane, showing the S\'ersic source, where the dashed curve delimits the radial caustic; the tangential caustic degenerates into a point at the centre due to the circular symmetry of the system. {\em Right panels}: Image plane, showing the formation of the Einstein ring over the tangential critical curve (outermost red dashed line); the radial critical curve (innermost red dashed line) exists due to the lens core ($r_{\rm c}>0$).}
    \label{caso_especial}
\end{figure}

The simulation of the optimal configuration for the two models, with the parameters of Sect.~\ref{Lens_Ger}, is shown in Fig.~\ref{caso_especial}. The simulator produced rings, which is the expected image output from a perfectly aligned system and a spherical lens, and the traced critical curves and the corresponding caustics are consistent with the strong lensing theory discussed in Sect.~\ref{Strong_Lensing}, with the tangential critical curves located at $\theta_{\rm crit}^{\Lambda\rm CDM}=0.715''$ and $\theta_{\rm crit}^{\Lambda\rm LTB}=0.838''$, recovering $\Delta\theta_{\rm crit}=+0.123''$. These radii fall within the range of Einstein radii found in observed samples, as can be confirmed by the lens modelling of the \textit{Euclid} DR1 catalogue, which indicates a median effective Einstein radius of $0.85''$ for its successfully modelled high-confidence lens candidates, with a 16th--84th percentile range between $0.61''$ and $1.29''$ \citep{Euclid_Nightingale_2026}, so that the choice of the configuration that maximises the difference between the models does not lead to rings outside those currently observed.

\begin{table}
    \centering
    \footnotesize
    \caption{Properties of the simulated images for the optimal configuration ($z_{\rm S}=1.031$, $z_{\rm L}=0.542$) in each cosmological model.}
    \label{tab:config_selecionada}
    \begin{tabular}{lccccc}
        \toprule
        Model & $D_{\rm L}$ & $D_{\rm S}$ & $D_{\rm LS}$ & $\theta_{\rm E}$ & $\theta_{\rm crit}$ \\
        \midrule
        $\Lambda$CDM & $1362.0$ & $1717.3$ & $683.1$ & $0.717''$ & $0.715''$ \\
        $\Lambda$LTB & $1261.0$ & $1701.8$ & $792.5$ & $0.839''$ & $0.838''$ \\
        \midrule
        Rel.\ diff. & $-7.4\%$ & $-0.9\%$ & $+16.0\%$ & $+17.1\%$ & $+17.2\%$ \\
        \bottomrule
    \end{tabular}
    \tablefoot{Columns indicate angular diameter distances (in Mpc), SIS Einstein radius $\theta_{\rm E}$ computed from Eq.~\ref{theta_E_SIS}, and tangential critical curve radius $\theta_{\rm crit}$, measured from the images.}
\end{table}
The critical radius determined by the image simulator is shown in Table~\ref{tab:config_selecionada}, together with the Einstein radius computed from Eq.~\eqref{theta_E_SIS}, using the velocity dispersion of the lens and the distances computed with the integrator, as well as the various  distances. Since the lens model is the same for both cosmological models, the difference in the radii is essentially due to the increase in $D_{\rm LS}$ in $\Lambda$LTB, as $D_{\rm S}$ practically does not vary between the models (Sect.~\ref{DA_integracao}). The increase in $D_{\rm LS}$ is, in turn, an effect of the $\Lambda$LTB underdensity, and the algebraic relation of Eq.~\eqref{Approx_D_Ls} would underestimate it by about $6\%$ (resulting in $744$~Mpc), on the same order of magnitude as the effect one intends to measure.

\begin{figure}
    \centering
    \includegraphics[width=\linewidth]{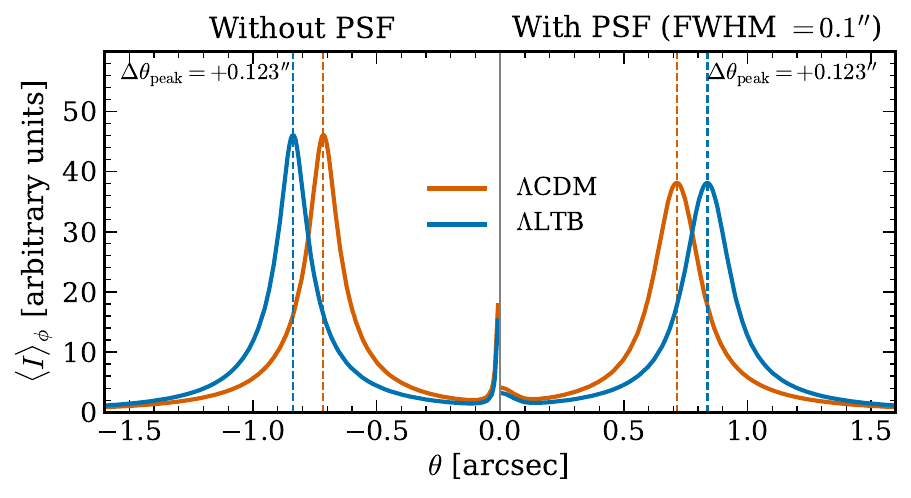}
    \caption{Mean radial brightness profile, in arbitrary units, of the Einstein ring for the selected configuration for the $\Lambda$CDM (orange) and $\Lambda$LTB (blue) models, shown without PSF for $\theta<0$ and convolved with a Gaussian PSF with an FWHM of $0.1''$ for $\theta>0$, since by the circular symmetry of the system $I(-\theta)=I(\theta)$. The dashed vertical lines mark $\pm\theta_{\rm crit}$, the tangential critical curve radius of each model, the separation $\Delta\theta_{\rm peak}$ between the two peaks is indicated in each half, and the central peak corresponds to the strongly demagnified central image.}
    \label{perfil_1D}
\end{figure}

The Einstein radius can also be estimated from the image, by extracting the mean radial brightness profile of the image plane for each model, that is, the azimuthal average of the intensity of the rings. This procedure allows us to estimate the peak of the ring with sub-pixel precision and is only valid due to the azimuthal symmetry tested in Appendix~\ref{app:implimg}, with results reported in Table~\ref{tab:validacao_lensing}. The profiles, with and without convolution by the PSF of Sect.~\ref{Lens_Ger}, are presented in Fig.~\ref{perfil_1D}. Without PSF, the ring is located at $\theta_{\rm peak}=0.716''$ for $\Lambda$CDM and $\theta_{\rm peak}=0.838''$ for $\Lambda$LTB, about $0.03$ pixels from the respective critical curves, so that $\Delta\theta_{\rm peak}=0.123''$ coincides with $\Delta\theta_{\rm crit}$, confirming that both measures describe the same effect. When convolved with the PSF, the ring thickness (FWHM) broadens from $0.17''$ to $0.23''$ and the peak of each model is displaced by about $0.1$ pixels inward, practically equally for both models, so that $\Delta\theta_{\rm peak}$ remains at $0.123''$, which corresponds to $9.8$ pixels of the simulation and $1.2$ times the FWHM of the PSF. This shows that the deviation induced by the $\Lambda$LTB underdensity does not constitute a numerical artefact, since it survives the resolution degradation imposed by the PSF. Nevertheless, we stress that these are clean simulated images without noise, an idealised scenario that does not account for the uncertainties inherent to a real observation.

We also considered additional image properties, and conclude that, in contrast with the Einstein radius, they do not discriminate between the models. The intensity of the peaks differs by less than $0.2\%$, as expected from the conservation of surface brightness (Sect.~\ref{GL}); the total flux is about $15\%$ higher in $\Lambda$LTB due to the larger area of the ring, but it can only be exploited with knowledge of the intrinsic luminosity of the source, which is generally inaccessible; and the peak at $\theta=0$, the odd image predicted by Burke's theorem for non-singular lenses (Sect.~\ref{Strong_Lensing}), is strongly demagnified and diluted by the pixel resolution. The radius of the ring is thus the primary geometric discriminator of the underdensity in a strong lensing analysis.

\subsection{Degeneracies between $\Lambda$LTB and $\Lambda$CDM}\label{SL_desg}

We found that the optimal lensing configuration produces Einstein radii for the two cosmological models that differ by $+0.123''$. It is worth investigating whether changing the values of the parameters of the lensing system would result in a $\Lambda$CDM Einstein radius identical to the value found for the $\Lambda$LTB model, in order to gain insight into the amplitude of the biases that could be made by assuming the $\Lambda$CDM model in a hypothetical $\Lambda$LTB universe.

According to Eq.~\eqref{theta_E_SIS}, the Einstein radius depends on the lens parameters and on the angular diameter distances in the lensing system, so that there are two sources of degeneracies that can produce biases: mass degeneracies, related to the velocity dispersion $\sigma_v$ of the lens galaxy, and geometric degeneracies, related to the lens and source redshifts. Since the observable is the Einstein ring, we consider that the models are indistinguishable at the $1\sigma$ level when $|\theta_{\rm E}^{\Lambda\mathrm{CDM}}-\theta_{\rm E}^{\Lambda\mathrm{LTB}}|\leq\sigma_{\theta_{\rm E}}$. For $\sigma_{\theta_{\rm E}}$, 
we adopt a value of  $2\%$ of $\theta_{\rm E}$,  
following \citet[Sect.~6.1]{Bolton_2008}. 

The geometric degeneracy requires redshift deviations, $\Delta z_{\rm L}=-0.07$ for a fixed source or $\Delta z_{\rm S}=+0.19$ for a fixed lens, that are sufficiently large when compared with high-precision spectroscopic redshifts \citep{Euclid_Q1_LeBrun_2026}, which are able to break it. We therefore focus on the mass degeneracy, keeping the redshifts fixed at the reference values in both models and letting the velocity dispersion of the lens vary in the $\Lambda$CDM simulation. 

\begin{figure}
    \centering
    \includegraphics[width=\linewidth]{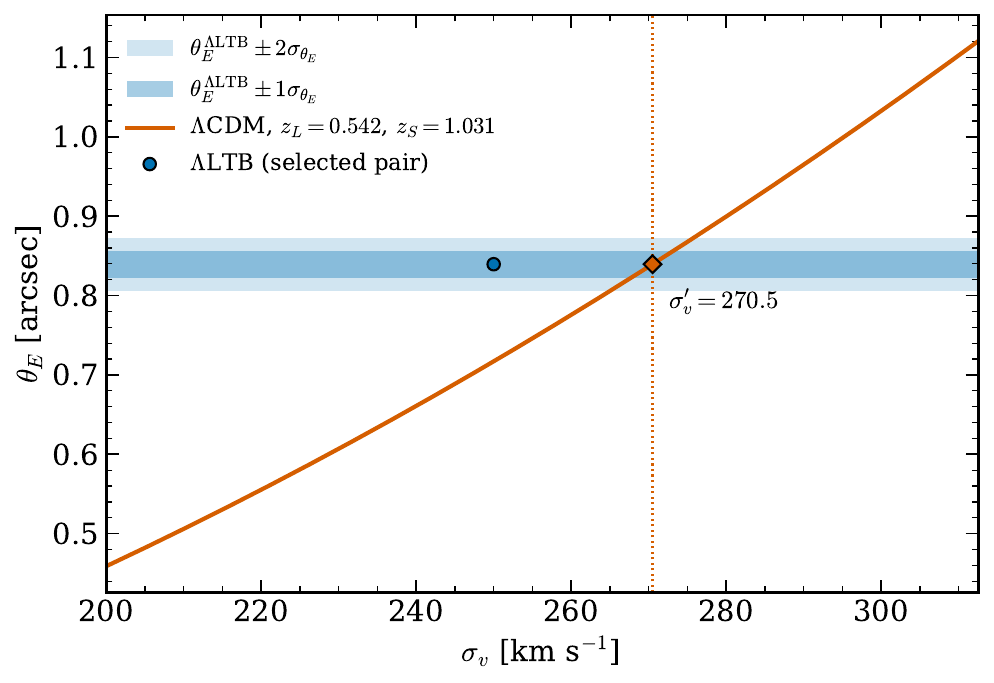}
    \caption{$\Lambda$CDM Einstein radius as a function of the lens velocity dispersion for the selected configuration ($z_{\rm S}=1.031$, $z_{\rm L}=0.542$). The horizontal bands indicate the adopted $1\sigma$ and $2\sigma$ levels around $\theta_{\rm E}^{\Lambda\mathrm{LTB}}$. The blue dot corresponds to the $\Lambda$LTB reference and the orange diamond, on the dotted vertical line, to the $\Lambda$CDM value $\sigma_v'$ needed to reproduce the reference $\theta_{\rm E}$.}
    \label{fig:degen_massa}
\end{figure}

Figure~\ref{fig:degen_massa} shows the resulting Einstein radius predicted by the $\Lambda$CDM model as a function of $\sigma_v$. We observe that the reference $\Lambda$LTB point lies above the $\Lambda$CDM curve, $7.3\,\sigma_{\theta_{\rm E}}$ away. Following the $\Lambda$CDM curve, we find that in order to reproduce the reference image in $\Lambda$CDM the lens would need to have $\sigma_v'=270.5$\,km\,s$^{-1}$, that is, a velocity dispersion $8.2\%$ higher.  
Moreover, the $\Lambda$CDM image simulated with $\sigma_v'$ coincides pixel by pixel with the $\Lambda$LTB image of the reference simulation, meaning the degeneracy is exact and not limited to the ring radius, extending to the entire image. The wrong assumption of a cosmological model can thus lead to a bias in the mass inferred for the lens. The Einstein mass usually reported in strong lensing studies, the projected mass inside the Einstein ring, satisfies $M_E\propto\theta_{\rm E}^2\,D_{\rm L}D_{\rm S}/D_{\rm LS}$, so that, since both models yield the same image and therefore the same $\theta_{\rm E}$, the difference in estimated mass depends entirely on the cosmological distance combination. A ring produced in a $\Lambda$LTB universe but interpreted as $\Lambda$CDM thus leads to an overestimation of the lens Einstein mass by about $26\%$ for this configuration, regardless of the assumed mass profile.

The question then follows whether the degeneracy can be broken by measurements independent of the cosmological model, such as the direct observation of $\sigma_v$ of the lens galaxy. Adopting a typical uncertainty of $7\%$ in $\sigma_v$ \citep{Bolton_2008}, the $\Lambda$CDM solution lies at $1.17\sigma$. Thus, spectroscopic measurements with this 
precision,
combined with the Einstein ring measurement, would not allow us to distinguish the two models. Recent  spectroscopic observations report, however, uncertainties in $\sigma_v$ of order $2\%$ to $4\%$ \citep{Mozumdar_2025}. Such data would place the $\Lambda$CDM prediction between $2\sigma$ and $4\sigma$ from the $\Lambda$LTB value, enabling one to disentangle the cosmological models from the measurement of the Einstein ring combined with independent measurements of the velocity dispersion. 

This discussion assumes that the measured velocity dispersion corresponds to that of the lens profile defined in the simulations, but the problem is lens-model-dependent. When comparing with real lenses, it is necessary to take into account the anisotropy of stellar orbits, the mass along the line of sight, and the fundamental  
mass-sheet degeneracy \citep{Falco_1985}, which can emulate or hide the differences to be tested, as well as degeneracies within each cosmological model, including those involving the $\Lambda$LTB parameters $\delta_0$ and $z_B$.


\section{Conclusions}\label{ch:conclusions}

We have presented a strong lensing image simulator that includes a solver for the lens-to-source angular diameter distance, based on the Sachs optical equations. 
This feature renders the simulator applicable to the study of non-homogeneous cosmologies, where the impossibility of defining a comoving distance prevents one to obtain the lens-to-source angular diameter distance algebraically. 
We describe the distance solver and the image simulator and present validation tests for both.

We applyied the simulator to the study of galaxy strong lensing systems in a non-homogeneous $\Lambda$LTB Universe. The $\Lambda$LTB background 
is computed with the \texttt{vd2020} code, which we interfaced with our simulator. 
In this first application of the simulator, we restrict ourselves to a $\Lambda$LTB model with a large underdense region (a void) centred at the observer.   

Unlike distances measured from the observer, the lens–source distance preserves the memory of the underdensity, up to large redshifts outside the void, which suggests the the Einstein ring as a potential observable to contain an imprint from the inhomogeneity. We make a thorough investigation of the lensing configurations that have the potential to produce meaningful differences between the Einstein rings predicted by $\Lambda$LTB and $\Lambda$CDM, while being at a redshift range where the lensing is efficient. We find that the Einstein ring can vary significantly in size between the two models, for the same lens mass and cosmological parameters. There is thus potential in the use of this observable, provided that the lens properties are independently constrained. This need can be circumvented by using ratios of Einstein radii with compound lenses data.

In future work, we will extend the formalism to observers located away from the centre of the inhomogeneity. The present framework was designed to accommodate this extension by generalising its distance module to propagate the full Jacobi matrix. This will allow us to investigate whether light propagation along non-radial paths produces direction-dependent distortions of rings and arcs that can be distinguished from the effects of lens ellipticity and structure along the line of sight. We will also extend the framework to predict time delays between multiple images and assess how radial inhomogeneity affects the expansion rate inferred from them, providing a complementary probe of the Hubble tension. Finally, we aim to study compound lens systems with sources at different redshifts and to assess which combinations of image observables, time delays, and independent lens-mass constraints can help break degeneracies and constrain the parameters of inhomogeneous cosmologies.


\bibliographystyle{aa}
\bibliography{Refs}

@ARTICLE{Mazo_2022,
       author = {{Mazo}, Brayan Yamid Del Valle and {Romano}, Antonio Enea and {Quintero}, Maryi Alejandra Carvajal},
        title = "{H0 tension or M overestimation?}",
      journal = {European Physical Journal C},
         year = 2022,
        month = jul,
       volume = {82},
       number = {7},
          eid = {610},
        pages = {610},
          doi = {10.1140/epjc/s10052-022-10526-3},
archivePrefix = {arXiv},
       eprint = {2202.11852},
 primaryClass = {astro-ph.CO},
       adsurl = {https://scixplorer.org/abs/2022EPJC...82..610M}
}

@article{Camarena2022Void,
  author        = {Camarena, David and Marra, Valerio and Sakr, Ziad and Clarkson, Chris},
  title         = {A void in the {Hubble} tension? The end of the line for the {Hubble} bubble},
  journal       = {Classical and Quantum Gravity},
  year          = {2022},
  volume        = {39},
  number        = {18},
  pages         = {184001},
  doi           = {10.1088/1361-6382/ac8635},
  eprint        = {2205.05422},
  archivePrefix = {arXiv},
  primaryClass  = {astro-ph.CO}
}

@article{Planck_2020,
   title={\textit{Planck} 2018 results: VII. Isotropy and statistics of the CMB},
   volume={641},
   ISSN={1432-0746},
   url={http://dx.doi.org/10.1051/0004-6361/201935201},
   DOI={10.1051/0004-6361/201935201},
   journal={\aap},
   publisher={EDP Sciences},
   author={{Planck Collaboration} and Akrami, Y. and Ashdown, M. and Aumont, J. and Baccigalupi, C. and Ballardini, M. and Banday, A. J. and Barreiro, R. B. and Bartolo, N. and Basak, S. and Benabed, K. and Bersanelli, M. and Bielewicz, P. and Bock, J. J. and Bond, J. R. and Borrill, J. and Bouchet, F. R. and Boulanger, F. and Bucher, M. and Burigana, C. and Butler, R. C. and Calabrese, E. and Cardoso, J.-F. and Casaponsa, B. and Chiang, H. C. and Colombo, L. P. L. and Combet, C. and Contreras, D. and Crill, B. P. and de Bernardis, P. and de Zotti, G. and Delabrouille, J. and Delouis, J.-M. and Di Valentino, E. and Diego, J. M. and Dor{\'e}, O. and Douspis, M. and Ducout, A. and Dupac, X. and Efstathiou, G. and Elsner, F. and En{\ss}lin, T. A. and Eriksen, H. K. and Fantaye, Y. and Fernandez-Cobos, R. and Finelli, F. and Frailis, M. and Fraisse, A. A. and Franceschi, E. and Frolov, A. and Galeotta, S. and Galli, S. and Ganga, K. and G{\'e}nova-Santos, R. T. and Gerbino, M. and Ghosh, T. and Gonz{\'a}lez-Nuevo, J. and G{\'o}rski, K. M. and Gruppuso, A. and Gudmundsson, J. E. and Hamann, J. and Handley, W. and Hansen, F. K. and Herranz, D. and Hivon, E. and Huang, Z. and Jaffe, A. H. and Jones, W. C. and Keih{\"a}nen, E. and Keskitalo, R. and Kiiveri, K. and Kim, J. and Krachmalnicoff, N. and Kunz, M. and Kurki-Suonio, H. and Lagache, G. and Lamarre, J.-M. and Lasenby, A. and Lattanzi, M. and Lawrence, C. R. and Le Jeune, M. and Levrier, F. and Liguori, M. and Lilje, P. B. and Lindholm, V. and L{\'o}pez-Caniego, M. and Ma, Y.-Z. and Mac{\'{i}}as-P{\'e}rez, J. F. and Maggio, G. and Maino, D. and Mandolesi, N. and Mangilli, A. and Marcos-Caballero, A. and Maris, M. and Martin, P. G. and Mart{\'{i}}nez-Gonz{\'a}lez, E. and Matarrese, S. and Mauri, N. and McEwen, J. D. and Meinhold, P. R. and Mennella, A. and Migliaccio, M. and Miville-Desch{\^e}nes, M.-A. and Molinari, D. and Moneti, A. and Montier, L. and Morgante, G. and Moss, A. and Natoli, P. and Pagano, L. and Paoletti, D. and Partridge, B. and Perrotta, F. and Pettorino, V. and Piacentini, F. and Polenta, G. and Puget, J.-L. and Rachen, J. P. and Reinecke, M. and Remazeilles, M. and Renzi, A. and Rocha, G. and Rosset, C. and Roudier, G. and Rubi{\~n}o-Mart{\'{i}}n, J. A. and Ruiz-Granados, B. and Salvati, L. and Savelainen, M. and Scott, D. and Shellard, E. P. S. and Sirignano, C. and Sunyaev, R. and Suur-Uski, A.-S. and Tauber, J. A. and Tavagnacco, D. and Tenti, M. and Toffolatti, L. and Tomasi, M. and Trombetti, T. and Valenziano, L. and Valiviita, J. and Van Tent, B. and Vielva, P. and Villa, F. and Vittorio, N. and Wandelt, B. D. and Wehus, I. K. and Zacchei, A. and Zibin, J. P. and Zonca, A.},
   year={2020},
   month=sep,
   pages={A7} }

@article{Cyburt_2016,
   title={Big bang nucleosynthesis: Present status},
   volume={88},
   ISSN={1539-0756},
   url={http://dx.doi.org/10.1103/RevModPhys.88.015004},
   DOI={10.1103/revmodphys.88.015004},
   number={1},
   journal={Rev. Mod. Phys.},
   publisher={American Physical Society (APS)},
   author={Cyburt, Richard H. and Fields, Brian D. and Olive, Keith A. and Yeh, Tsung-Han},
   year={2016},
   month=feb }

@article{Eisenstein_2005,
   title={Detection of the Baryon Acoustic Peak in the Large-Scale Correlation Function of SDSS Luminous Red Galaxies},
   volume={633},
   ISSN={1538-4357},
   url={http://dx.doi.org/10.1086/466512},
   DOI={10.1086/466512},
   number={2},
   journal={\apj},
   publisher={American Astronomical Society},
   author={Eisenstein, Daniel J. and Zehavi, Idit and Hogg, David W. and Scoccimarro, Roman and Blanton, Michael R. and Nichol, Robert C. and Scranton, Ryan and Seo, Hee-Jong and Tegmark, Max and Zheng, Zheng and Anderson, Scott F. and Annis, Jim and Bahcall, Neta and Brinkmann, Jon and Burles, Scott and Castander, Francisco J. and Connolly, Andrew and Csabai, Istvan and Doi, Mamoru and Fukugita, Masataka and Frieman, Joshua A. and Glazebrook, Karl and Gunn, James E. and Hendry, John S. and Hennessy, Gregory and Ivezi{\'c}, Zeljko and Kent, Stephen and Knapp, Gillian R. and Lin, Huan and Loh, Yeong-Shang and Lupton, Robert H. and Margon, Bruce and McKay, Timothy A. and Meiksin, Avery and Munn, Jeffery A. and Pope, Adrian and Richmond, Michael W. and Schlegel, David and Schneider, Donald P. and Shimasaku, Kazuhiro and Stoughton, Christopher and Strauss, Michael A. and SubbaRao, Mark and Szalay, Alexander S. and Szapudi, Istvan and Tucker, Douglas L. and Yanny, Brian and York, Donald G.},
   year={2005},
   month=nov, pages={560--574} }

@article{Perlmutter_1998,
       author = {{Perlmutter}, S. and {Aldering}, G. and {della Valle}, M. and {Deustua}, S. and {Ellis}, R.~S. and {Fabbro}, S. and {Fruchter}, A. and {Goldhaber}, G. and {Groom}, D.~E. and {Hook}, I.~M. and et al.},
        title = {Discovery of a supernova explosion at half the age of the Universe},
      journal = {\nat},
         year = 1998,
        month = jan,
       volume = {391},
       number = {6662},
        pages = {51-54},
          doi = {10.1038/34124},
archivePrefix = {arXiv},
       eprint = {astro-ph/9712212},
 primaryClass = {astro-ph},
       url = {https://ui.adsabs.harvard.edu/abs/1998Natur.391...51P}
}

@article{Riess_1998,
   title={Observational Evidence from Supernovae for an Accelerating Universe and a Cosmological Constant},
   volume={116},
   ISSN={0004-6256},
   url={http://dx.doi.org/10.1086/300499},
   DOI={10.1086/300499},
   number={3},
   journal={\aj},
   publisher={American Astronomical Society},
   author={Riess, Adam G. and Filippenko, Alexei V. and Challis, Peter and Clocchiatti, Alejandro and Diercks, Alan and Garnavich, Peter M. and Gilliland, Ron L. and Hogan, Craig J. and Jha, Saurabh and Kirshner, Robert P. and Leibundgut, B. and Phillips, M. M. and Reiss, David and Schmidt, Brian P. and Schommer, Robert A. and Smith, R. Chris and Spyromilio, J. and Stubbs, Christopher and Suntzeff, Nicholas B. and Tonry, John},
   year={1998},
   month=sep, pages={1009--1038} }

@article{Riess_2016,
   title={A 2.4\% Determination of the Local Value of the Hubble Constant},
   volume={826},
   ISSN={1538-4357},
   url={http://dx.doi.org/10.3847/0004-637X/826/1/56},
   DOI={10.3847/0004-637x/826/1/56},
   number={1},
   journal={\apj},
   publisher={American Astronomical Society},
   author={Riess, Adam G. and Macri, Lucas M. and Hoffmann, Samantha L. and Scolnic, Dan and Casertano, Stefano and Filippenko, Alexei V. and Tucker, Brad E. and Reid, Mark J. and Jones, David O. and Silverman, Jeffrey M. and Chornock, Ryan and Challis, Peter and Yuan, Wenlong and Brown, Peter J. and Foley, Ryan J.},
   year={2016},
   month=jul, pages={56} }

@article{Riess_2019,
   title={Large Magellanic Cloud Cepheid Standards Provide a 1\% Foundation for the Determination of the Hubble Constant and Stronger Evidence for Physics beyond {\ensuremath{\Lambda}}CDM},
   volume={876},
   ISSN={1538-4357},
   url={http://dx.doi.org/10.3847/1538-4357/ab1422},
   DOI={10.3847/1538-4357/ab1422},
   number={1},
   journal={\apj},
   publisher={American Astronomical Society},
   author={Riess, Adam G. and Casertano, Stefano and Yuan, Wenlong and Macri, Lucas M. and Scolnic, Dan},
   year={2019},
   month=may, pages={85} }

@article{Riess_2022,
   title={A Comprehensive Measurement of the Local Value of the Hubble Constant with 1 km s$^{-1}$ Mpc$^{-1}$ Uncertainty from the Hubble Space Telescope and the SH0ES Team},
   volume={934},
   ISSN={2041-8213},
   url={http://dx.doi.org/10.3847/2041-8213/ac5c5b},
   DOI={10.3847/2041-8213/ac5c5b},
   number={1},
   journal={\apjl},
   publisher={American Astronomical Society},
   author={Riess, Adam G. and Yuan, Wenlong and Macri, Lucas M. and Scolnic, Dan and Brout, Dillon and Casertano, Stefano and Jones, David O. and Murakami, Yukei and Anand, Gagandeep S. and Breuval, Louise and Brink, Thomas G. and Filippenko, Alexei V. and Hoffmann, Samantha and Jha, Saurabh W. and D'arcy Kenworthy, W. and Mackenty, John and Stahl, Benjamin E. and Zheng, WeiKang},
   year={2022},
   month=jul, pages={L7}}

@ARTICLE{Mellier_2025,
       author = {{Euclid Collaboration} and {Mellier}, Y. and {Abdurro'uf} and {Acevedo Barroso}, J.~A. and {Ach{\'u}carro}, A. and {Adamek}, J. and {Adam}, R. and {Addison}, G.~E. and {Aghanim}, N. and {Aguena}, M. and et al.},
        title = {Euclid: I. Overview of the Euclid mission},
      journal = {\aap},
         year = 2025,
        month = may,
       volume = {697},
          eid = {A1},
        pages = {A1},
          doi = {10.1051/0004-6361/202450810},
archivePrefix = {arXiv},
       eprint = {2405.13491},
 primaryClass = {astro-ph.CO},
       url = {https://ui.adsabs.harvard.edu/abs/2025A&A...697A...1E}
}

@ARTICLE{DiValentino_2021,
       author = {{Di Valentino}, Eleonora and {Mena}, Olga and {Pan}, Supriya and {Visinelli}, Luca and {Yang}, Weiqiang and {Melchiorri}, Alessandro and {Mota}, David F. and {Riess}, Adam G. and {Silk}, Joseph},
        title = "{In the realm of the Hubble tension-a review of solutions}",
      journal = {Class. Quantum Grav.},
         year = 2021,
        month = jul,
       volume = {38},
       number = {15},
          eid = {153001},
        pages = {153001},
          doi = {10.1088/1361-6382/ac086d},
archivePrefix = {arXiv},
      eprint = {2103.01183},
 primaryClass = {astro-ph.CO},
       url = {https://ui.adsabs.harvard.edu/abs/2021CQGra..38o3001D}
}

@article{Dunsby_2010,
   title={How close can an inhomogeneous universe mimic the concordance model?},
   volume={2010},
   ISSN={1475-7516},
   url={http://dx.doi.org/10.1088/1475-7516/2010/06/017},
   DOI={10.1088/1475-7516/2010/06/017},
   number={06},
   journal={\jcap},
   publisher={IOP Publishing},
   author={Dunsby, Peter and Goheer, Naureen and Osano, Bob and Uzan, Jean-Philippe},
   year={2010},
   month=jun, pages={017} }

@ARTICLE{Sakr_2024,
       author = {{Sakr}, Z. and {Carvalho}, A. and {Da Silva}, A. and {Garc{\'{i}}a-Bellido}, J. and {Mimoso}, J.~P. and {Camarena}, D. and {Nesseris}, S. and {Martins}, C.~J.~A.~P. and {Nunes}, N. and {Sapone}, D.},
        title = {Constraining {\ensuremath{\Lambda}}LTB models with galaxy cluster counts from next-generation surveys},
      journal = {\aap},
         year = 2024,
        month = mar,
       volume = {683},
          eid = {A230},
        pages = {A230},
          doi = {10.1051/0004-6361/202348200},
    archivePrefix = {arXiv},
       eprint = {2309.17151},
 primaryClass = {astro-ph.CO},
       url = {https://ui.adsabs.harvard.edu/abs/2024A&A...683A.230S}
}

@article{Camarena_2021,
   title={The Copernican principle in light of the latest cosmological data},
   volume={509},
   ISSN={1365-2966},
   url={http://dx.doi.org/10.1093/mnras/stab3077},
   DOI={10.1093/mnras/stab3077},
   number={1},
   journal={\mnras},
   publisher={Oxford University Press (OUP)},
   author={Camarena, David and Marra, Valerio and Sakr, Ziad and Clarkson, Chris},
   year={2022},
   month=jan, pages={1291--1302} }

@PHDTHESIS{Camarena_2022,
       author = {{Camarena}, David},
        title = "{Pushing the boundaries of modern cosmology: physics beyond the Copernican principle}",
       school = {Universidade Federal do Esp{\'{i}}rito Santo},
         year = 2022,
        month = jul,
       adsurl = {https://ui.adsabs.harvard.edu/abs/2022PhDT........27C}
}

@article{Lemaitre:1933gd,
    author = "Lema{\^{i}}tre, G.",
    title = "{The expanding universe}",
    doi = "10.1023/A:1018855621348",
    journal = "Annales Soc. Sci. Bruxelles A",
    volume = "53",
    pages = "51--85",
    year = "1933"
}

@article{Tolman:1934pn,
    author = "Tolman, R. C.",
    title = "{Effect of Inhomogeneity on Cosmological Models}",
    doi = "10.1073/pnas.20.3.169",
    journal = "Proc. Nat. Acad. Sci.",
    volume = "20",
    pages = "169--176",
    year = "1934"
}

@article{Bondi_1947,
    author = {Bondi, H.},
    title = {Spherically Symmetrical Models in General Relativity},
    journal = {\mnras},
    volume = {107},
    number = {5-6},
    pages = {410-425},
    year = {1947},
    month = dec,
    issn = {0035-8711},
    doi = {10.1093/mnras/107.5-6.410},
    url = {https://doi.org/10.1093/mnras/107.5-6.410}
}

@article{Valkenburg_2012,
   title={Complete solutions to the metric of spherically collapsing dust in an expanding spacetime with a cosmological constant},
   volume={44},
   ISSN={1572-9532},
   url={http://dx.doi.org/10.1007/s10714-012-1405-9},
   DOI={10.1007/s10714-012-1405-9},
   number={10},
   journal={Gen. Relativ. Gravit.},
   publisher={Springer Science and Business Media LLC},
   author={Valkenburg, Wessel},
   year={2012},
   month=jul, pages={2449--2476} }

@BOOK{Meneghetti_2022,
       author = {{Meneghetti}, Massimo},
        title = "{Introduction to Gravitational Lensing: With Python Examples}",
        publisher = {Springer},
         year = 2022,
       adsurl = {https://ui.adsabs.harvard.edu/abs/2022iglp.book.....M}
}

@phdthesis{Fleury2015thesis,
  author        = {Fleury, Pierre},
  title         = {Light propagation in inhomogeneous and anisotropic cosmologies},
  school        = {Universit\'e Pierre et Marie Curie},
  year          = {2015},
  note          = {},
  eprint        = {1511.03702},
  archivePrefix = {arXiv},
  primaryClass  = {gr-qc},
  url           = {https://arxiv.org/abs/1511.03702}
}

@article{LEWIS20061,
title = {Weak gravitational lensing of the CMB},
journal = {\physrep},
volume = {429},
number = {1},
pages = {1-65},
year = {2006},
issn = {0370-1573},
doi = {10.1016/j.physrep.2006.03.002},
url = {https://www.sciencedirect.com/science/article/pii/S0370157306000810},
author = {Antony Lewis and Anthony Challinor}
}

@article{Bartelmann_2010,
   title={Gravitational lensing},
   volume={27},
   ISSN={1361-6382},
   url={http://dx.doi.org/10.1088/0264-9381/27/23/233001},
   DOI={10.1088/0264-9381/27/23/233001},
   number={23},
   journal={Class. Quantum Grav.},
   publisher={IOP Publishing},
   author={Bartelmann, Matthias},
   year={2010},
   month=nov, pages={233001}
}

@misc{narayan_1997,
      title={Lectures on Gravitational Lensing},
      author={Ramesh Narayan and Matthias Bartelmann},
      year={1997},
      eprint={astro-ph/9606001},
      archivePrefix={arXiv},
      primaryClass={astro-ph},
      url={https://arxiv.org/abs/astro-ph/9606001}
}

@ARTICLE{Schneider_1985,
       author = {{Schneider}, P.},
        title = "{A new formulation of gravitational lens theory, time-delay, and Fermat's principle}",
      journal = {\aap},
         year = 1985,
        month = feb,
       volume = {143},
       number = {2},
        pages = {413-420},
       url = {https://ui.adsabs.harvard.edu/abs/1985A&A...143..413S}
}

@article{Sarma_2026,
   title={Covariant cosmography in the presence of local structures: comparing exact solutions and perturbation theory},
   volume={2026},
   ISSN={1475-7516},
   url={http://dx.doi.org/10.1088/1475-7516/2026/04/069},
   DOI={10.1088/1475-7516/2026/04/069},
   number={04},
   journal={\jcap},
   publisher={IOP Publishing},
   author={Sarma, Maharshi and Marinoni, Christian and Kalbouneh, Basheer and Clarkson, Chris and Maartens, Roy},
   year={2026},
   month=apr, pages={069} }

@article{R_s_nen_2010,
   title={Light propagation in statistically homogeneous and isotropic universes with general matter content},
   volume={2010},
   ISSN={1475-7516},
   url={http://dx.doi.org/10.1088/1475-7516/2010/03/018},
   DOI={10.1088/1475-7516/2010/03/018},
   number={03},
   journal={\jcap},
   publisher={IOP Publishing},
   author={R{\"a}s{\"a}nen, Syksy},
   year={2010},
   month=mar, pages={018} }

@inproceedings{Fike2011,
  author    = {Fike, Jeffrey A. and Alonso, Juan J.},
  title     = {The Development of Hyper-Dual Numbers for Exact Second-Derivative Calculations},
  booktitle = {49th AIAA Aerospace Sciences Meeting including the New Horizons Forum and Aerospace Exposition},
  month     = jan,
  address   = {Orlando, Florida},
  year      = {2011},
  doi       = {10.2514/6.2011-886}
}

@article{Wong_2019,
   title={H0LiCOW -- XIII. A 2.4 per cent measurement of $H_0$ from lensed quasars: $5.3\sigma$ tension between early- and late-Universe probes},
   volume={498},
   ISSN={1365-2966},
   url={http://dx.doi.org/10.1093/mnras/stz3094},
   DOI={10.1093/mnras/stz3094},
   number={1},
   journal={\mnras},
   publisher={Oxford University Press (OUP)},
   author={Wong, Kenneth C and Suyu, Sherry H and Chen, Geoff C-F and Rusu, Cristian E and Millon, Martin and Sluse, Dominique and Bonvin, Vivien and Fassnacht, Christopher D and Taubenberger, Stefan and Auger, Matthew W and Birrer, Simon and Chan, James H H and Courbin, Frederic and Hilbert, Stefan and Tihhonova, Olga and Treu, Tommaso and Agnello, Adriano and Ding, Xuheng and Jee, Inh and Komatsu, Eiichiro and Shajib, Anowar J and Sonnenfeld, Alessandro and Blandford, Roger D and Koopmans, L{\'e}on V E and Marshall, Philip J and Meylan, Georges},
   year={2019},
   month=sep, pages={1420--1439} }

@ARTICLE{McKenzie_1985,
       author = {{McKenzie}, R.~H.},
        title = "{A gravitational lens produces an odd number of images.}",
      journal = {J. Math. Phys.},
         year = 1985,
        month = jan,
       volume = {26},
       number = {7},
        pages = {1592-1597},
          doi = {10.1063/1.526923},
       adsurl = {https://ui.adsabs.harvard.edu/abs/1985JMP....26.1592M}
}

@ARTICLE{Burke_1981,
       author = {{Burke}, W.~L.},
        title = "{Multiple Gravitational Imaging by Distributed Masses}",
      journal = {\apjl},
         year = 1981,
        month = feb,
       volume = {244},
        pages = {L1},
          doi = {10.1086/183466},
       adsurl = {https://ui.adsabs.harvard.edu/abs/1981ApJ...244L...1B}
}

@misc{jax2018github,
  author  = {Bradbury, James and Frostig, Roy and Hawkins, Peter and Johnson, Matthew James and Leary, Chris and Maclaurin, Dougal and Necula, George and Paszke, Adam and Vander{P}las, Jake and Wanderman-Milne, Skye and Zhang, Qiao},
  title   = {{JAX}: composable transformations of {P}ython+{N}um{P}y programs},
  howpublished = {\url{http://github.com/jax-ml/jax}},
  note    = {Version 0.6.2},
  year    = {2018}
}

@book{Pareto1906,
  author    = {Pareto, Vilfredo},
  title     = {Manuale di Economia Politica con una Introduzione alla Scienza Sociale},
  publisher = {Società Editrice Libraria},
  address   = {Milano},
  year      = {1906},
  url       = {https://archive.org/details/manualedieconomi00pareuoft}
}

@book{Miettinen1998,
  author    = {Miettinen, Kaisa},
  title     = {Nonlinear Multiobjective Optimization},
  publisher = {Springer},
  address   = {Boston, MA},
  year      = {1998},
  doi       = {10.1007/978-1-4615-5563-6}
}

@article{Das1999,
  author  = {Das, I.},
  title   = {On characterizing the ``knee'' of the {P}areto curve based on {N}ormal-{B}oundary {I}ntersection},
  journal = {Struct. Optim.},
  volume  = {18},
  number  = {2},
  pages   = {107--115},
  year    = {1999},
  month   = oct,
  issn    = {1615-1488},
  doi     = {10.1007/BF01195985},
  url     = {https://doi.org/10.1007/BF01195985}
}

@manual{ACS_IHB,
  author       = {{Stark}, D. V. and others},
  title        = {{ACS} Instrument Handbook, Version 25.0},
  organization = {Space Telescope Science Institute},
  address      = {Baltimore},
  year         = {2025}
}

@ARTICLE{Euclid_Vis2025,
       author = {{Euclid Collaboration} and {Cropper}, M.~S. and {Al-Bahlawan}, A. and {Amiaux}, J. and {Awan}, S. and {Azzollini}, R. and {Benson}, K. and {Berthe}, M. and {Boucher}, J. and {Bozzo}, E. and {Brockley-Blatt}, C. and {Candini}, G.~P. and {Cara}, C. and {Chaudery}, R.~A. and {Cole}, R.~E. and {Danto}, P. and {Denniston}, J. and {Di Giorgio}, A.~M. and {Dryer}, B. and {Dubois}, J.-P. and {Endicott}, J. and {Farina}, M. and {Galli}, E. and {Genolet}, L. and {Gow}, J.~P.~D. and {Guttridge}, P. and {Hailey}, M. and {Hall}, D. and {Harper}, C. and {Hoekstra}, H. and {Holland}, A.~D. and {Horeau}, B. and {Hu}, D. and {James}, R.~E. and {Khalil}, A. and {King}, R. and {Kitching}, T. and {Kohley}, R. and {Larcheveque}, C. and {Lawrenson}, A. and {Liebing}, P. and {Liu}, S.~J. and {Martignac}, J. and {Massey}, R. and {McCracken}, H.~J. and {Miller}, L. and {Murray}, N. and {Nakajima}, R. and {Niemi}, S.-M. and {Nightingale}, J.~W. and {Paltani}, S. and {Pendem}, A. and {Philippon}, A. and {Plana}, C. and {Pool}, P. and {Pottinger}, S. and {Racca}, G.~D. and {Rhodes}, J. and {Rousseau}, A. and {Ruane}, K. and {Salatti}, M. and {Salvignol}, J.-C. and {Sciortino}, A. and {Short}, A. and {Skottfelt}, J. and {Smit}, S.~J.~A. and {Swindells}, I. and {Szafraniec}, M. and {Thomas}, P.~D. and {Thomas}, W. and {Tommasi}, E. and {Tosti}, S. and {Visticot}, F. and {Walton}, D.~M. and {Willis}, G. and {Winter}, B. and {Aghanim}, N. and {Altieri}, B. and {Amara}, A. and {Andreon}, S. and {Auricchio}, N. and {Aussel}, H. and {Baccigalupi}, C. and {Baldi}, M. and {Balestra}, A. and {Bardelli}, S. and {Basset}, A. and {Bender}, R. and {Bernardeau}, F. and {Bodendorf}, C. and {Boenke}, T. and {Bonino}, D. and {Branchini}, E. and {Brescia}, M. and {Brinchmann}, J. and {Camera}, S. and {Capobianco}, V. and {Carbone}, C. and {Cardone}, V.~F. and {Carretero}, J. and {Casas}, R. and {Casas}, S. and {Castander}, F.~J. and {Castellano}, M. and {Castignani}, G. and {Cavuoti}, S. and {Cimatti}, A. and {Colodro-Conde}, C. and {Congedo}, G. and {Conselice}, C.~J. and {Conversi}, L. and {Copin}, Y. and {Courbin}, F. and {Courtois}, H.~M. and {Crocce}, M. and {Cuby}, J.-G. and {Cuillandre}, J.-C. and {Da Silva}, A. and {Degaudenzi}, H. and {De Lucia}, G. and {Dinis}, J. and {Dolding}, C. and {Douspis}, M. and {Duncan}, C.~A.~J. and {Dupac}, X. and {Dusini}, S. and {Ealet}, A. and {Fabricius}, M. and {Farrens}, S. and {Ferriol}, S. and {Fosalba}, P. and {Fotopoulou}, S. and {Frailis}, M. and {Franceschi}, E. and {Franzetti}, P. and {Frugier}, P.-A. and {Fumana}, M. and {Galeotta}, S. and {Garilli}, B. and {George}, K. and {Gillard}, W. and {Gillis}, B. and {Giocoli}, C. and {G{\'o}mez-Alvarez}, P. and {Granett}, B.~R. and {Grazian}, A. and {Grupp}, F. and {Guzzo}, L. and {Haugan}, S.~V.~H. and {Herent}, O. and {Hoar}, J. and {Holliman}, M.~S. and {Holmes}, W. and {Hook}, I. and {Hormuth}, F. and {Hornstrup}, A. and {Hudelot}, P. and {Ili{\'c}}, S. and {Jahnke}, K. and {Jhabvala}, M. and {Joachimi}, B. and {Keih{\"a}nen}, E. and {Kermiche}, S. and {Kiessling}, A. and {Kilbinger}, M. and {Kubik}, B. and {Kuijken}, K. and {K{\"u}mmel}, M. and {Kunz}, M. and {Kurki-Suonio}, H. and {Lahav}, O. and {Laureijs}, R. and {Ligori}, S. and {Lilje}, P.~B. and {Lindholm}, V. and {Lloro}, I. and {Lorenzo Alvarez}, J. and {Mainetti}, G. and {Maino}, D. and {Maiorano}, E. and {Mansutti}, O. and {Marcin}, S. and {Marggraf}, O. and {Markovic}, K. and {Martinelli}, M. and {Martinet}, N. and {Marulli}, F. and {Masters}, D.~C. and {Maurogordato}, S. and {Medinaceli}, E. and {Mei}, S. and {Melchior}, M. and {Mellier}, Y. and {Meneghetti}, M. and {Merlin}, E. and {Meylan}, G. and {Mohr}, J.~J. and {Moresco}, M. and {Moscardini}, L. and {Neissner}, C.},
        title = "{Euclid: II. The VIS instrument}",
      journal = {\aap},
         year = 2025,
        month = may,
       volume = {697},
          eid = {A2},
        pages = {A2},
          doi = {10.1051/0004-6361/202450996},
archivePrefix = {arXiv},
       eprint = {2405.13492},
 primaryClass = {astro-ph.IM},
       url = {https://ui.adsabs.harvard.edu/abs/2025A&A...697A...2E}
}

@article{Bartelmann_2001,
   title={Weak gravitational lensing},
   volume={340},
   ISSN={0370-1573},
   url={http://dx.doi.org/10.1016/S0370-1573(00)00082-X},
   DOI={10.1016/s0370-1573(00)00082-x},
   number={4-5},
   journal={\physrep},
   publisher={Elsevier BV},
   author={Bartelmann, Matthias and Schneider, Peter},
   year={2001},
   month=Jan, pages={291–472} }

@article{Bolton_2008,
   title={The Sloan Lens ACS Survey. V. The Full ACS Strong‐Lens Sample},
   volume={682},
   ISSN={1538-4357},
   url={http://dx.doi.org/10.1086/589327},
   DOI={10.1086/589327},
   number={2},
   journal={\apj},
   publisher={American Astronomical Society},
   author={Bolton, Adam S. and Burles, Scott and Koopmans, Léon V. E. and Treu, Tommaso and Gavazzi, Raphaël and Moustakas, Leonidas A. and Wayth, Randall and Schlegel, David J.},
   year={2008},
   month=Aug, pages={964–984} }

@article{Mozumdar_2025,
   title={TDCOSMO: XXI. Accurate stellar velocity dispersions of the SL2S lens sample and the fundamental plane of the lensing mass},
   volume={705},
   ISSN={1432-0746},
   url={http://dx.doi.org/10.1051/0004-6361/202555583},
   DOI={10.1051/0004-6361/202555583},
   journal={\aap},
   publisher={EDP Sciences},
   author={Mozumdar, Pritom and Knabel, Shawn and Treu, Tommaso and Sonnenfeld, Alessandro and Shajib, Anowar J. and Cappellari, Michele and Nipoti, Carlo},
   year={2025},
   month=Dec, pages={A13} }

@article{Euclid_Q1_LeBrun_2026,
  author        = {{Euclid Collaboration} and Le Brun, V. and Bethermin, M. and Moresco, M. and
                   Vibert, D. and Vergani, D. and Surace, C. and Zamorani, G. and others},
  title         = {{Euclid} {Quick Data Release} ({Q1}): {VII}. {Characteristics} and limitations
                   of the spectroscopic measurements},
  journal       = {\aap},
  volume        = {711},
  pages         = {A7},
  year          = {2026},
  doi           = {10.1051/0004-6361/202554607},
  eprint        = {2503.15308},
  archivePrefix = {arXiv},
  primaryClass  = {astro-ph.CO}
}

@ARTICLE{Falco_1985,
       author = {{Falco}, E.~E. and {Gorenstein}, M.~V. and {Shapiro}, I.~I.},
        title = "{On model-dependent bounds on $H_0$ from gravitational images : application to Q 0957+561 A, B.}",
      journal = {\apjl},
         year = 1985,
        month = feb,
       volume = {289},
        pages = {L1-L4},
          doi = {10.1086/184422},
       adsurl = {https://ui.adsabs.harvard.edu/abs/1985ApJ...289L...1F}
}

@article{Graham_2005,
   title={A Concise Reference to (Projected) Sérsic
                    $\rm{R}^{1/{\rm n}}$
                    Quantities, Including Concentration, Profile Slopes, Petrosian Indices, and Kron Magnitudes},
   volume={22},
   ISSN={1448-6083},
   url={http://dx.doi.org/10.1071/AS05001},
   DOI={10.1071/as05001},
   number={2},
   journal={PASA},
   publisher={Cambridge University Press (CUP)},
   author={Graham, Alister W. and Driver, Simon P.},
   year={2005},
   pages={118–127} }

@article{Lines_2025,
   title={The revolution in strong lensing discoveries from Euclid},
   volume={9},
   ISSN={2397-3366},
   url={http://dx.doi.org/10.1038/s41550-025-02616-5},
   DOI={10.1038/s41550-025-02616-5},
   number={8},
   journal={Nat. Astron.},
   publisher={Springer Science and Business Media LLC},
   author={Lines, Natalie E. P. and Li, Tian and Collett, Thomas E. and Holloway, Philip and Nightingale, James W. and Rojas, Karina and Verma, Aprajita and Walmsley, Mike},
   year={2025},
   month=Aug, pages={1116–1122} }

@ARTICLE{Euclid_Walmsley_2025,
       author = {{Euclid Collaboration} and {Walmsley}, M. and {Holloway}, P. and {Lines}, N.~E.~P. and {Rojas}, K. and {Collett}, T.~E. and {Verma}, A. and {Li}, T. and {Nightingale}, J.~W. and {Despali}, G. and et al.},
        title = {Euclid Quick Data Release (Q1): XXVI. The Strong Lensing Discovery Engine A - System overview and first lens sample},
      journal = {\aap},
         year = 2026,
        month = jun,
       volume = {711},
          eid = {A26},
        pages = {A26},
          doi = {10.1051/0004-6361/202555141},
archivePrefix = {arXiv},
       eprint = {2503.15324},
 primaryClass = {astro-ph.GA},
       url = {https://ui.adsabs.harvard.edu/abs/2026A&A...711A...26E}
}

@article{Cao_2015,
   title={Cosmology with strong-lensing systems},
   volume={806},
   ISSN={1538-4357},
   url={http://dx.doi.org/10.1088/0004-637X/806/2/185},
   DOI={10.1088/0004-637x/806/2/185},
   number={2},
   journal={\apj},
   publisher={American Astronomical Society},
   author={Cao, Shuo and Biesiada, Marek and Gavazzi, Raphaël and Piórkowska, Aleksandra and Zhu, Zong-Hong},
   year={2015},
   month=jun, 
   pages={185} }

@article{Ghassemi_2009,
   title={Lensing effects in inhomogeneous cosmological models},
   volume={79},
   ISSN={1550-2368},
   url={http://dx.doi.org/10.1103/PhysRevD.79.102002},
   DOI={10.1103/physrevd.79.102002},
   number={10},
   journal={\prd},
   publisher={American Physical Society (APS)},
   author={Ghassemi, Sima and Khoeini-Moghaddam, Salomeh and Mansouri, Reza},
   year={2009},
   month=May }

@ARTICLE{Sersic_1963,
       author = {{S{\'e}rsic}, J.~L.},
        title = "{Influence of the atmospheric and instrumental dispersion on the brightness distribution in a galaxy}",
      journal = {Bol. Asoc. Argent. Astron.},
         year = 1963,
        month = feb,
       volume = {6},
        pages = {41-43},
       adsurl = {https://ui.adsabs.harvard.edu/abs/1963BAAA....6...41S}
}

@article{Abdalla_2022,
   title={Cosmology intertwined: A review of the particle physics, astrophysics, and cosmology associated with the cosmological tensions and anomalies},
   volume={34},
   ISSN={2214-4048},
   url={http://dx.doi.org/10.1016/j.jheap.2022.04.002},
   DOI={10.1016/j.jheap.2022.04.002},
   journal={J. High Energy Astrophys.},
   publisher={Elsevier BV},
   author={Abdalla, Elcio and Abellán, Guillermo Franco and Aboubrahim, Amin and Agnello, Adriano and Akarsu, Özgür and Akrami, Yashar and Alestas, George and Aloni, Daniel and Amendola, Luca and Anchordoqui, Luis A. and Anderson, Richard I. and Arendse, Nikki and Asgari, Marika and Ballardini, Mario and Barger, Vernon and Basilakos, Spyros and Batista, Ronaldo C. and Battistelli, Elia S. and Battye, Richard and Benetti, Micol and Benisty, David and Berlin, Asher and de Bernardis, Paolo and Berti, Emanuele and Bidenko, Bohdan and Birrer, Simon and Blakeslee, John P. and Boddy, Kimberly K. and Bom, Clecio R. and Bonilla, Alexander and Borghi, Nicola and Bouchet, François R. and Braglia, Matteo and Buchert, Thomas and Buckley-Geer, Elizabeth and Calabrese, Erminia and Caldwell, Robert R. and Camarena, David and Capozziello, Salvatore and Casertano, Stefano and Chen, Geoff C.-F. and Chluba, Jens and Chen, Angela and Chen, Hsin-Yu and Chudaykin, Anton and Cicoli, Michele and Copi, Craig J. and Courbin, Fred and Cyr-Racine, Francis-Yan and Czerny, Bożena and Dainotti, Maria and D’Amico, Guido and Davis, Anne-Christine and de Cruz Pérez, Javier and de Haro, Jaume and Delabrouille, Jacques and Denton, Peter B. and Dhawan, Suhail and Dienes, Keith R. and Di Valentino, Eleonora and Du, Pu and Eckert, Dominique and Escamilla-Rivera, Celia and Ferté, Agnès and Finelli, Fabio and Fosalba, Pablo and Freedman, Wendy L. and Frusciante, Noemi and Gaztañaga, Enrique and Giarè, William and Giusarma, Elena and Gómez-Valent, Adrià and Handley, Will and Harrison, Ian and Hart, Luke and Hazra, Dhiraj Kumar and Heavens, Alan and Heinesen, Asta and Hildebrandt, Hendrik and Hill, J. Colin and Hogg, Natalie B. and Holz, Daniel E. and Hooper, Deanna C. and Hosseininejad, Nikoo and Huterer, Dragan and Ishak, Mustapha and Ivanov, Mikhail M. and Jaffe, Andrew H. and Jang, In Sung and Jedamzik, Karsten and Jimenez, Raul and Joseph, Melissa and Joudaki, Shahab and Kamionkowski, Marc and Karwal, Tanvi and Kazantzidis, Lavrentios and Keeley, Ryan E. and Klasen, Michael and Komatsu, Eiichiro and Koopmans, Léon V.E. and Kumar, Suresh and Lamagna, Luca and Lazkoz, Ruth and Lee, Chung-Chi and Lesgourgues, Julien and Levi Said, Jackson and Lewis, Tiffany R. and L’Huillier, Benjamin and Lucca, Matteo and Maartens, Roy and Macri, Lucas M. and Marfatia, Danny and Marra, Valerio and Martins, Carlos J.A.P. and Masi, Silvia and Matarrese, Sabino and Mazumdar, Arindam and Melchiorri, Alessandro and Mena, Olga and Mersini-Houghton, Laura and Mertens, James and Milaković, Dinko and Minami, Yuto and Miranda, Vivian and Moreno-Pulido, Cristian and Moresco, Michele and Mota, David F. and Mottola, Emil and Mozzon, Simone and Muir, Jessica and Mukherjee, Ankan and Mukherjee, Suvodip and Naselsky, Pavel and Nath, Pran and Nesseris, Savvas and Niedermann, Florian and Notari, Alessio and Nunes, Rafael C. and Ó Colgáin, Eoin and Owens, Kayla A. and Özülker, Emre and Pace, Francesco and Paliathanasis, Andronikos and Palmese, Antonella and Pan, Supriya and Paoletti, Daniela and Perez Bergliaffa, Santiago E. and Perivolaropoulos, Leandros and Pesce, Dominic W. and Pettorino, Valeria and Philcox, Oliver H.E. and Pogosian, Levon and Poulin, Vivian and Poulot, Gaspard and Raveri, Marco and Reid, Mark J. and Renzi, Fabrizio and Riess, Adam G. and Sabla, Vivian I. and Salucci, Paolo and Salzano, Vincenzo and Saridakis, Emmanuel N. and Sathyaprakash, Bangalore S. and Schmaltz, Martin and Schöneberg, Nils and Scolnic, Dan and Sen, Anjan A. and Sehgal, Neelima and Shafieloo, Arman and Sheikh-Jabbari, M.M. and Silk, Joseph and Silvestri, Alessandra and Skara, Foteini and Sloth, Martin S. and Soares-Santos, Marcelle and Solà Peracaula, Joan and Songsheng, Yu-Yang and Soriano, Jorge F. and Staicova, Denitsa and Starkman, Glenn D. and Szapudi, István and Teixeira, Elsa M. and Thomas, Brooks and Treu, Tommaso and Trott, Emery and van de Bruck, Carsten and Vazquez, J. Alberto and Verde, Licia and Visinelli, Luca and Wang, Deng and Wang, Jian-Min and Wang, Shao-Jiang and Watkins, Richard and Watson, Scott and Webb, John K. and Weiner, Neal and Weltman, Amanda and Witte, Samuel J. and Wojtak, Radosław and Yadav, Anil Kumar and Yang, Weiqiang and Zhao, Gong-Bo and Zumalacárregui, Miguel},
   year={2022},
   month=jun, 
   pages={49–211} }

@article{Bull_2012,
   title={Kinematic Sunyaev-Zel’dovich effect as a test of general radial inhomogeneity in Lemaître-Tolman-Bondi cosmology},
   volume={85},
   ISSN={1550-2368},
   url={http://dx.doi.org/10.1103/PhysRevD.85.024002},
   DOI={10.1103/physrevd.85.024002},
   number={2},
   journal={\prd},
   publisher={American Physical Society (APS)},
   author={Bull, Philip and Clifton, Timothy and Ferreira, Pedro G.},
   year={2012},
   month=Jan }

@article{Zumalac_2012,
   title={Tension in the void: cosmic rulers strain inhomogeneous cosmologies},
   volume={2012},
   ISSN={1475-7516},
   url={http://dx.doi.org/10.1088/1475-7516/2012/10/009},
   DOI={10.1088/1475-7516/2012/10/009},
   number={10},
   journal={\jcap},
   publisher={IOP Publishing},
   author={Zumalacárregui, Miguel and García-Bellido, Juan and Ruiz-Lapuente, Pilar},
   year={2012},
   month=Oct, pages={009–009} }

@article{Camarena_2025,
   title={Designing concordant distances in the age of precision cosmology: The impact of density fluctuations},
   volume={112},
   ISSN={2470-0029},
   url={http://dx.doi.org/10.1103/1gj5-3x4m},
   DOI={10.1103/1gj5-3x4m},
   number={8},
   journal={\prd},
   publisher={American Physical Society (APS)},
   author={Camarena, David and Greene, Kylar and Houghteling, John and Cyr-Racine, Francis-Yan},
   year={2025},
   month=Oct }

@article{Caldwell_2008,
   title={A Test of the Copernican Principle},
   volume={100},
   ISSN={1079-7114},
   url={http://dx.doi.org/10.1103/PhysRevLett.100.191302},
   DOI={10.1103/physrevlett.100.191302},
   number={19},
   journal={Phys. Rev. Lett.},
   publisher={American Physical Society (APS)},
   author={Caldwell, R. R. and Stebbins, A.},
   year={2008},
   month=May }

@ARTICLE{Adam_2025,
       author = {{Adam}, James and {Maartens}, Roy and {Larena}, Julien and {Clarkson}, Chris},
        title = {Probing the Cosmological Principle with weak lensing shear},
      journal = {\jcap},
         year = 2025,
        month = feb,
       volume = {2025},
       number = {2},
          eid = {016},
        pages = {016},
          doi = {10.1088/1475-7516/2025/02/016},
archivePrefix = {arXiv},
       eprint = {2411.08560},
 primaryClass = {astro-ph.CO},
       url = {https://ui.adsabs.harvard.edu/abs/2025JCAP...02..016A}
}

@article{Goto_2011,
   title={The Gravitational Lensing Effect on the CMB Polarisation Anisotropy in the $\Lambda$-LTB Model},
   volume={125},
   ISSN={1347-4081},
   url={http://dx.doi.org/10.1143/PTP.125.815},
   DOI={10.1143/ptp.125.815},
   number={4},
   journal={Prog. Theor. Phys.},
   publisher={Oxford University Press (OUP)},
   author={Goto, Hajime and Kodama, Hideo},
   year={2011},
   month=Apr, 
   pages={815–836} }

@article{Seitz_1994,
   title={Light propagation in arbitrary spacetimes and the gravitational lens approximation},
   volume={11},
   ISSN={1361-6382},
   url={http://dx.doi.org/10.1088/0264-9381/11/9/016},
   DOI={10.1088/0264-9381/11/9/016},
   number={9},
   journal={Class. Quantum Grav.},
   publisher={IOP Publishing},
   author={Seitz, Stella and Schneider, Peter and Ehlers, Jürgen},
   year={1994},
   month=sep, 
   pages={2345–2373} }

@ARTICLE{Euclid_Nersesian_2026,
      journal={arXiv e-prints},
      pages={arXiv:2611.15877},
      title={Euclid Data Release 1 (DR1). Euclid Lens Identification Pipeline and Scientific Exploration (ECLIPSE) D: Photometric redshifts and the first census of strong lenses in the galaxy evolution framework}, 
      author={{Euclid Collaboration} and A. Nersesian and J. W. Nightingale and D. Sluse and M. Bolzonella and N. E. P. Lines and P. Holloway and S. Schuldt and A. Sonnenfeld and C. Tortora and T. E. Collett and G. Despali and M. Fogliardi and S. H. Vincken and R. Cañameras and R. B. Metcalf and D. J. Bowden and L. R. Ecker and A. L. Faisst and A. Galan and R. Gavazzi and Q. He and T. Li and L. A. Moustakas and L. Quilley and K. Rojas and D. Scott and G. Smadja and C. J. Storfer and B. Altieri and S. Andreon and N. Auricchio and C. Baccigalupi and M. Baldi and A. Balestra and S. Bardelli and P. Battaglia and A. Biviano and E. Branchini and M. Brescia and S. Camera and V. Capobianco and C. Carbone and V. F. Cardone and J. Carretero and M. Castellano and G. Castignani and S. Cavuoti and K. C. Chambers and A. Cimatti and C. Colodro-Conde and G. Congedo and C. J. Conselice and L. Conversi and Y. Copin and A. Costille and F. Courbin and H. M. Courtois and M. Cropper and H. Degaudenzi and G. De Lucia and C. Dolding and H. Dole and F. Dubath and X. Dupac and S. Escoffier and M. Farina and R. Farinelli and F. Faustini and S. Ferriol and M. Frailis and E. Franceschi and M. Fumana and L. Gabarra and S. Galeotta and K. George and B. Gillis and C. Giocoli and P. Gómez-Alvarez and J. Gracia-Carpio and A. Grazian and F. Grupp and S. Gwyn and W. G. Hartley and S. V. H. Haugan and S. Hemmati and H. Hoekstra and W. Holmes and F. Hormuth and A. Hornstrup and M. Huertas-Company and K. Jahnke and M. Jhabvala and B. Joachimi and S. Kermiche and A. Kiessling and B. Kubik and M. Kümmel and M. Kunz and H. Kurki-Suonio and A. M. C. Le Brun and S. Ligori and P. B. Lilje and V. Lindholm and I. Lloro and M. Magliocchetti and G. Mainetti and D. Maino and O. Mansutti and S. Marcin and O. Marggraf and M. Martinelli and N. Martinet and F. Marulli and R. J. Massey and N. Mauri and E. Medinaceli and S. Mei and M. Meneghetti and E. Merlin and G. Meylan and P. Monaco and A. Mora and M. Moresco and C. Moretti and L. Moscardini and R. Nakajima and C. Neissner and R. C. Nichol and S. -M. Niemi and C. Padilla and S. Paltani and F. Pasian and W. J. Percival and V. Pettorino and A. Pezzotta and G. Polenta and M. Poncet and L. A. Popa and L. Pozzetti and F. Raison and R. Rebolo and A. Renzi and J. Rhodes and G. Riccio and I. Risso and F. Rizzo and E. Romelli and M. Roncarelli and B. Rusholme and R. Saglia and Z. Sakr and D. Sapone and B. Sartoris and P. Schneider and T. Schrabback and A. Secroun and E. Sihvola and P. Simon and C. Sirignano and G. Sirri and L. Stanco and P. Tallada-Crespí and A. N. Taylor and H. I. Teplitz and I. Tereno and N. Tessore and S. Toft and R. Toledo-Moreo and F. Torradeflot and A. Tsyganov and I. Tutusaus and J. Valiviita and T. Vassallo and Y. Wang and J. Weller and A. Zacchei and G. Zamorani and F. M. Zerbi and E. Zucca and M. Ballardini and K. Benson and P. Bergamini and E. Bozzo and C. Burigana and R. Cabanac and A. Cappi and F. Caro and T. Castro and J. A. Escartin Vigo and J. García-Bellido and T. Gasparetto and V. Gautard and C. M. Gutierrez and J. Macias-Perez and R. Maoli and M. Pöntinen and E. Sarpa and V. Scottez and M. Sereno and M. Tenti and M. Tucci and M. Viel and M. Wiesmann and J. A. Acevedo Barroso and Y. Akrami and I. T. Andika and G. Angora and S. Anselmi and M. Archidiacono and G. Aricò and F. Atrio-Barandela and M. Baes and L. Bazzanini and D. Bertacca and M. Bethermin and F. Beutler and A. Blanchard and L. Blot and M. L. Brown and S. Bruton and A. Calabro and B. Camacho Quevedo and C. S. Carvalho and F. Cogato and A. R. Cooray and P. Corcho-Caballero and B. Csizi and O. Cucciati and H. Dannerbauer and T. de Boer and G. Desprez and A. Díaz-Sánchez and S. Di Domizio and J. M. Diego and P. -A. Duc and V. Duret and M. Y. Elkhashab and A. Enia and Y. Fang and A. Farina and A. Finoguenov and A. Franco and Y. Fu and K. Ganga and E. Gaztanaga and Z. Ghaffari and F. Giacomini and F. Gianotti and G. Gozaliasl and A. Gruppuso and M. Guidi and A. Hall and N. A. Hatch and C. Hernández-Monteagudo and T. Heuberger and H. Hildebrandt and J. Hjorth and M. Jauzac and J. J. E. Kajava and Y. Kang and V. Kansal and D. Karagiannis and J. Kim and C. C. Kirkpatrick and A. Kovács and I. Kova{č}i{ć} and K. Koyama and S. Kruk and M. C. Lam and M. Lattanzi and L. Legrand and G. Leroy and G. F. Lesci and J. Lesgourgues and L. Linke and S. J. Liu and G. Maggio and E. A. Magnier and F. Mannucci and F. R. Marleau and C. J. A. P. Martins and M. Migliaccio and M. Miluzio and G. Morgante and K. Naidoo and A. Navarro-Alsina and S. Nesseris and L. Nicastro and M. Oguri and F. Oppizzi and F. Pace and D. Paoletti and F. Passalacqua and K. Paterson and C. Pattison and A. Pisani and D. Potter and G. W. Pratt and S. Quai and M. Radovich and G. Rodighiero and W. Roster and M. Sahlén and D. B. Sanders and A. Schneider and D. Sciotti and E. Sellentin and S. Serjeant and F. Shankar and L. C. Smith and I. Szapudi and M. Talia and K. Tanidis and C. Tao and F. Tarsitano and G. Testera and R. Teyssier and S. Tosi and A. Troja and C. Uhlemann and C. Valieri and A. Venhola and D. Vergani and G. Vernardos and P. Vielzeuf and E. Vilenius and S. Vinciguerra and M. von Wietersheim-Kramsta and N. A. Walton and L. Wang and J. Wilde and A. H. Wright},
      year={2026},
      eprint={2611.15877},
      archivePrefix={arXiv},
      primaryClass={astro-ph.GA},
      url={https://arxiv.org/abs/2611.15877}, 
}

@ARTICLE{Euclid_Nightingale_2026,
      journal={arXiv e-prints},
      pages={arXiv:2611.15888},
      title={Euclid Data Release 1 (DR1). Euclid Lens Identification Pipeline and Scientific Exploration (ECLIPSE) C: Lens modelling pipeline overview}, 
      author={{Euclid Collaboration} and J. W. Nightingale and S. H. Vincken and L. R. Ecker and A. Nersesian and D. J. Bowden and C. M. O'Riordan and N. E. P. Lines and T. E. Collett and G. Despali and S. Schuldt and F. Courbin and R. B. Metcalf and A. Verma and A. Angress and W. J. R. Enzi and K. Finner and P. Fleury and M. Fogliardi and A. Galan and R. Gavazzi and Q. He and N. B. Hogg and P. Holloway and D. P. Johnson and S. K. Li and T. Li and N. Massari and A. Melo and M. Meneghetti and M. Millon and L. Moustakas and C. Olave and H. Qu and K. Rojas and C. Saulder and D. Sluse and G. Smadja and H. Srivastava and A. Sonnenfeld and C. J. Storfer and T. T. Thai and C. Tortora and N. Wielders and Y. Xie and B. Altieri and A. Amara and S. Andreon and N. Auricchio and C. Baccigalupi and M. Baldi and S. Bardelli and P. Battaglia and R. Bender and A. Biviano and M. Bolzonella and E. Branchini and M. Brescia and S. Camera and V. Capobianco and C. Carbone and V. F. Cardone and J. Carretero and M. Castellano and G. Castignani and S. Cavuoti and K. C. Chambers and A. Cimatti and C. Colodro-Conde and G. Congedo and C. J. Conselice and L. Conversi and Y. Copin and A. Costille and H. M. Courtois and M. Cropper and H. Degaudenzi and G. De Lucia and C. Dolding and H. Dole and F. Dubath and X. Dupac and S. Escoffier and M. Farina and R. Farinelli and F. Faustini and S. Ferriol and M. Frailis and E. Franceschi and M. Fumana and L. Gabarra and S. Galeotta and K. George and B. Gillis and C. Giocoli and P. Gómez-Alvarez and J. Gracia-Carpio and A. Grazian and F. Grupp and L. Guzzo and W. G. Hartley and S. V. H. Haugan and S. Hemmati and H. Hoekstra and M. S. Holliman and W. Holmes and I. M. Hook and F. Hormuth and A. Hornstrup and M. Huertas-Company and K. Jahnke and M. Jhabvala and S. Kermiche and A. Kiessling and B. Kubik and M. Kümmel and M. Kunz and H. Kurki-Suonio and A. M. C. Le Brun and D. Le Mignant and S. Ligori and P. B. Lilje and V. Lindholm and I. Lloro and M. Magliocchetti and G. Mainetti and D. Maino and O. Mansutti and S. Marcin and O. Marggraf and M. Martinelli and N. Martinet and F. Marulli and R. J. Massey and N. Mauri and E. Medinaceli and S. Mei and M. Melchior and E. Merlin and G. Meylan and P. Monaco and A. Mora and M. Moresco and C. Moretti and L. Moscardini and R. Nakajima and C. Neissner and R. C. Nichol and S. -M. Niemi and C. Padilla and S. Paltani and F. Pasian and W. J. Percival and V. Pettorino and A. Pezzotta and G. Polenta and L. A. Popa and L. Pozzetti and F. Raison and A. Renzi and J. Rhodes and G. Riccio and I. Risso and E. Romelli and M. Roncarelli and B. Rusholme and R. Saglia and Z. Sakr and D. Sapone and B. Sartoris and M. Schirmer and P. Schneider and T. Schrabback and A. Secroun and E. Sihvola and C. Sirignano and G. Sirri and J. Skottfelt and L. Stanco and P. Tallada-Crespí and A. N. Taylor and I. Tereno and N. Tessore and S. Toft and R. Toledo-Moreo and F. Torradeflot and A. Tsyganov and I. Tutusaus and J. Valiviita and T. Vassallo and A. Veropalumbo and Y. Wang and J. Weller and A. Zacchei and G. Zamorani and E. Zucca and M. Ballardini and P. Bergamini and S. Borgani and E. Bozzo and C. Burigana and R. Cabanac and M. Calabrese and A. Cappi and F. Caro and T. Castro and D. Di Ferdinando and J. A. Escartin Vigo and J. García-Bellido and T. Gasparetto and C. M. Gutierrez and A. Loureiro and J. Macias-Perez and R. Maoli and A. A. Nucita and M. Pöntinen and E. Sarpa and V. Scottez and M. Sereno and M. Tenti and M. Tucci and M. Wiesmann and J. A. Acevedo Barroso and Y. Akrami and I. T. Andika and G. Angora and S. Anselmi and M. Archidiacono and G. Aricò and F. Atrio-Barandela and M. Baes and L. Bazzanini and D. Bertacca and M. Bethermin and F. Beutler and L. Blot and M. L. Brown and S. Bruton and A. Calabro and B. Camacho Quevedo and C. S. Carvalho and F. Cogato and A. R. Cooray and P. Corcho-Caballero and B. Csizi and O. Cucciati and T. de Boer and G. Desprez and A. Díaz-Sánchez and S. Di Domizio and J. M. Diego and P. -A. Duc and V. Duret and M. Y. Elkhashab and A. Enia and Y. Fang and A. Farina and A. Finoguenov and A. Fontana and A. Franco and Y. Fu and K. Ganga and E. Gaztanaga and Z. Ghaffari and F. Giacomini and F. Gianotti and G. Gozaliasl and A. Gruppuso and M. Guidi and A. Hall and N. A. Hatch and T. Heuberger and H. Hildebrandt and J. Hjorth and M. Jauzac and J. J. E. Kajava and Y. Kang and V. Kansal and D. Karagiannis and J. Kim and C. C. Kirkpatrick and A. Kovács and I. Kovačić and K. Koyama and S. Kruk and M. C. Lam and M. Lattanzi and G. Leroy and G. F. Lesci and J. Lesgourgues and L. Linke and S. J. Liu and G. Maggio and E. A. Magnier and F. Mannucci and F. R. Marleau and C. J. A. P. Martins and M. Migliaccio and M. Miluzio and G. Morgante and K. Naidoo and A. Navarro-Alsina and S. Nesseris and L. Nicastro and M. Oguri and F. Oppizzi and F. Pace and D. Paoletti and F. Passalacqua and K. Paterson and L. Patrizii and C. Pattison and A. Pisani and D. Potter and G. W. Pratt and S. Quai and M. Radovich and G. Rodighiero and W. Roster and M. Sahlén and D. B. Sanders and A. Schneider and D. Sciotti and E. Sellentin and S. Serjeant and L. C. Smith and I. Szapudi and M. Talia and K. Tanidis and C. Tao and G. Testera and R. Teyssier and S. Tosi and A. Troja and C. Uhlemann and C. Valieri and D. Vergani and G. Vernardos and P. Vielzeuf and E. Vilenius and S. Vinciguerra and M. von Wietersheim-Kramsta and N. A. Walton and L. Wang and J. Wilde and A. H. Wright and D. Scott},
      year={2026},
      eprint={2611.15888},
      archivePrefix={arXiv},
      primaryClass={astro-ph.GA},
      url={https://arxiv.org/abs/2611.15888}, 
}

@ARTICLE{Euclid_Lines_2026,
      journal={arXiv e-prints},
      pages={arXiv:2611.15883},
      title={Euclid Data Release 1 (DR1). Euclid Lens Identification Pipeline and Scientific Exploration (ECLIPSE) A: Strong lens catalogue overview}, 
      author={{Euclid Collaboration} and N. E. P. Lines and P. Holloway and T. Li and T. E. Collett and S. Schuldt and S. H. Vincken and J. W. Nightingale and A. Nersesian and M. Fogliardi and A. Verma and C. J. Storfer and A. Manjón-García and J. Wilde and W. J. R. Enzi and J. A. Acevedo Barroso and A. Sonnenfeld and K. Rojas and E. Baeten and P. Bergamini and C. Cornen and F. Courbin and H. Degaudenzi and G. Despali and R. Gavazzi and C. Macmillan and R. B. Metcalf and D. J. Bowden and R. Cañameras and J. H. -H. Chan and L. R. Ecker and M. Grespan and Q. He and D. P. Johnson and C. Krawczyk and N. Massari and A. Melo and L. A. Moustakas and C. Olave and A. Paulino-Afonso and C. Saulder and D. Sluse and G. Smadja and H. Srivastava and D. Stern and T. T. Thai and C. Tortora and N. Wielders and I. T. Andika and A. Angress and D. J. Ballard and M. Bradac and D. Delley and P. Fleury and A. Galan and G. Granata and T. Harvey and N. B. Hogg and K. Jahnke and K. Kuijken and T. Li and P. J. Marshall and M. Meneghetti and A. More and B. C. Nagam and D. O'Ryan and J. Pearson and H. Qu and D. Ryczanowski and T. Saifollahi and S. Serjeant and C. Spiniello and L. Trobbiani and X. Xu and M. Zumalacarregui and A. L. Faisst and B. Altieri and S. Andreon and N. Auricchio and C. Baccigalupi and M. Baldi and A. Balestra and S. Bardelli and P. Battaglia and R. Bender and A. Biviano and M. Bolzonella and E. Branchini and M. Brescia and S. Camera and G. Cañas-Herrera and V. Capobianco and C. Carbone and V. F. Cardone and J. Carretero and M. Castellano and G. Castignani and S. Cavuoti and K. C. Chambers and A. Cimatti and C. Colodro-Conde and G. Congedo and C. J. Conselice and L. Conversi and Y. Copin and A. Costille and H. M. Courtois and M. Cropper and G. De Lucia and C. Dolding and H. Dole and F. Dubath and X. Dupac and S. Dusini and S. Escoffier and M. Farina and R. Farinelli and F. Faustini and S. Ferriol and S. Fotopoulou and M. Frailis and E. Franceschi and M. Fumana and L. Gabarra and S. Galeotta and K. George and B. Gillis and C. Giocoli and J. Gracia-Carpio and A. Grazian and F. Grupp and S. Gwyn and W. G. Hartley and S. V. H. Haugan and S. Hemmati and H. Hoekstra and W. Holmes and I. M. Hook and F. Hormuth and A. Hornstrup and M. Jhabvala and B. Joachimi and S. Kermiche and A. Kiessling and B. Kubik and M. Kümmel and M. Kunz and H. Kurki-Suonio and A. M. C. Le Brun and S. Ligori and P. B. Lilje and V. Lindholm and I. Lloro and M. Magliocchetti and G. Mainetti and D. Maino and O. Mansutti and O. Marggraf and M. Martinelli and N. Martinet and F. Marulli and R. J. Massey and N. Mauri and E. Medinaceli and M. Melchior and E. Merlin and G. Meylan and P. Monaco and A. Mora and M. Moresco and C. Moretti and L. Moscardini and R. Nakajima and C. Neissner and R. C. Nichol and S. -M. Niemi and C. Padilla and S. Paltani and F. Pasian and W. J. Percival and V. Pettorino and A. Pezzotta and G. Polenta and M. Poncet and L. A. Popa and L. Pozzetti and F. Raison and R. Rebolo and A. Renzi and J. Rhodes and G. Riccio and I. Risso and F. Rizzo and E. Romelli and M. Roncarelli and B. Rusholme and R. Saglia and Z. Sakr and D. Sapone and B. Sartoris and M. Schirmer and P. Schneider and T. Schrabback and A. Secroun and E. Sihvola and P. Simon and C. Sirignano and G. Sirri and L. Stanco and J. -L. Starck and P. Tallada-Crespí and A. N. Taylor and H. I. Teplitz and I. Tereno and N. Tessore and S. Toft and R. Toledo-Moreo and F. Torradeflot and A. Tsyganov and I. Tutusaus and J. Valiviita and T. Vassallo and Y. Wang and J. Weller and A. Zacchei and G. Zamorani and F. M. Zerbi and E. Zucca and M. Ballardini and S. Borgani and E. Bozzo and C. Burigana and R. Cabanac and A. Cappi and F. Caro and T. Castro and J. A. Escartin Vigo and J. García-Bellido and T. Gasparetto and V. Gautard and C. M. Gutierrez and J. Macias-Perez and R. Maoli and M. Pöntinen and E. Sarpa and V. Scottez and M. Sereno and M. Tenti and M. Tucci and M. Wiesmann and Y. Akrami and G. Angora and S. Anselmi and M. Archidiacono and G. Aricò and F. Atrio-Barandela and M. Baes and L. Bazzanini and D. Bertacca and M. Bethermin and F. Beutler and L. Blot and M. L. Brown and S. Bruton and A. Calabro and B. Camacho Quevedo and C. S. Carvalho and F. Cogato and S. Conseil and A. R. Cooray and P. Corcho-Caballero and B. Csizi and O. Cucciati and H. Dannerbauer and T. de Boer and F. De Paolis and G. Desprez and A. Díaz-Sánchez and S. Di Domizio and J. M. Diego and P. -A. Duc and V. Duret and M. Y. Elkhashab and A. Enia and Y. Fang and A. Farina and A. Finoguenov and A. Franco and Y. Fu and K. Ganga and E. Gaztanaga and Z. Ghaffari and F. Giacomini and F. Gianotti and G. Gozaliasl and A. Gruppuso and M. Guidi and A. Hall and C. Hernández-Monteagudo and T. Heuberger and H. Hildebrandt and J. Hjorth and M. Jauzac and J. J. E. Kajava and Y. Kang and V. Kansal and D. Karagiannis and J. Kim and C. C. Kirkpatrick and A. Kovács and I. Kova{č}i{ć} and K. Koyama and S. Kruk and M. C. Lam and M. Lattanzi and L. Legrand and G. Leroy and G. F. Lesci and J. Lesgourgues and L. Linke and S. J. Liu and G. Maggio and E. A. Magnier and F. Mannucci and F. R. Marleau and C. J. A. P. Martins and M. Migliaccio and M. Miluzio and G. Morgante and S. Nadathur and K. Naidoo and P. Natoli and A. Navarro-Alsina and S. Nesseris and L. Nicastro and M. Oguri and F. Oppizzi and F. Pace and D. Paoletti and F. Passalacqua and K. Paterson and L. Patrizii and C. Pattison and A. Pisani and D. Potter and G. W. Pratt and S. Quai and M. Radovich and G. Rodighiero and W. Roster and S. Sacquegna and M. Sahlén and D. B. Sanders and J. Schaye and A. Schneider and D. Sciotti and E. Sellentin and F. Shankar and L. C. Smith and M. Talia and K. Tanidis and C. Tao and G. Testera and R. Teyssier and S. Tosi and A. Troja and C. Uhlemann and C. Valieri and A. Venhola and D. Vergani and G. Vernardos and P. Vielzeuf and E. Vilenius and S. Vinciguerra and M. von Wietersheim-Kramsta and N. A. Walton and L. Wang and A. H. Wright and D. Scott},
      year={2026},
      eprint={2611.15883},
      archivePrefix={arXiv},
      primaryClass={astro-ph.GA},
      url={https://arxiv.org/abs/2611.15883}, 
}

@article{Koopmans_2009,
   title={THE STRUCTURE AND DYNAMICS OF MASSIVE EARLY-TYPE GALAXIES: ON HOMOLOGY, ISOTHERMALITY, AND ISOTROPY INSIDE ONE EFFECTIVE RADIUS},
   volume={703},
   ISSN={1538-4357},
   url={http://dx.doi.org/10.1088/0004-637X/703/1/L51},
   DOI={10.1088/0004-637x/703/1/l51},
   number={1},
   journal={\apj},
   publisher={American Astronomical Society},
   author={Koopmans, L. V. E. and Bolton, A. and Treu, T. and Czoske, O. and Auger, M. W. and Barnabè, M. and Vegetti, S. and Gavazzi, R. and Moustakas, L. A. and Burles, S.},
   year={2009},
   month=sep, pages={L51–L54} }

@misc{CLASS_Lesgourgues_2011,
      title={The Cosmic Linear Anisotropy Solving System (CLASS) I: Overview}, 
      author={Julien Lesgourgues},
      year={2011},
      eprint={1104.2932},
      archivePrefix={arXiv},
      primaryClass={astro-ph.IM},
      url={https://arxiv.org/abs/1104.2932}, 
}

@ARTICLE{rotac,
       author = {{Zasowski}, Gail and {Suarabh}, Ja W and et al},
        title = "{Roman Observations Time Allocation Committee: Final Report and Recommendations}",
      journal = {arXiv e-prints},
         year = 2025,
        month = may,
          eid = {arXiv:2505.10574},
        pages = {arXiv:2505.10574},
          doi = {10.48550/arXiv.2505.10574},
archivePrefix = {arXiv},
       eprint = {2505.10574},
 primaryClass = {astro-ph.IM},
       adsurl = {https://scixplorer.org/abs/2025arXiv250510574O}
}

@ARTICLE{lsst,
       author = {{Ivezi{\'c}}, {\v{Z}}eljko and {Kahn}, Steven M. and {Tyson}, J. Anthony and {Abel}, Bob and {Acosta}, Emily and {Allsman}, Robyn and {Alonso}, David and {AlSayyad}, Yusra and {Anderson}, Scott F. and {Andrew}, John and et al.},
        title = "{LSST: From Science Drivers to Reference Design and Anticipated Data Products}",
      journal = {\apj},
         year = 2019,
        month = mar,
       volume = {873},
       number = {2},
          eid = {111},
        pages = {111},
          doi = {10.3847/1538-4357/ab042c},
archivePrefix = {arXiv},
       eprint = {0805.2366},
 primaryClass = {astro-ph},
       adsurl = {https://scixplorer.org/abs/2019ApJ...873..111I}
}

@ARTICLE{Saha_2024,
       author = {{Saha}, Prasenjit and {Sluse}, Dominique and {Wagner}, Jenny and {Williams}, Liliya L.~R.},
        title = "{Essentials of Strong Gravitational Lensing}",
      journal = {\ssr},
         year = 2024,
        month = feb,
       volume = {220},
       number = {1},
          eid = {12},
        pages = {12},
          doi = {10.1007/s11214-024-01041-w},
archivePrefix = {arXiv},
       eprint = {2401.04165},
 primaryClass = {astro-ph.CO},
       adsurl = {https://scixplorer.org/abs/2024SSRv..220...12S}
}

@misc{Kochanek_2006,
        title = "{Gravitational Lensing: Strong, Weak and Micro}",
        author={C. S. Kochanek},
    booktitle = {Saas-Fee Advanced Course 33: Gravitational Lensing: Strong, Weak and Micro},
         year = 2006,
       editor = {{Meylan}, Georges and {Jetzer}, Philippe and {North}, Pierre and {Schneider}, Peter and {Kochanek}, Christopher S. and {Wambsganss}, Joachim},
        month = jan,
          doi = {10.48550/arXiv.astro-ph/0407232},
archivePrefix = {arXiv},
       eprint = {astro-ph/0407232},
 primaryClass = {astro-ph},
       adsurl = {https://scixplorer.org/abs/2006glsw.conf.....M}
}

@ARTICLE{Treu_2010,
       author = {{Treu}, Tommaso},
        title = "{Strong Lensing by Galaxies}",
      journal = {\araa},
         year = 2010,
        month = sep,
       volume = {48},
        pages = {87-125},
          doi = {10.1146/annurev-astro-081309-130924},
archivePrefix = {arXiv},
       eprint = {1003.5567},
 primaryClass = {astro-ph.CO},
       adsurl = {https://scixplorer.org/abs/2010ARA&A..48...87T}
}

@ARTICLE{Treu_2026,
       author = {{Treu}, Tommaso},
        title = "{Cosmology with strong gravitational lenses}",
      journal = {arXiv e-prints},
         year = 2026,
        month = oct,
          eid = {arXiv:2610.19598},
        pages = {arXiv:2610.19598},
          doi = {10.48550/arXiv.2610.19598},
archivePrefix = {arXiv},
       eprint = {2610.19598},
 primaryClass = {astro-ph.CO},
       adsurl = {https://scixplorer.org/abs/2026arXiv261019598T}
}

@ARTICLE{Birrer_2024,
       author = {{Birrer}, S. and {Millon}, M. and {Sluse}, D. and {Shajib}, A.~J. and {Courbin}, F. and {Erickson}, S. and {Koopmans}, L.~V.~E. and {Suyu}, S.~H. and {Treu}, T.},
        title = "{Time-Delay Cosmography: Measuring the Hubble Constant and Other Cosmological Parameters with Strong Gravitational Lensing}",
      journal = {\ssr},
         year = 2024,
        month = aug,
       volume = {220},
       number = {5},
          eid = {48},
        pages = {48},
          doi = {10.1007/s11214-024-01079-w},
archivePrefix = {arXiv},
       eprint = {2210.10833},
 primaryClass = {astro-ph.CO},
       adsurl = {https://scixplorer.org/abs/2024SSRv..220...48B}
}

@ARTICLE{Grillo_2008,
       author = {{Grillo}, C. and {Lombardi}, M. and {Bertin}, G.},
        title = "{Cosmological parameters from strong gravitational lensing and stellar dynamics in elliptical galaxies}",
      journal = {\aap},
         year = 2008,
        month = jan,
       volume = {477},
       number = {2},
        pages = {397-406},
          doi = {10.1051/0004-6361:20077534},
archivePrefix = {arXiv},
       eprint = {0711.0882},
 primaryClass = {astro-ph},
       adsurl = {https://scixplorer.org/abs/2008A&A...477..397G}
}

@ARTICLE{Gavazzi_2008,
       author = {{Gavazzi}, Rapha{\"e}l and {Treu}, Tommaso and {Koopmans}, L{\'e}on V.~E. and {Bolton}, Adam S. and {Moustakas}, Leonidas A. and {Burles}, Scott and {Marshall}, Philip J.},
        title = "{The Sloan Lens ACS Survey. VI. Discovery and Analysis of a Double Einstein Ring}",
      journal = {\apj},
         year = 2008,
        month = apr,
       volume = {677},
       number = {2},
        pages = {1046-1059},
          doi = {10.1086/529541},
archivePrefix = {arXiv},
       eprint = {0801.1555},
 primaryClass = {astro-ph},
       adsurl = {https://scixplorer.org/abs/2008ApJ...677.1046G}
}

@ARTICLE{CollettA_2014,
       author = {{Collett}, Thomas E. and {Auger}, Matthew W.},
        title = "{Cosmological constraints from the double source plane lens SDSSJ0946+1006}",
      journal = {\mnras},
         year = 2014,
        month = sep,
       volume = {443},
       number = {2},
        pages = {969-976},
          doi = {10.1093/mnras/stu1190},
archivePrefix = {arXiv},
       eprint = {1403.5278},
 primaryClass = {astro-ph.CO},
       adsurl = {https://scixplorer.org/abs/2014MNRAS.443..969C}
}

@ARTICLE{Schneider_2014,
       author = {{Schneider}, Peter},
        title = "{Can one determine cosmological parameters from multi-plane strong lens systems?}",
      journal = {\aap},
         year = 2014,
        month = aug,
       volume = {568},
          eid = {L2},
        pages = {L2},
          doi = {10.1051/0004-6361/201424450},
archivePrefix = {arXiv},
       eprint = {1406.6152},
 primaryClass = {astro-ph.CO},
       adsurl = {https://scixplorer.org/abs/2014A&A...568L...2S}
}

@ARTICLE{Marra_2011,
       author = {{Marra}, Valerio and {Notari}, Alessio},
        title = "{Observational constraints on inhomogeneous cosmological models without dark energy}",
      journal = {Classical and Quantum Gravity},
         year = 2011,
        month = aug,
       volume = {28},
       number = {16},
          eid = {164004},
        pages = {164004},
          doi = {10.1088/0264-9381/28/16/164004},
archivePrefix = {arXiv},
       eprint = {1102.1015},
 primaryClass = {astro-ph.CO},
       adsurl = {https://scixplorer.org/abs/2011CQGra..28p4004M}
}

@ARTICLE{Bolejko_2011,
       author = {{Bolejko}, Krzysztof and {C{\'e}l{\'e}rier}, Marie-No{\"e}lle and {Krasi{\'n}ski}, Andrzej},
        title = "{Inhomogeneous cosmological models: exact solutions and their applications}",
      journal = {Classical and Quantum Gravity},
         year = 2011,
        month = aug,
       volume = {28},
       number = {16},
          eid = {164002},
        pages = {164002},
          doi = {10.1088/0264-9381/28/16/164002},
archivePrefix = {arXiv},
       eprint = {1102.1449},
 primaryClass = {astro-ph.CO},
       adsurl = {https://scixplorer.org/abs/2011CQGra..28p4002B}
}

@ARTICLE{Redlich_2014,
       author = {{Redlich}, M. and {Bolejko}, K. and {Meyer}, S. and {Lewis}, G.~F. and {Bartelmann}, M.},
        title = "{Probing spatial homogeneity with LTB models: a detailed discussion}",
      journal = {\aap},
         year = 2014,
        month = oct,
       volume = {570},
          eid = {A63},
        pages = {A63},
          doi = {10.1051/0004-6361/201424553},
archivePrefix = {arXiv},
       eprint = {1408.1872},
 primaryClass = {astro-ph.CO},
       adsurl = {https://scixplorer.org/abs/2014A&A...570A..63R}
}

@ARTICLE{Wu_2017,
       author = {{Wu}, Hao-Yi and {Huterer}, Dragan},
        title = "{Sample variance in the local measurements of the Hubble constant}",
      journal = {\mnras},
         year = 2017,
        month = nov,
       volume = {471},
       number = {4},
        pages = {4946-4955},
          doi = {10.1093/mnras/stx1967},
archivePrefix = {arXiv},
       eprint = {1706.09723},
 primaryClass = {astro-ph.CO},
       adsurl = {https://scixplorer.org/abs/2017MNRAS.471.4946W}
}

@ARTICLE{Kenworthy_2019,
       author = {{Kenworthy}, W. D'Arcy and {Scolnic}, Dan and {Riess}, Adam},
        title = "{The Local Perspective on the Hubble Tension: Local Structure Does Not Impact Measurement of the Hubble Constant}",
      journal = {\apj},
         year = 2019,
        month = apr,
       volume = {875},
       number = {2},
          eid = {145},
        pages = {145},
          doi = {10.3847/1538-4357/ab0ebf},
archivePrefix = {arXiv},
       eprint = {1901.08681},
 primaryClass = {astro-ph.CO},
       adsurl = {https://scixplorer.org/abs/2019ApJ...875..145K}
}

@ARTICLE{Castello_2022,
       author = {{Castello}, Sveva and {H{\"o}g{\r{a}}s}, Marcus and {M{\"o}rtsell}, Edvard},
        title = "{A cosmological underdensity does not solve the Hubble tension}",
      journal = {\jcap},
         year = 2022,
        month = jul,
       volume = {2022},
       number = {7},
          eid = {003},
        pages = {003},
          doi = {10.1088/1475-7516/2022/07/003},
archivePrefix = {arXiv},
       eprint = {2110.04226},
 primaryClass = {astro-ph.CO},
       adsurl = {https://scixplorer.org/abs/2022JCAP...07..003C}
}

@ARTICLE{Collett_2015,
       author = {{Collett}, Thomas E.},
        title = "{The Population of Galaxy-Galaxy Strong Lenses in Forthcoming Optical Imaging Surveys}",
      journal = {\apj},
         year = 2015,
        month = sep,
       volume = {811},
       number = {1},
          eid = {20},
        pages = {20},
          doi = {10.1088/0004-637X/811/1/20},
archivePrefix = {arXiv},
       eprint = {1507.02657},
 primaryClass = {astro-ph.CO},
       adsurl = {https://scixplorer.org/abs/2015ApJ...811...20C}
}

@ARTICLE{Fleury_2015,
       author = {{Fleury}, Pierre and {Larena}, Julien and {Uzan}, Jean-Philippe},
        title = "{The theory of stochastic cosmological lensing}",
      journal = {\jcap},
         year = 2015,
        month = nov,
       volume = {2015},
       number = {11},
        pages = {022-022},
          doi = {10.1088/1475-7516/2015/11/022},
archivePrefix = {arXiv},
       eprint = {1508.07903},
 primaryClass = {gr-qc},
       adsurl = {https://scixplorer.org/abs/2015JCAP...11..022F}
}

@ARTICLE{Fanizza_2015,
       author = {{Fanizza}, G. and {Nugier}, F.},
        title = "{Lensing in the geodesic light-cone coordinates and its (exact) illustration to an off-center observer in Lemaȋtre-Tolman-Bondi models}",
      journal = {\jcap},
         year = 2015,
        month = feb,
       volume = {2015},
       number = {2},
        pages = {002-002},
          doi = {10.1088/1475-7516/2015/02/002},
archivePrefix = {arXiv},
       eprint = {1408.1604},
 primaryClass = {astro-ph.CO},
       adsurl = {https://scixplorer.org/abs/2015JCAP...02..002F}
}

@ARTICLE{Tomita_2000,
       author = {{Tomita}, Kenji},
        title = "{Distances and Lensing in Cosmological Void Models}",
      journal = {\apj},
         year = 2000,
        month = jan,
       volume = {529},
       number = {1},
        pages = {38-46},
          doi = {10.1086/308277},
archivePrefix = {arXiv},
       eprint = {astro-ph/9906027},
 primaryClass = {astro-ph},
       adsurl = {https://scixplorer.org/abs/2000ApJ...529...38T}
}

@ARTICLE{Rasanan_2015,
       author = {{R{\"a}s{\"a}nen}, Syksy and {Bolejko}, Krzysztof and {Finoguenov}, Alexis},
        title = "{New Test of the Friedmann-Lema{\^\i}tre-Robertson-Walker Metric Using the Distance Sum Rule}",
      journal = {\prl},
         year = 2015,
        month = sep,
       volume = {115},
       number = {10},
          eid = {101301},
        pages = {101301},
          doi = {10.1103/PhysRevLett.115.101301},
archivePrefix = {arXiv},
       eprint = {1412.4976},
 primaryClass = {astro-ph.CO},
       adsurl = {https://scixplorer.org/abs/2015PhRvL.115j1301R}
}

@ARTICLE{Camarena_Euclid,
       author = {{Camarena}, D. and {Marra}, V. and {Sakr}, Z. and {Nesseris}, S. and {Da Silva}, A. and {Garcia-Bellido}, J. and {Fleury}, P. and {Lombriser}, L. and {Martinelli}, M. and {Martins}, C.~J.~A.~P. and et al.},
        title = "{Euclid: Testing the Copernican principle with next-generation surveys}",
      journal = {\aap},
         year = 2023,
        month = mar,
       volume = {671},
          eid = {A68},
        pages = {A68},
          doi = {10.1051/0004-6361/202244557},
archivePrefix = {arXiv},
       eprint = {2207.09995},
 primaryClass = {astro-ph.CO},
       adsurl = {https://scixplorer.org/abs/2023A&A...671A..68C}
}

@ARTICLE{Marra_2010,
       author = {{Marra}, Valerio and {P{\"a}{\"a}kk{\"o}nen}, Mikko},
        title = "{Observational constraints on the {\ensuremath{\Lambda}}LTB model}",
      journal = {\jcap},
         year = 2010,
        month = dec,
       volume = {2010},
       number = {12},
          eid = {021},
        pages = {021},
          doi = {10.1088/1475-7516/2010/12/021},
archivePrefix = {arXiv},
       eprint = {1009.4193},
 primaryClass = {astro-ph.CO},
       adsurl = {https://scixplorer.org/abs/2010JCAP...12..021M}
}

\begin{appendix}

\nolinenumbers

\section{Algebraic relation for the lens--source distance}\label{app:DLS_approx}

\begin{figure}[h!]
    \centering
    \includegraphics[width=\linewidth]{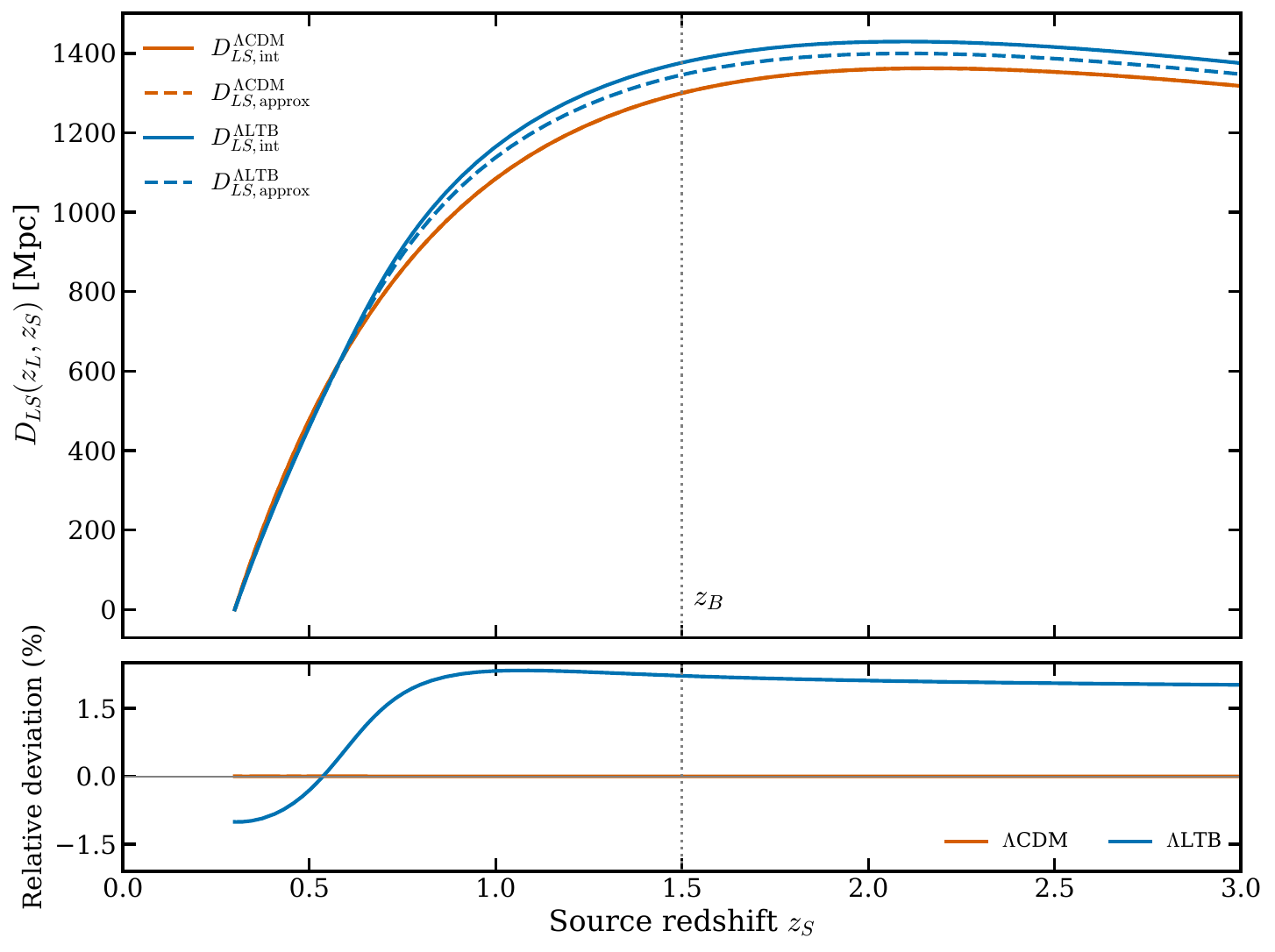}
    \caption{Lens--source angular diameter distances $D_{\rm LS}(z_{\rm L},z_{\rm S})$ for a fixed lens at $z_{\rm L}=0.3$ as a function of the source redshift $z_{\rm S}$, obtained by numerical integration (solid lines) and by the algebraic relation of Eq.~\eqref{Approx_D_Ls} (dashed lines) for the $\Lambda$CDM (orange) and $\Lambda$LTB (blue) models. {\em Top panel}: Distances in Mpc. {\em Bottom panel}: Percentage relative deviation between the two approaches for each model, with the $\Lambda$CDM curve lying on the zero line. The dotted vertical line marks the boundary of the spherical underdensity at $z_B=1.5$.}
    \label{fig:DLS_approx}
\end{figure}

To confirm that Eq.~\eqref{Approx_D_Ls} does not extend to the $\Lambda$LTB model, we compare the values of $D_{\rm LS}$ obtained via numerical integration (Sect.~\ref{DA_integracao}) with those obtained by the algebraic relation for both models, with the lens fixed at $z_{\rm L}=0.3$, well inside the underdensity (Fig.~\ref{fig:DLS_approx}). For Flat-$\Lambda$CDM, the deviation is below $1.4\times10^{-4}$~Mpc, so that Eq.~\eqref{Approx_D_Ls} is mathematically equivalent to the numerical integration, with this residual deviation reflecting only the numerical precision of the integration. For $\Lambda$LTB, in contrast, the approximation fails systematically: the deviation is $-1.0\%$ for sources located immediately behind the lens, changes sign, and reaches a maximum of $2.3\%$ at $z_{\rm S}\approx1.09$, falling to near $2.0\%$ at $z_{\rm S}=3$, while in Mpc it reaches a maximum of $\sim31$~Mpc at $z_{\rm S}\approx1.61$.

\section{Angular diameter distance integrator}\label{app:impl}

\subsection{Implementation}

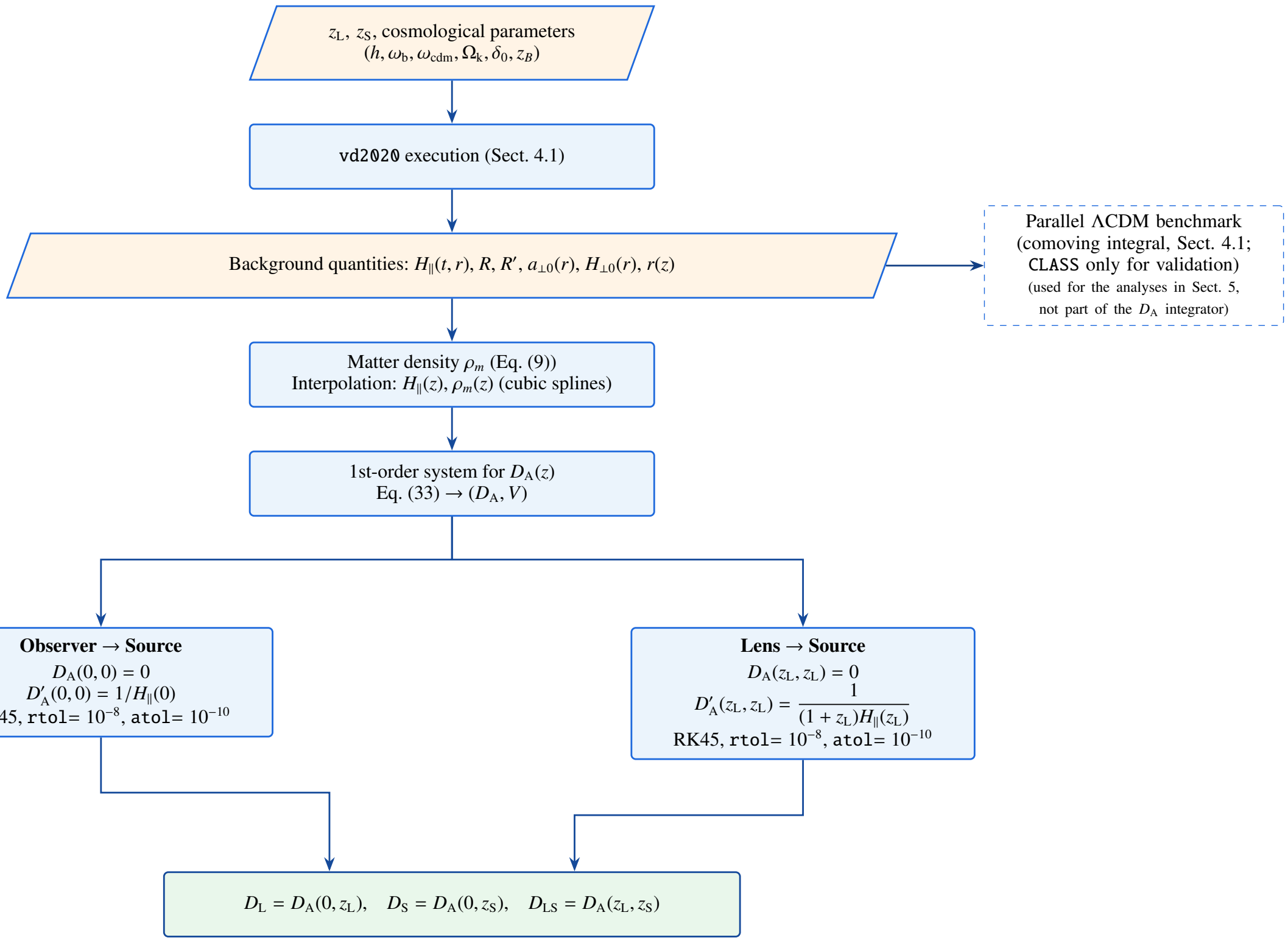
\begin{figure*}[h!]
    \centering
    \resizebox{0.8\textwidth}{!}
    {%
    \begin{tikzpicture}[
        font=\small,
        node distance=7mm and 10mm,
        io/.style={trapezium, trapezium left angle=70, trapezium right angle=110,
            draw=myblue, thick, fill=iofill, minimum width=6.6cm, minimum height=1.05cm,
            align=center, inner sep=4pt},
        proc/.style={rectangle, rounded corners=2pt, draw=myblue, thick, fill=procfill,
            minimum width=6.6cm, minimum height=1.05cm, align=center, inner sep=5pt},
        branch/.style={rectangle, rounded corners=2pt, draw=myblue, thick, fill=procfill,
            minimum width=5.6cm, minimum height=1.5cm, align=center, inner sep=5pt},
        outbox/.style={rectangle, rounded corners=2pt, draw=myblue!70!black, thick, fill=outfill,
            minimum width=6.6cm, minimum height=1.05cm, align=center, inner sep=5pt},
        aux/.style={rectangle, rounded corners=2pt, draw=myblue, dashed, fill=white,
            minimum width=4.6cm, minimum height=1.2cm, align=center, inner sep=4pt},
        arr/.style={-{Stealth[length=2.3mm]}, thick, myblue!70!black}
    ]

    \node[io] (input) {$z_{\rm L},\, z_{\rm S}$, cosmological parameters\\ ($h,\omega_{\text{b}},\omega_{\text{cdm}},\Omega_{\text{k}},\delta_0,z_B$)};

    \node[proc, below=of input] (vd2020) {\texttt{vd2020} execution (Sect.~\ref{Distance})};

    \node[io, below=of vd2020] (bg) {Background quantities: $H_\parallel(t,r)$, $R$, $R'$, $a_{\perp0}(r)$, $H_{\perp0}(r)$, $r(z)$};

    \node[proc, below=of bg] (interp) {Matter density $\rho_m$ (Eq.~\eqref{rho}) \\ Interpolation: $H_\parallel(z)$, $\rho_m(z)$ (cubic splines)};

    \node[proc, below=of interp] (sys) {1st-order system for $D_{\rm A}(z)$ \\ Eq.~\eqref{eq_DA} $\to$ $(D_{\rm A},V)$};

    \node[branch, below left=18mm and -4mm of sys] (branchA)
        {\textbf{Observer $\to$ Source}\\[2pt]
         $D_{\rm A}(0,0)=0$\\
         $D_{\rm A}'(0,0)=1/H_\parallel(0)$\\
         RK45, \texttt{rtol}$=10^{-8}$, \texttt{atol}$=10^{-10}$};

    \node[branch, below right=18mm and -4mm of sys] (branchB)
        {\textbf{Lens $\to$ Source}\\[2pt]
         $D_{\rm A}(z_{\rm L},z_{\rm L})=0$\\
         $D_{\rm A}'(z_{\rm L},z_{\rm L})=\dfrac{1}{(1+z_{\rm L})H_\parallel(z_{\rm L})}$\\
         RK45, \texttt{rtol}$=10^{-8}$, \texttt{atol}$=10^{-10}$};

    \coordinate (midbranch) at ($(branchA.south)!0.5!(branchB.south)$);
    \node[outbox, below=20mm of midbranch, minimum width=9.4cm] (output)
        {$D_{\rm L}=D_{\rm A}(0,z_{\rm L})$,\quad $D_{\rm S}=D_{\rm A}(0,z_{\rm S})$,\quad $D_{\rm LS}=D_{\rm A}(z_{\rm L},z_{\rm S})$};

    \node[aux, right=16mm of bg, text width=4.6cm] (lcdm)
        {Parallel $\Lambda$CDM benchmark \\ (comoving integral, Sect.~\ref{Distance}; \texttt{CLASS} only for validation) \\ \scriptsize (used for the analyses in Sect.~\ref{ch:results}, not part of the $D_{\rm A}$ integrator)};

    \draw[arr] (input) -- (vd2020);
    \draw[arr] (vd2020) -- (bg);
    \draw[arr] (bg) -- (interp);
    \draw[arr] (interp) -- (sys);
    \draw[arr] (sys.south) -- ++(0,-7mm) -| (branchA.north);
    \draw[arr] (sys.south) -- ++(0,-7mm) -| (branchB.north);
    \draw[arr] (branchA.south) -- ++(0,-9mm) -| ($(output.north)+(-2.0,0)$);
    \draw[arr] (branchB.south) -- ++(0,-9mm) -| ($(output.north)+(2.0,0)$);
    \draw[arr, myblue!60!black] (bg.east) -- (lcdm.west);
    \end{tikzpicture}%
    }
    \caption{Flowchart of the numerical procedure used in the integration of $D_{\rm A}$. The background quantities obtained with \texttt{vd2020} are interpolated and used to construct the first-order system associated with Eq.~\eqref{eq_DA}, which is subsequently integrated with the RK45 method in two branches: from the observer to the source and from the lens to the source. The dashed branch corresponds to the parallel reference calculation in $\Lambda$CDM.}
    \label{fig:fluxo_DA}
\end{figure*}

Figure~\ref{fig:fluxo_DA} summarises the numerical procedure used in the integration of $D_{\rm A}$.

The input quantities are defined on a discrete redshift grid and interpolated with cubic splines, so that the integrator can evaluate them at arbitrary $z$. Since the matter density is not an output of \texttt{vd2020}, it is built through numerical differentiation of the mass profile $m(r)$, reconstructed from Eq.~\eqref{Omega_m_k}, following Eq.~\eqref{rho}, and is also interpolated as a function of $z$. 

The code then rewrites Eq.~\eqref{eq_DA} as a system of two first-order equations for $D_{\rm A}$ and $D_{\rm A}'$ and solves it in parallel for the two optical paths, from the observer to the source, $D_{\rm A}(0,z)$, and from the lens to the source, $D_{\rm A}(z_{\rm L},z)$, each with its own initial conditions.
In the first case, the initial conditions of Eq.~\eqref{eq_jacobi_matricial}, $D_{\rm A}=0$ and $dD_{\rm A}/d\lambda=1$ at $\lambda=0$, together with the operator $d/d\lambda=H_\parallel(1+z)^2\,d/dz$, give $D_{\rm A}(0,0)=0$ and $D_{\rm A}'(0,0)=1/H_\parallel(0)$. At the centre the expansion is isotropic, $H_\parallel(0)=H_{\perp0}(0)$, which is the quantity returned by \texttt{vd2020}. In the second case, the integration is restarted at $z_{\rm L}$, defining it as the new vertex of the congruence. This requires a renormalisation of the affine parameter such that $k_\mu u^\mu=1$ at the lens, with the renormalised redshift $1+z^\ast=(1+z)/(1+z_{\rm L})$. The same procedure is then repeated, now centred on the lens, and the chain rule gives the initial conditions in terms of the global variable $z$,
\begin{equation}\label{DA_prime_zL}
    D_{\rm A}(z_{\rm L},z_{\rm L})=0, \qquad D_{\rm A}'(z_{\rm L},z_{\rm L})=\frac{1}{(1+z_{\rm L})\,H_\parallel(z_{\rm L})}.
\end{equation}

These integrations yield the distances $D_{\rm L}=D_{\rm A}(0,z_{\rm L})$, $D_{\rm S}=D_{\rm A}(0,z_{\rm S})$, and $D_{\rm LS}=D_{\rm A}(z_{\rm L},z_{\rm S})$, which are used as inputs to our lensing image simulator described in Sect.~\ref{Lens_Ger}. The two integrations follow the same numerical method, an adaptive Runge--Kutta integrator (RK45) with relative and absolute tolerances $\texttt{rtol}=10^{-8}$ and $\texttt{atol}=10^{-10}$.

In parallel, the code computes the distances of the reference Flat-$\Lambda$CDM model, with the same content as the $\Lambda$LTB background ($\Omega_{\rm r}=0$), through the comoving integral of Eq.~\eqref{D_C}, converted to angular diameter distances, and $D_{\rm LS}$ through the same integral between $z_{\rm L}$ and $z_{\rm S}$, using only the background module of \texttt{CLASS} \citep{CLASS_Lesgourgues_2011}, on order to validate these values.

\subsection{Validation}

\begin{figure}[h!]
    \centering
    \includegraphics[width=\linewidth]{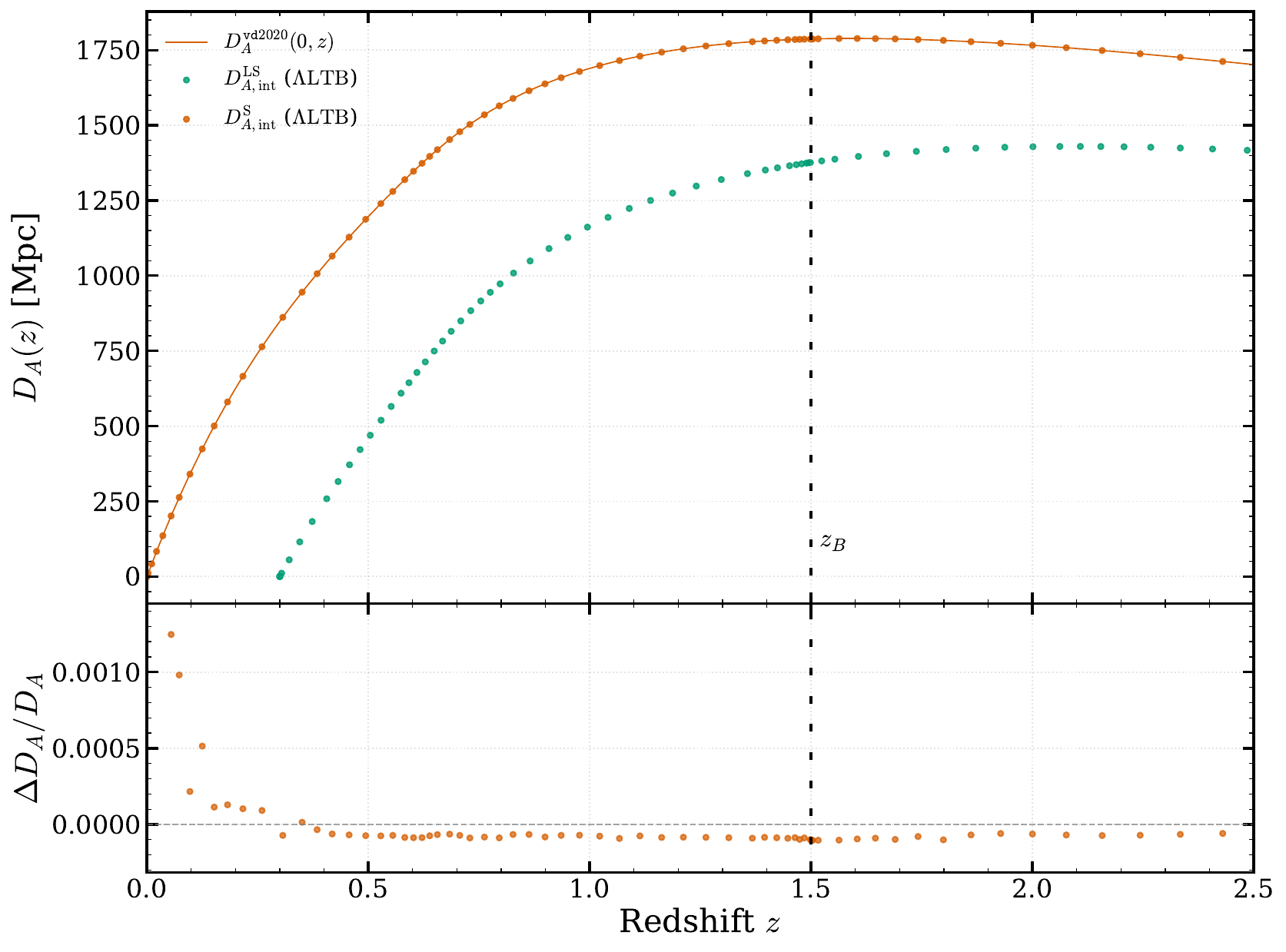}
    \caption{Cross-validation of the numerical integration of $D_{\rm A}$ with the results from \texttt{vd2020}. \textit{Top panel}: Angular diameter distances obtained for the observer--source (orange dots) and lens--source (green dots) branches of the integrator, and reference observer--source distance from \texttt{vd2020} (line). \textit{Bottom panel}: Relative deviation $\Delta D_{\rm A}/D_{\rm A}$ between the integrator results and the reference values. The vertical dashed line marks the underdensity boundary $z_B$.}
    \label{D_A_val}
\end{figure}

 The integrator was validated against a second-order linear ODE of the same mathematical class as Eq.~\eqref{eq_DA} with a known analytical solution, with a maximum absolute difference of $2.5\times10^{-10}$, and against the distances $D_{\rm A}(0,z)$ calculated directly by \texttt{vd2020}, with a relative deviation below $0.0013$ for $z>0.05$ (Fig.~\ref{D_A_val}). At lower redshift, the relative measure diverges because $D_{\rm A}\to0$; at the first grid point, where $D_{\rm A}\approx0.01$~Mpc, the absolute deviation is $2\times10^{-4}$~Mpc.
Given that both branches of the integrator share the same system and numerical method, diverging only in the initial conditions, this agreement validates the implementation as a whole, including the calculation of $D_{\rm LS}$.

\section{Strong lensing image simulator}\label{app:implimg}

\subsection{Implementation}

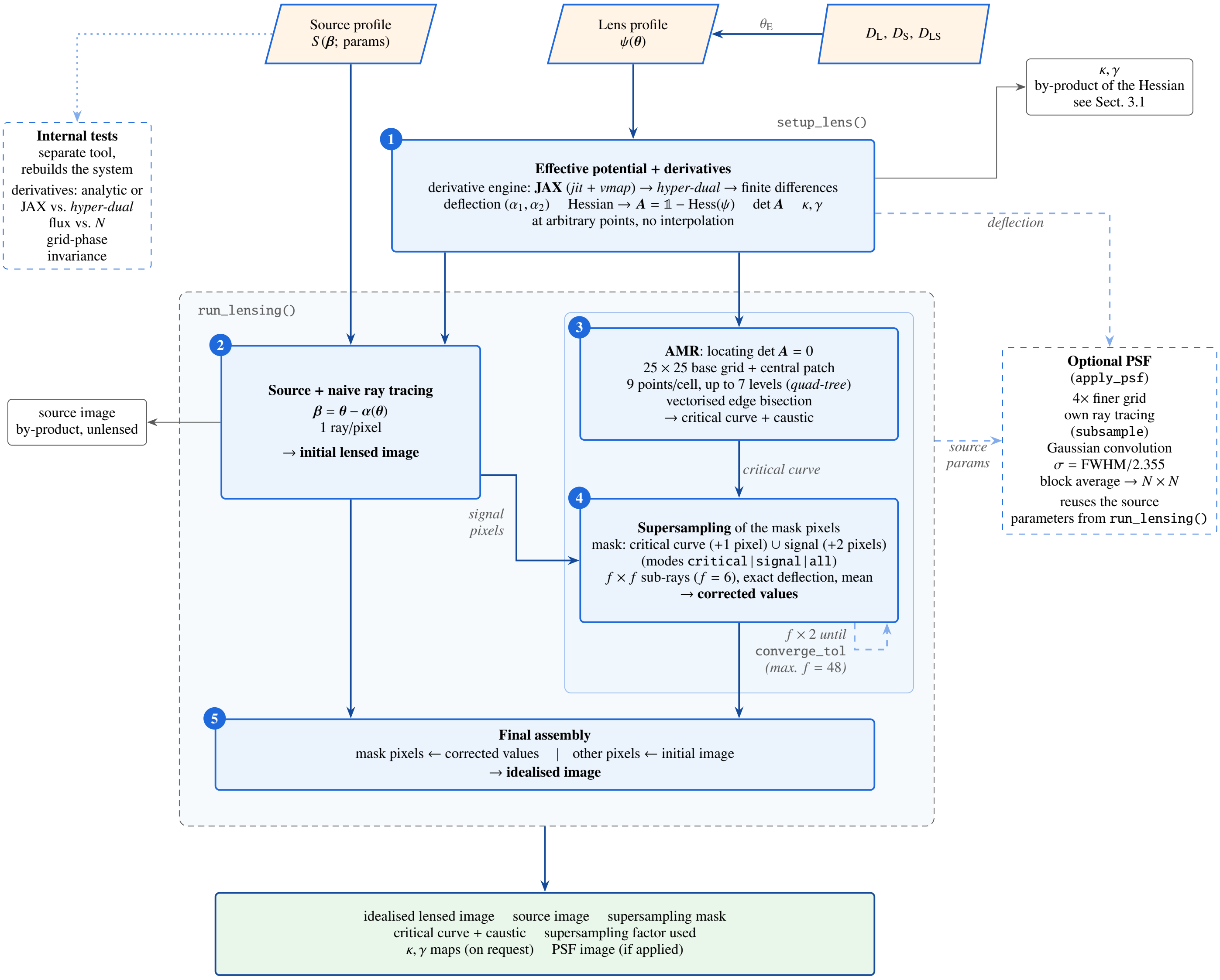
\begin{figure*}[h!]
    \centering
    \resizebox{\textwidth}{!}{%
    \begin{tikzpicture}[
        font=\scriptsize,
        io/.style={trapezium, trapezium left angle=70, trapezium right angle=110, trapezium stretches body,
            draw=myblue, thick, fill=iofill, minimum height=1.0cm, minimum width=2.9cm, align=center, inner xsep=2pt},
        proc/.style={rectangle, rounded corners=2pt, draw=myblue, thick, fill=procfill,
            align=center, inner sep=5pt},
        outbox/.style={rectangle, rounded corners=2pt, draw=myblue!70!black, thick, fill=outfill,
            align=center, inner sep=6pt},
        aux/.style={rectangle, rounded corners=2pt, draw=myblue, dashed, fill=white,
            align=center, inner sep=4pt},
        sideout/.style={rectangle, rounded corners=2pt, draw=sideln, thin, fill=white,
            align=center, inner sep=4pt},
        tag/.style={circle, fill=myblue, text=white, font=\scriptsize\bfseries,
            inner sep=0pt, minimum size=3.8mm},
        arr/.style={-{Stealth[length=2.0mm]}, thick, myblue!70!black},
        darr/.style={-{Stealth[length=2.0mm]}, thick, myblue!55, dashed},
        cutarr/.style={-{Stealth[length=2.0mm]}, thick, myblue!55, dash pattern=on 1pt off 2pt},
        sarr/.style={-{Stealth[length=1.8mm]}, thin, sideln},
        lbl/.style={font=\scriptsize\itshape, text=sideln, inner sep=1.5pt},
        fn/.style={font=\scriptsize\ttfamily, text=sideln},
    ]


    \node[io] (in1) at (0,0)   {Source profile \\ $S(\bm\beta;\,\text{params})$};
    \node[io] (in2) at (4.8,0) {Lens profile \\ $\psi(\bm\theta)$};
    \node[io] (in3) at (9.4,0) {$D_{\rm L},\,D_{\rm S},\,D_{\rm LS}$};

    \node[proc, minimum width=8.2cm, minimum height=1.9cm] (box1) at (4.8,-2.75)
        {\textbf{Effective potential + derivatives} \\[1pt]
         derivative engine: \textbf{JAX} (\textit{jit} + \textit{vmap}) $\to$ \textit{hyper-dual} $\to$ finite differences \\
         deflection $(\alpha_1,\alpha_2)$ \quad Hessian $\to\bm A=\mathbb{1}-\mathrm{Hess}(\psi)$ \quad $\det\bm A$ \quad $\kappa,\gamma$ \\
         at arbitrary points, no interpolation};
    \node[fn, anchor=south east] (lsetup) at ($(box1.north east)+(0,0.05)$) {setup\_lens()};

    \node[proc, minimum width=4.4cm, minimum height=2.6cm] (box2) at (0,-6.6)
        {\textbf{Source + naive ray tracing} \\[2pt]
         $\bm\beta=\bm\theta-\bm\alpha(\bm\theta)$ \\
         1 ray/pixel \\[4pt]
         $\to$ \textbf{initial lensed image}};

    \node[proc, minimum width=5.4cm, minimum height=1.9cm] (box3) at (6.6,-5.95)
        {\textbf{AMR}: locating $\det\bm A=0$ \\
         $25\times25$ base grid + central patch \\
         9 points/cell, up to 7 levels (\textit{quad-tree}) \\
         vectorised edge bisection \\
         $\to$ critical curve + caustic};

    \node[proc, minimum width=5.4cm, minimum height=2.1cm] (box4) at (6.6,-8.95)
        {\textbf{Supersampling} of the mask pixels \\
         mask: critical curve ($+1$ pixel) $\cup$ signal ($+2$ pixels) \\
         (modes \texttt{critical}\,|\,\texttt{signal}\,|\,\texttt{all}) \\
         $f\times f$ sub-rays ($f=6$), exact deflection, mean \\
         $\to$ \textbf{corrected values}};

    \coordinate (lpA) at ($(box4.south east)+(-0.75,0)$);
    \coordinate (lpB) at ($(box4.south east)+(-0.2,0)$);
    \draw[darr] (lpA) -- ++(0,-0.45) -| (lpB);
    \node[lbl, anchor=north east, align=right] (lpL) at ($(lpA)+(-0.08,-0.04)$) {$f\times2$ until\\\texttt{\upshape converge\_tol}\\(max.\ $f=48$)};
    \coordinate (lpBot) at ($(lpA)+(0,-0.5)$);

    \node[proc, minimum width=11.2cm, minimum height=1.2cm] (box5) at (3.3,-12.25)
        {\textbf{Final assembly} \\[1pt]
         mask pixels $\leftarrow$ corrected values \quad\vline\quad other pixels $\leftarrow$ initial image \\[1pt]
         $\to$ \textbf{idealised image}};

    \begin{scope}[on background layer]
    \node[draw=sideln, dash pattern=on 3pt off 2pt, rounded corners=4pt,
          fit=(box2)(box3)(box4)(lpBot)(box5), inner sep=6mm, fill=wrapfill] (wrap) {};
    \node[draw=myblue!45, rounded corners=3pt, fit=(box3)(box4)(lpL)(lpBot), inner sep=2.5mm, fill=rcolfill] (rcol) {};
    \end{scope}
    \node[fn, anchor=north west] (lrun) at ($(wrap.north west)+(2mm,-1mm)$) {run\_lensing()};

    \node[outbox, minimum width=11.2cm, minimum height=1.4cm] (output) at (3.3,-15.3)
        {idealised lensed image \quad source image \quad supersampling mask \\
         critical curve + caustic \quad supersampling factor used \\
         $\kappa,\gamma$ maps (on request) \quad PSF image (if applied)};

    \node[aux, anchor=west] (psf) at ($(wrap.east |- box3.south)+(1.15,0)$)
        {\textbf{Optional PSF} \\ (\texttt{apply\_psf}) \\[2pt]
         $4\times$ finer grid \\ own ray tracing \\ (\texttt{subsample}) \\
         Gaussian convolution \\ $\sigma=\mathrm{FWHM}/2.355$ \\ block average $\to N\times N$ \\[2pt]
         reuses the source \\ parameters from \texttt{run\_lensing()}};

    \node[sideout] (kg) at (psf.north |- 0,-0.9)
        {$\kappa,\gamma$ \\ by-product of the Hessian \\ see Sect.~\ref{Strong_Lensing}};

    \node[aux] (tests) at (-4.65,-2.75)
        {\textbf{Internal tests} \\ separate tool, \\ rebuilds the system \\[2pt]
         derivatives: analytic or \\ JAX vs.\ \textit{hyper-dual} \\ flux vs.\ $N$ \\ grid-phase \\ invariance};

    \node[sideout] (src) at (-4.65,-6.6)
        {source image \\ by-product, unlensed};

    \node[tag] (tg1) at (box1.north west) {1};
    \node[tag] (tg2) at (box2.north west) {2};
    \node[tag] (tg3) at (box3.north west) {3};
    \node[tag] (tg4) at (box4.north west) {4};
    \node[tag] (tg5) at (box5.north west) {5};

    \draw[arr] (in3) -- node[lbl, above] (lthE) {$\theta_{\rm E}$} (in2);
    \draw[arr] (in2) -- (box1.north -| in2);
    \draw[arr] (in1) -- (box2);

    \draw[arr] (box1.south -| 1.6,0) -- (box2.north -| 1.6,0);
    \draw[arr] (box1.south -| box3.north) -- (box3.north);

    \draw[arr] (box3) -- node[lbl, right] (lcc) {critical curve} (box4);

    \coordinate (mk) at ($(box2.east)+(0,-0.9)$);
    \draw[arr] (mk) -- ($(mk)+(0.6,0)$) |- (box4.west);
    \node[lbl, anchor=north east, align=right] (lsig) at ($(box2.south east)+(0.45,-0.1)$) {signal\\pixels};

    \draw[arr] (box2.south) -- (box2.south |- box5.north);
    \draw[arr] (box4.south) -- (box4.south |- box5.north);

    \draw[arr] (wrap.south -| output.north) -- (output.north);

    \draw[sarr] ($(box1.east)+(0,0.3)$) -| ($(kg.west)+(-0.5,0)$) -- (kg.west);
    \draw[sarr] (box2.west) -- (src.east);
    \draw[darr] ($(box1.east)+(0,-0.3)$) -| node[lbl, pos=0.3, below] (ldef) {deflection} (psf.north);
    \draw[darr] (wrap.east |- psf.west) -- node[lbl, below, align=center] (lpar) {source\\params} (psf.west);
    \draw[cutarr] (in1.west) -| (tests.north);

    \end{tikzpicture}%
    }
    \caption{Flowchart of the strong lensing image-mapping procedure. The derivatives of the effective potential are computed by automatic differentiation (JAX, with hyper-dual numbers \citep{Fike2011} and finite differences as alternatives) at arbitrary points, feeding the initial ray tracing, the location of the critical curves by AMR, the supersampling and the PSF. Supersampling is applied to the pixels of a mask defined by the critical curve and, by default, by the pixels with signal in the initial image, followed by the assembly of the final image. Numbers (1)--(5) label the processing blocks in data-flow order. The PSF application and the internal consistency tests are auxiliary branches to the main flow of \texttt{run\_lensing()}.}
    \label{fig:fluxo_lensing}
\end{figure*}

Figure~\ref{fig:fluxo_lensing} illustrates the data flow and the architecture of the simulator.

\begin{table}[ht]
    \centering
    \caption{Validation tests of the image simulator}
    \label{tab:validacao_lensing}
    \begin{tabular}{p{6.0cm}c}
        \toprule
        Test (reference) & Error \\
        \midrule
        \multicolumn{2}{l}{\textit{Lens} (analytical expressions)} \\
        \quad Deflection, $\kappa$, $\gamma$, and $\det\bm A$ & $5.3\times10^{-14}$ \\
        \quad Critical and caustic curves & $\ang{;;2.8}\times10^{-15}$ \\
        \quad Ring position relative to the tangential critical curve & $\ang{;;3.0}\times10^{-4}$ \\
        \addlinespace
        \multicolumn{2}{l}{\textit{Image} (exact solution)} \\
        \quad Flux & $8.2\times10^{-7}$ \\
        \quad Pixel-by-pixel brightness, w/o centre & $6.3\times10^{-5}$ \\
        \quad Pixel-by-pixel brightness, with centre & $3.7\times10^{-4}$ \\
        \addlinespace
        \multicolumn{2}{l}{\textit{Discretisation}} \\
        \quad Flux with $N=800$ vs.\ $N=1600$ & $2.7\times10^{-10}$ \\
        \quad Flux with half-pixel shifted grid & $8.0\times10^{-8}$ \\
        \quad Image maximum, half-pixel shifted grid & $1.1\times10^{-6}$ \\
        \quad Azimuthal asymmetry, $|a_4|/a_0$ & $8.1\times10^{-4}$ \\
        \quad PSF with $4\times$ vs.\ $8\times$ finer grid & $5.9\times10^{-5}$ \\
        \bottomrule
    \end{tabular}
     \tablefoot{Tests are applied to the optimal $\Lambda$LTB lensing configuration from Table~\ref{tab:config_selecionada} ($\theta_{\rm E}=0.839''$) with the parameters of Sect.~\ref{Lens_Ger}. The lens errors are absolute, while the others are relative, with the brightness errors normalised to the maximum of the exact image. Pixel-by-pixel brightness without centre refers to discarding the central region and considering only $\theta>0.15''$.}
\end{table}

The derivatives of the effective lens potential are evaluated at any point without interpolation, by automatic differentiation \citep{jax2018github}, giving the deflection field and the amplification matrix $\bm A$ of Eq.~\eqref{amplificacao}. The code treats the lens potential in a general way, allowing for any lens mass profile.
The resulting ray-tracing calculation applies the lens equation, Eq.~\eqref{Eq.Lente}, to each point of a Cartesian grid of $N\times N$ pixels in the lens plane, while the critical curves, where $\det\bm A=0$, are located with high precision via adaptive mesh refinement (AMR), independently of the pixel grid, and mapped onto the source plane to trace the caustic curves. Since the magnification diverges at the critical curves, the pixels with signal and those of the critical curves are corrected by supersampling, being described by the average brightness over the pixel instead of the point sample at its centre. 

The images are simulated on a field of side $L=10''$ with $800\times800$ pixels ($0.0125''$ per pixel), a supersampling factor $f=6$, and, when applied, a Gaussian PSF with an FWHM of $0.1''$, computed on a grid four times finer. This width 
is of the order of the FWHM of the PSF of the Hubble Space Telescope's Advanced Camera for Surveys \citep[Sect.~5.6.6]{ACS_IHB}, which is a higher resolution than the $0.18''$ resolution of the visible camera on \textit{Euclid} \citep{Euclid_Vis2025}. These parameters, common to the $\Lambda$CDM and $\Lambda$LTB models and used in the code validation and in all simulated configurations, ensure that the ring is resolved by multiple pixels.

The generic utility layer that computes $\kappa$, $\gamma$, and $\det\bm A$, as well as the ray-tracing routine, stems from a Python computational application example presented by \citet{Meneghetti_2022}, while all the other methods in the code, which constitute the core of this implementation, are original to this work.

\subsection{Validation}

Table~\ref{tab:validacao_lensing} summarises the validation of the image simulator for the optimal $\Lambda$LTB lensing configuration of Sect.~\ref{SL_map}. 

Regarding the lens, the deflection, $\kappa$, $\gamma$, and $\det\bm A$ are compared with the analytical expressions of the NIS at 2000 random grid points, with the largest absolute difference among all these quantities being reported, and the radius of each point of the critical curves found by AMR, as well as its mapping onto the source plane, is compared with the corresponding analytical one. The position of the ring is obtained from the mean radial profile of the image, calculated by the azimuthal average of the brightness in radial intervals half a pixel wide, with its maximum located with sub-pixel precision. 
Regarding the image, the reference is the exact lensed solution, whose integral serves as a comparison for the image flux, while for the brightness in each pixel the exact average of the lensed source over its area is calculated with $24\times24$ sub-rays. Since the strongly demagnified central odd image is smaller than one pixel, this deviation is reported with and without the central region. 
To test the dependence on the grid discretisation, the same system was recalculated on a grid with twice the resolution and on a grid shifted by half a pixel in both directions. The azimuthal symmetry of the ring was tested by decomposing the brightness along the ring into azimuthal harmonics $a_m$: since a Cartesian grid is invariant only to $90^\circ$ rotations, a square pixel signature remains in $|a_4|/a_0$, its value below $10^{-3}$ legitimises the use of the mean radial profile in Sect.~\ref{SL_map}. 
Finally, the PSF computed on a grid four times finer was tested against an eight times finer reference grid.

\section{Selection of the optimal configuration}\label{app:selecao}

This appendix describes the selection of the optimal lensing configuration used in the analysis of Sect.~\ref{ch:results}, that is the 
combination of lens and source redshifts that maximises the difference
between the lensing signal of the two models without compromising the efficiency of the lens. 

We adopted the following five methods to select the optimal configuration. They are constructed from the two metrics defined in Sect.~\ref{Deff_res}, namely the efficiency $\mathcal{G}$ that can be defined  globally $(E)$ or locally $(\epsilon)$ and the  difference in $D_{\rm eff}$ between the two cosmological models $(\Delta D_{\rm eff})$, which has a positive and a negative regime.

\begin{figure*}[h!]
    \centering
    \includegraphics[width=0.9\textwidth]{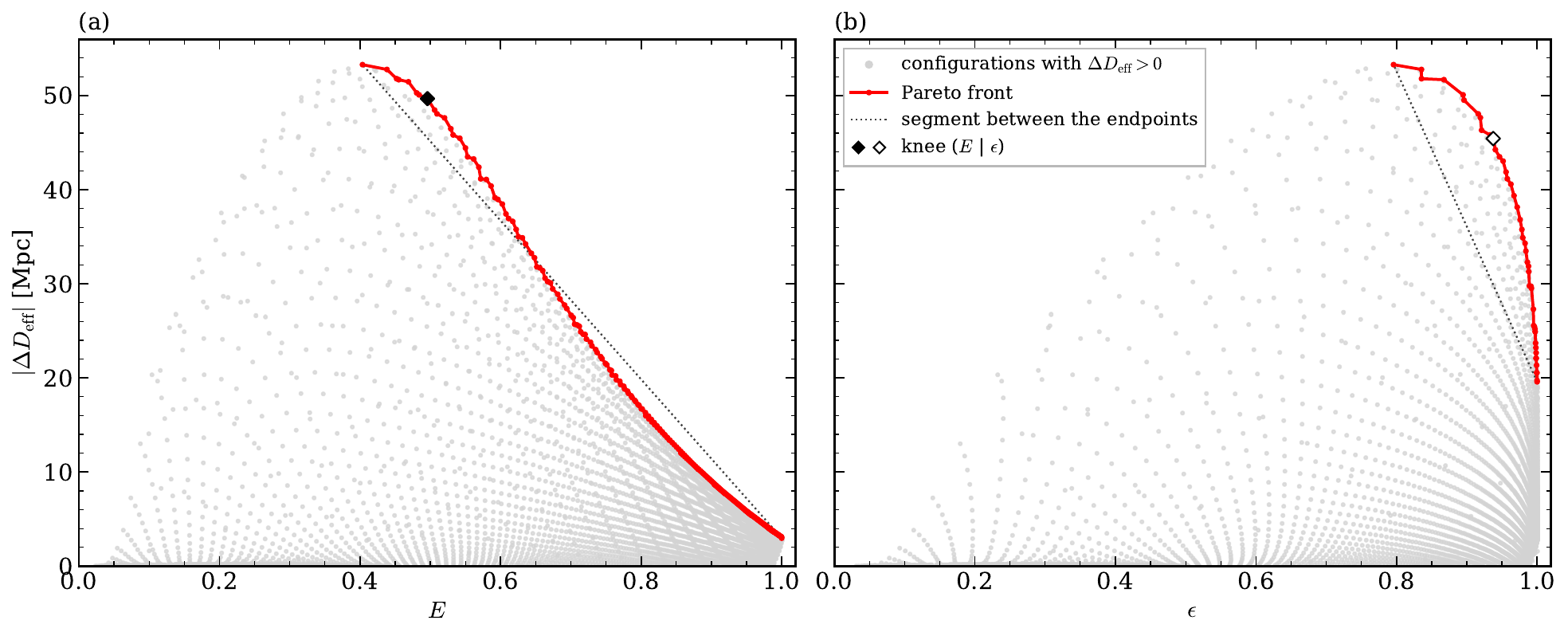}
    \caption{Pareto fronts in the positive regime ($\Delta D_{\rm eff}>0$) on the $(E,|\Delta D_{\rm eff}|)$ plane ({\em left panel}) and on the $(\epsilon,|\Delta D_{\rm eff}|)$ plane ({\em right panel}). The dots correspond to the grid configurations, the red line is the front of non-dominated configurations, and the diamond marks the knee (black: using $E$; white: using $\epsilon$).}
    \label{fig:pareto}
\end{figure*}

\begin{itemize}
\item Maximum difference: The global extremum of $\Delta D_{\rm eff}$ for each regime, 
$
    (z_{\rm S},z_{\rm L})_{\pm} = \arg\max_{(z_{\rm S},z_{\rm L})} \left(\pm\,\Delta D_{\rm eff}\right).
$
\item Efficiency ridge: Considers that, for each source, the lens is at the optimal position on the ridge where $\epsilon=1$, that is,
$
    z_{\rm L}^{\max}(z_{\rm S})=\arg\max_{z_{\rm L}}D_{\rm eff}^{\Lambda\mathrm{LTB}}\,.
$
\item Weighting: Combines both metrics into a scalar objective function
$
    F_{\mathcal{G}}(z_{\rm S},z_{\rm L}) = |\Delta D_{\rm eff}| \;\mathcal{G}(z_{\rm S}, z_{\rm L})
$
The candidates are chosen by the maximum of $F_{\mathcal{G}}(z_{\rm S},z_{\rm L})$ for each regime. Thus $F_{\mathcal{G}}(z_{\rm S},z_{\rm L})$ has units of $\Delta D_{\rm eff}$ and penalises poor efficient configurations. 

\item Restriction: In this method, instead of weighting the metrics against each other, a restriction is made in the parameter space by imposing a threshold $\eta$ on the efficiency $\mathcal{G}(z_{\rm S},z_{\rm L})$. In this subspace, the extremum of $\Delta D_{\rm eff}$ is sought
$
    (z_{\rm S},z_{\rm L})_{\pm} = \arg\max_{\mathcal{G}(z_{\rm S},z_{\rm L})\,\geq\,0.8} \left(\pm\,\Delta D_{\rm eff}\right).
   $ 
The choice of the threshold value $0.8$ corresponds to a moderate tolerance. Values considerably below this would lead the method to reproduce the maximum difference method, whereas for values considerably above it, the method tends to collapse onto the efficiency ridge. 

\item Pareto's knee: 
This method attempts to maximise $\mathcal{G}$ and $|\Delta D_{\rm eff}|$ simultaneously. In this regard, each configuration $(z_{\rm S}, z_{\rm L})$ on the grid represents a point on the $(\mathcal G,|\Delta D_{\rm eff}|)$ plane, as shown in Fig.~\ref{fig:pareto} separately for $E$ and $\epsilon$, always for the positive regime, since in the negative regime $|\Delta D_{\rm eff}|$ grows up to the limit $z_{\rm S}=10$ of the grid, as discussed below. A configuration is classified as ``dominated'' if there is another one that is not inferior in either of the two metrics and is strictly superior in at least one. These points form the cloud of grey dots in the figure and are excluded from the choice since there is at least one configuration that is not worse in either metric. The remaining set of points, that is, the ``non-dominated'' ones, forms the Pareto front \citep{Pareto1906, Miettinen1998}, marked in red in the figure, where any gain in one metric implies a loss in the other. Its two extremes are the configuration points of highest $\mathcal{G}$ and highest $|\Delta D_{\rm eff}|$, connected by the dotted line segment seen in the figure. From it, the candidate is chosen by taking Pareto's ``knee'' \citep{Das1999}, that is, the point on the front with the greatest perpendicular distance to the segment connecting the two extremes of the front. Note that the method is invariant to independent scale changes on each axis and therefore does not depend on the units of the metrics.

\end{itemize}

\begin{figure*}[h!]
    \centering
    \includegraphics[width=0.9\textwidth]{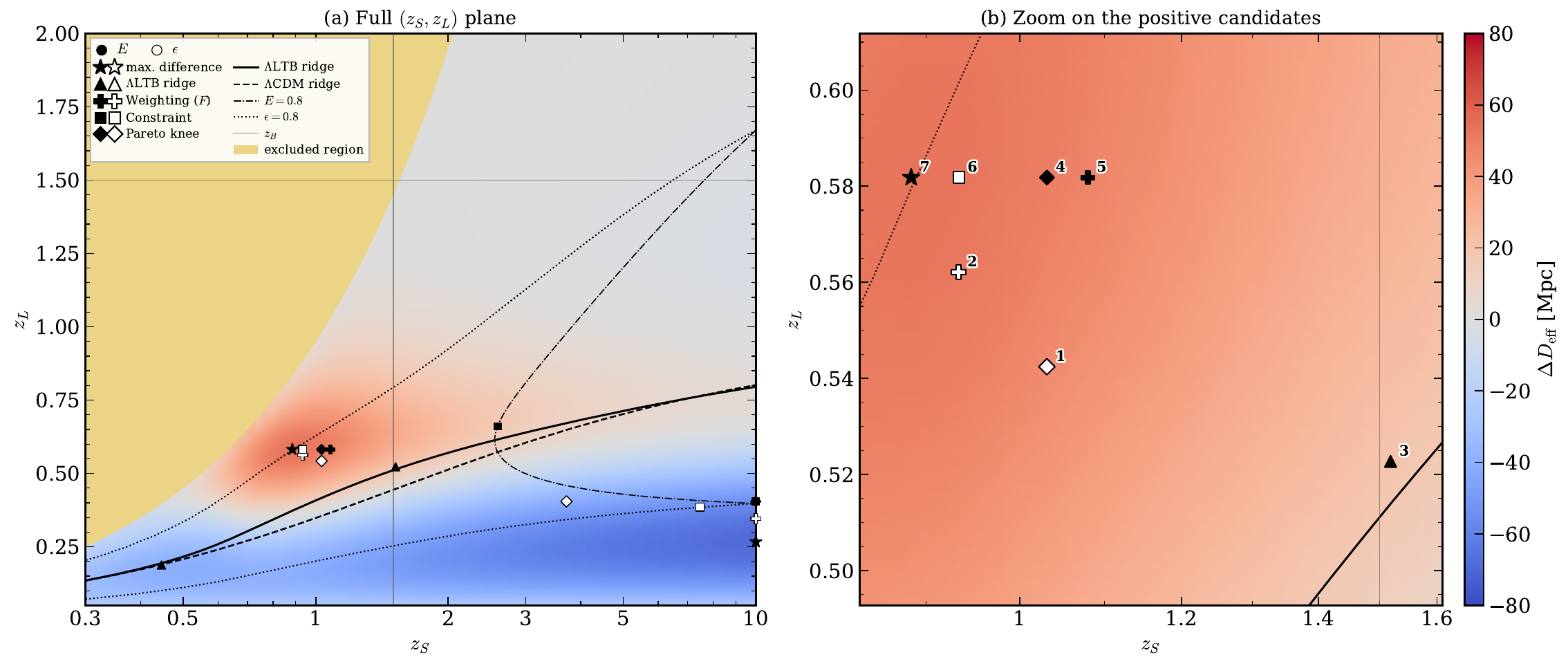}
    \caption{Difference of the effective lensing distance $\Delta D_{\rm eff}$ (in Mpc) between the two models in the $(z_{\rm S},z_{\rm L})$ plane, with $z_{\rm S}$ on a logarithmic scale. The best candidates resulting from the sixteen search strategies are overlaid on the map. {\em Left panel}: Full plane probed. {\em Right panel}: Detail of the high positive region, with the candidates labelled by the ranking of Table~\ref{tab:decisao}. The marker's shape identifies the method and its colour identifies the efficiency parameter (black: $E$; white: $\epsilon$). The solid and dashed lines are the $\Lambda$LTB and $\Lambda$CDM ridges, respectively; the dash-dot and dotted lines delimit the $E\geq0.8$ and $\epsilon\geq0.8$ regions. The yellow region, $z_{\rm L}\geq z_{\rm S}-0.05$, in the upper-left corner of the left panel, was excluded and the three candidates at $z_{\rm S}=10$ were excluded from the final decision process. The thin grey lines mark $z_B=1.5$.}
    \label{fig:candidatos}
\end{figure*}

This results in 16 distinct criteria. Some of these criteria lead to the same point in the parameter space. In the negative regime, the restriction and Pareto's knee with $E$ coincide, and the weighting yields the same point with both efficiencies, a candidate at the edge $z_{\rm S}=10$, where the column maximum is also the grid maximum and therefore $E=\epsilon$. This reduces the number of distinct candidates to the 14 shown in Fig.~\ref{fig:candidatos}, where they are overlaid on the map of $\Delta D_{\rm eff}$. 

For the positive regime, all candidates but one tend to occupy the same region, delimited by $z_{\rm S}\approx0.9$--$1.5$ and $z_{\rm L}\approx0.52$--$0.58$, with lens and source inside the underdensity, which reinforces the confidence in the strategy implemented in the search for candidates. 
The negative regime ($\Delta D_{\rm eff}<0$) corresponds to lenses close to the observer, a result of the deficit $D_{\rm L}^{\Lambda\mathrm{LTB}}<D_{\rm L}^{\Lambda\mathrm{CDM}}$ inside the underdensity, while the positive regime ($\Delta D_{\rm eff}>0$) appears above it, peaking at $+53.3$~Mpc at $(z_{\rm S},z_{\rm L})=(0.885,\,0.582)$. For lenses close to the observer, the relative deviation of $D_{\rm eff}$ remains around $-12\%$ regardless of the source redshift, so that, since $D_{\rm eff}$ grows with $z_{\rm S}$, $|\Delta D_{\rm eff}|$ also grows until reaching the edge of the grid, which explains the three negative candidates at $z_{\rm S}=10$, the upper limit of the grid, an extremum imposed by the grid limits and not by the physics of the problem, and were therefore excluded.

 The eleven remaining candidates were then tested at the image level, simulating the gravitational lens images for each cosmological model with the simulator and parameters of Sect.~\ref{Lens_Ger}, so that the difference between the obtained images is solely due to the different redshift values of the candidates. The radius of the tangential critical curve, $\theta_{\rm crit}$, which coincides with the Einstein ring, is calculated with sub-pixel precision and constitutes the ultimate selection criterion, the best candidate being the one that produces the largest $|\Delta\theta_{\rm crit}|=|\theta_{\rm crit}^{\Lambda\mathrm{LTB}}-\theta_{\rm crit}^{\Lambda\mathrm{CDM}}|$.
 They are shown in Table~\ref{tab:decisao}, sorted by $\Delta\theta_{\rm crit}$. The top ranked candidate is the one selected as the optimal configuration for the analysis of Sect.~\ref{ch:results}. 
 This order does not follow the sorting by $\Delta D_{\rm eff}$, which shows that a difference in $D_{\rm eff}$ does not, on its own, translate into an observable difference. 
 
 We note that the seven best candidates, with $\Delta\theta_{\rm crit}$ between $0.108''$ and $0.123''$, belong to the positive regime. We also note that the highest rank in the ridge criteria has the smallest difference in distance ($+19.6$~Mpc), but the third largest difference in ring size difference, while the candidate with the largest difference in distance only appears in the seventh position. Despite their $\Delta D_{\rm eff}<0$, the negative regime candidates have a $\Lambda$LTB Einstein ring that is angularly larger than that of $\Lambda$CDM, even though it is physically smaller, since for an isothermal lens with a fixed $\sigma_v$ the angular radius depends only on $D_{\rm LS}/D_{\rm S}$.  

\begin{table}[h!]
    \centering
    \footnotesize
    \setlength{\tabcolsep}{3.5pt}
    \caption{The eleven final candidates ranked by $|\Delta\theta_{\rm crit}|$}
    \label{tab:decisao}
    \begin{tabular}{rlccrcr}
        \hline
        \# & Strategy & $z_{\rm S}$ & $z_{\rm L}$ & $\Delta D_{\rm eff}$ & $\epsilon$ & $\Delta\theta_{\rm crit}$ \\
        \hline
        \textbf{1} & \textbf{Pareto, $\epsilon$ ($+$)} & \textbf{1.031} & \textbf{0.542} & $\mathbf{+45.4}$ & \textbf{0.938} & $\mathbf{+0.123}$ \\
        2  & Weighting, $\epsilon$ ($+$)  & 0.934 & 0.562 & $+51.7$ & 0.867 & $+0.117$ \\
        3  & Ridge ($+$)                  & 1.519 & 0.523 & $+19.6$ & 1.000 & $+0.114$ \\
        4  & Pareto, $E$ ($+$)             & 1.031 & 0.582 & $+49.7$ & 0.892 & $+0.113$ \\
        5  & Weighting, $E$ ($+$)         & 1.080 & 0.582 & $+47.6$ & 0.911 & $+0.113$ \\
        6  & Restriction, $\epsilon$ ($+$)   & 0.934 & 0.582 & $+52.8$ & 0.835 & $+0.111$ \\
        7  & Maximum difference ($+$)        & 0.885 & 0.582 & $+53.3$ & 0.796 & $+0.108$ \\
        8  & Pareto, $\epsilon$ ($-$)      & 3.712 & 0.405 & $-40.9$ & 0.869 & $+0.086$ \\
        9  & Restriction, $\epsilon$ ($-$)   & 7.465 & 0.385 & $-55.0$ & 0.800 & $+0.075$ \\
        10 & Restriction, $E$ ($+$)          & 2.591 & 0.661 & $+16.7$ & 0.996 & $+0.071$ \\
        11 & Ridge ($-$)                  & 0.446 & 0.188 & $-39.3$ & 1.000 & $+0.014$ \\
        \hline
    \end{tabular}
     \tablefoot{Each case is the best candidate of a strategy (that consists of a method, efficiency type, and regime).  Redshifts, $\Delta D_{\rm eff}$ in Mpc, positional efficiency value, and $\Delta\theta_{\rm crit}$ in arcseconds are given. The selected configuration is highlighted.}
\end{table}

\end{appendix}

\end{document}